\documentclass[a4paper,10pt]{modreport}
\usepackage{feynmf}
\usepackage{epsfig}
\usepackage{mychapter}

\begin{document}

\pagenumbering{Roman}
\setcounter{page}{1}

\begin{center}
UNIVERSITY OF HELSINKI \hspace{2.2cm} REPORT SERIES IN PHYSICS 
\vspace{1cm} \\
HU-P-D100 
\vspace{3.5cm} \\
{\Large \bf STRONG AND ELECTROMAGNETIC}
\vspace{.1cm} \\
{\Large \bf TRANSITIONS IN HEAVY FLAVOR MESONS}
\vspace{1.5cm} \\
{\large \bf TIMO L\"AHDE}
\vspace{1.5cm} \\
{\normalsize \it Faculty of Science} \\
{\normalsize \it Helsinki Institute of Physics and Department of
Physical Sciences,} \\
{\normalsize \it PL 64 University of Helsinki, 00014 Helsinki, Finland} \\
{\normalsize e-mail: talahde@pcu.helsinki.fi}

\vspace{4cm} 

{\it ACADEMIC DISSERTATION} \vspace{.3cm} \\
\parbox{10cm}
{\it To be presented, with the permission of the Faculty of Science of the 
University of Helsinki, for public criticism in Auditorium E204 of the 
Department of Physical Sciences, on December 13th, 2002, at noon.}
\vspace{2cm} \\
Helsinki 2002
\end{center}

\thispagestyle{empty}

\newpage

\thispagestyle{empty}

~\vspace{2cm}
\begin{center}
\parbox{12cm}
{\large This thesis is dedicated to all those who have shown interest in 
my work and encouraged me when progress has been slow.}
\vspace{1cm}

\hspace{-5cm}\parbox{7cm}
{\it Ut desint vires, \\
tamen est laudanda voluntas.}
\vspace{10cm}

ISBN 952-10-0563-7 \\
ISSN 0356-0961 \\
Helsinki 2002 \\
Yliopistopaino

\vspace{1cm}
ISBN 952-10-0564-5 (PDF version) \\
Helsinki 2002 \\
Helsingin yliopiston verkkojulkaisut \\
{\texttt http://ethesis.helsinki.fi/}
\end{center}

\newpage
\setcounter{page}{1}
\chapter*{Preface}
\addcontentsline{toc}{chapter}{\bf Preface}
\vspace{-.1cm}

This thesis is a summary of research done at the Department of Physics of 
the University of Helsinki, and later at the Department of Physical 
Sciences and the Helsinki Institute of Physics~(HIP), during the years 
1998 - 2002. This research has mainly been funded by the University of 
Helsinki, the Academy of Finland and~HIP. Fund grants by the V.~K.~\&~Y. 
V\"ais\"ala, M. Ehrnrooth and W. von~Frenckell foundations are also 
gratefully acknowledged.

\noindent
First and foremost, the author wishes to express his thanks to the 
esteemed colleague and supervisor of this thesis, Prof. Dan-Olof Riska, 
who in spite of numerous other pressing duties, including the 
directorship of HIP, has provided invaluable guidance, without which the 
completion of this thesis would have been a formidable task. Prof. 
Dan-Olof Riska and Doc. Mikko Sainio are also gratefully acknowledged by 
the author for providing a postgraduate researcher position at HIP. 
Special thanks are due to the chairman of the Department of Physical 
Sciences, Prof. Juhani Keinonen, for suggesting the referees for this 
thesis, and to Profs. Anthony Green and Jukka Maalampi for agreeing to 
referee the manuscript. 

\noindent
The author also wishes to thank all the colleagues at HIP and elsewhere 
that have in any way contributed to the research presented in this thesis. 
Special thanks are due to Prof. Anthony Green for his many suggestions and 
remarks, to Dr. Tomas Lind\'{e}n for assistance with the typesetting of 
the manuscript, and to Dr. Christina Helminen, with whom the author has 
had many useful conversations. Instructive discussions with 
Profs. Keijo Kajantie, Paul Hoyer, Masud Chaichian, Andy Jackson, Carl 
Carlson, Doc. Claus Montonen, Doc. Jouni Niskanen, Dr. Gunnar Bali and Dr. 
M.R. Robilotta are also gratefully acknowledged.

\noindent
Several other colleagues have also contributed indirectly, most 
notably Christer Nyf\"alt, Krister Henriksson, Lars-Erik Hannelius, Mikko 
Jahma, Jonna Koponen and Pekko \mbox{Piirola.} The author has also enjoyed 
many instructive off-topic conversations with his colleagues at the 
Department of Theoretical Physics~(TFO), the Laser Physics and 
\mbox{X-ray} Research Units and the group of Doc. Kai Nordlund.

\noindent
The thorough nitpicking by Christer Nyfält, although a time-consuming 
source of irritation, is gratefully acknowledged by the author and has led 
to a substantial improvement in the mathematical quality of the 
manuscript, and in many cases to a better understanding of the underlying 
physics. Finally, all colleagues at HIP are acknowledged for the creation 
of a pleasant working atmosphere. \\

\vspace{-.5cm}
\noindent
\hspace{7cm} Helsingfors, October 2002 \\ 

\vspace{-.6cm}
\noindent
\hspace{7cm} {\it Timo L\"ahde}

\newpage
\tableofcontents
\newpage
\newcommand{\vectr}[1]{\ensuremath{\mbox{\boldmath #1}}}

\chapter*{~\\ Abstract}
\addcontentsline{toc}{chapter}{Abstract}

\vspace{-6.1cm}
\small
T. L\"ahde: Strong and Electromagnetic Transitions in Heavy Flavor Mesons, 
University of \mbox{Helsinki,} 2002, VII, 65 p. + reprints, University of 
Helsinki Report Series in Physics, \\ HU-P-D100, ISSN 0356-0961, ISBN 
952-10-0563-7, ISBN 952-10-0564-5 (PDF version) \\ {\texttt 
http://ethesis.helsinki.fi/}.

\noindent
Classification (INSPEC): A1240Q, A1325, A1340F, A1340H, A1440L, A1440N \\
Keywords: charmed mesons, bottom mesons, pionic transitions, 
electromagnetic transitions, potential models

\vspace{3.7cm}
\noindent \normalsize
The electromagnetic and pionic transitions in mesons with heavy flavor 
charm~($c$) or bottom~($b$) quarks are calculated within the framework of 
the covariant Blankenbecler-Sugar (BSLT) equation. The magnetic dipole (M1) 
transitions in the charmonium~($c\bar c$) system are shown to be sensitive 
to the relativistic aspect of the spin-flip magnetic moment operator, and 
the Lorentz coupling structure of the $Q\bar Q$ interaction. The observed 
rate for the M1 transition $J/\psi\rightarrow \eta_c \gamma$ is shown to 
provide strong evidence for a scalar confining interaction. On the other 
hand, the electric dipole (E1) transitions are shown to be sensitive to 
the hyperfine splittings in the $Q\bar Q$ system, and to require a 
nonperturbative treatment of the hyperfine components in the $Q\bar Q$ 
interaction.

\noindent
In addition to the spin-flip M1 transitions, the single pion~($\pi$) and 
dipion~($\pi\pi$) widths are calculated for the heavy-light ($Q\bar 
q$) $D$ mesons, by employment of the pseudovector pion-quark coupling 
suggested by chiral perturbation theory. The pionic transitions 
$D^*\rightarrow D\pi$ are shown to provide useful and constraining 
information on the pion-quark axial coupling $g_A^q$. It is also shown 
that axial exchange charge contributions associated with the $Q\bar q$ 
interaction suppress the axial charge amplitude for pion emission by an 
order of magnitude. The models for $\pi$ and M1 transitions also make it 
possible to estimate the $\eta$-nucleon coupling from the transition 
$D_s^*\rightarrow D_s\pi^0$, once the value of the $\pi^0 - \eta$ mixing 
angle is known.

\noindent
Finally, the $\pi\pi$ dipion transition rates of the $L=1$ 
$D$ mesons are calculated, and are shown to make up a significant fraction 
of their total widths for strong decay. The $\pi\pi$ transitions between 
$S$-wave charmonium ($c\bar c$) and bottomonium ($b\bar b$) states are 
modeled in terms of a broad $\sigma$ meson or a glueball, with derivative 
couplings to pions. The effects of pion rescattering by the spectator 
quark are also investigated.

\newpage
\chapter*{List of Publications}
\addcontentsline{toc}{chapter}{\bf List of Publications}

This thesis consists of two parts. The first and main part of the thesis is 
a summary and discussion of the results obtained in the 
published, peer-reviewed research papers listed below. Where available, 
newer and more accurate results are also presented. The second part 
consists of reprints of selected papers that have been signed by the 
author. These papers are based on research done at the Department 
of Physical Sciences of the University of Helsinki and the Helsinki 
Institute of Physics in 1998 - 2002.

\begin{itemize}

\item[{\bf I}] T.A. L\"ahde, C.J. Nyf\"alt and D.O. Riska, \\
{\it The Confining Interaction and Radiative Decays of Heavy Quarkonia}, \\
Published in: Nucl.Phys.{\bf A645:587-603} (1999), eprint hep-ph/9808438

\item[{\bf II}] K.O.E. Henriksson, T.A. L\"ahde, C.J. Nyf\"alt and D.O. 
Riska, \\
{\it Pion Decay Widths of $D$ Mesons}, \\
Published in: Nucl.Phys.{\bf A686:355-378} (2001), eprint hep-ph/0009095

\item[{\bf III}] T.A. L\"ahde and D.O. Riska, \\
{\it Two-Pion Decay Widths of Excited Charm Mesons}, \\
Published in: Nucl.Phys.{\bf A693:755-774} (2001), eprint hep-ph/0102039

\item[{\bf IV}] T.A. L\"ahde and D.O. Riska, \\
{\it Pion Rescattering in Two-Pion Decay of Heavy Quarkonia}, \\
Published in: Nucl.Phys.{\bf A707:425-451} (2002), eprint hep-ph/0112131

\item[{\bf V}] T.A. L\"ahde and D.O. Riska, \\
{\it The Coupling of $\eta$ Mesons to Quarks and Baryons from 
$D_s^*\rightarrow D_s\pi^0$ Decay}, \\
Published in: Nucl.Phys.{\bf A710:99-116} (2002), eprint hep-ph/0204230 

\item[{\bf VI}] T.A. L\"ahde, \\
{\it Exchange Current Operators and Electromagnetic Dipole Transitions \\ 
in Heavy Quarkonia}, \\
Published in: to be published in Nucl.Phys.{\bf A}, eprint hep-ph/0208110

\end{itemize}

\newpage
\chapter*{Short Introduction to the Papers}
\addcontentsline{toc}{chapter}{\bf Short Introduction to the Papers}

\begin{itemize}

\item[{\bf I}]
This paper presents a calculation of the M1 transition rates in the $c\bar 
c$ and $b\bar b$ systems within the framework of the nonrelativistic 
Schr\"odinger equation. A relativistic version of the single quark 
spin-flip magnetic moment operator is derived, along with the two-quark 
exchange current operators for M1 transitions. It is shown that the 
two-quark operator associated with a scalar confining interaction may 
provide, together with the relativistic single quark operator, a possible 
explanation of the empirically measured width of $\sim 1$ keV for the 
transition $J/\psi\rightarrow \eta_c\gamma$.

\item[{\bf II}]
The framework of the covariant Blankenbecler-Sugar (BSLT) equation is used 
together with the pseudovector pion-quark coupling suggested by chiral 
perturbation theory to predict the widths for pionic transitions in the 
heavy-light ($c\bar q$) $D$ meson systems. It is found that useful and 
constraining information on the pion-quark axial coupling $g_A^q$ is 
provided by the $D^*\rightarrow D\pi$ transitions. A satisfactory 
description of the empirically measured pion widths of the $L=1$ $D_2^*$ 
meson is obtained. Also, the axial charge component of the amplitude 
for pion emission is shown to be suppressed by axial exchange charge 
contributions associated with the $Q\bar q$ interaction.

\item[{\bf III}]
The chiral pseudovector Lagrangian, augmented by a Weinberg-Tomozawa term 
for dipion emission, is used to predict the widths for $\pi\pi$ transitions 
from the $L=1$ $D$ mesons. It is found that widths of several MeV are 
expected for these transitions, in analogy with the experimentally 
well-studied decays of the strange $K_2^*$ meson. It is thus expected that 
the $\pi\pi$ modes should constitute a significant fraction of the total 
widths of the $L=1$ $D$ mesons.

\item[{\bf IV}]
The dipion transitions between $S$-wave states in the charmonium~($c\bar 
c$) and bottomonium~($b\bar b$) systems are studied using a 
phenomenological model with derivative couplings to pions. The 
dipions are modeled in terms of a broad $\sigma$ meson or a glueball. 
Effects of pion rescattering by the spectator quark are investigated and 
shown to be small for $2S\rightarrow 1S$ transitions. The present 
experimental data on these transitions is shown to constrain the $\sigma$ 
meson mass to about 500 MeV. Finally, it is demonstrated that the anomalous 
double-peaked $\pi\pi$ spectrum of the $\Upsilon(3S)\rightarrow 
\Upsilon(1S)\,\pi\pi$ transition may be modeled in terms of a heavier $\sim 
1500$ MeV scalar meson.

\newpage
\thispagestyle{plain}

\item[{\bf V}]
The empirically measured branching ratios for $D_s^*\rightarrow D_s\,\pi^0$ 
and $D_s^*\rightarrow D_s\,\gamma$ are shown to provide a means of 
determining the strength of the $\eta$ coupling to quarks and baryons. This 
requires that the value of the $\pi^0 - \eta$ mixing angle is available, 
along with realistic models for the M1 and pionic transitions in heavy-light 
mesons. The value thus obtained for the $\eta$-nucleon pseudovector 
coupling $f_{\eta NN}$ is shown to be much smaller than that suggested by 
$SU(3)$ symmetry, which is consistent with other recent phenomenological 
analyses. It is also shown that a significant $\eta$-charm coupling, if 
present, serves to increase the estimated value of $f_{\eta NN}$.

\item[{\bf VI}]
The two-quark exchange current operators that arise from the 
elimination of the negative energy components of the Bethe-Salpeter 
equation in the BSLT quasipotential reduction, are calculated for 
electromagnetic E1 and M1 transitions in heavy quarkonium systems. Although 
the exchange charge operators that contribute to E1 transitions are shown 
to be mostly negligible, the corresponding exchange current operators for 
M1 transitions are shown to be crucial, if agreement with the empirical 
width for $J/\psi\rightarrow \eta_c\,\gamma$ is to be achieved. This 
requires that the effective confining interaction couples as a Lorentz 
scalar, since an effective vector interaction is shown to yield a 
spin-flip magnetic moment operator only if the constituent quark masses 
are unequal. Consequently, in the $B_c^\pm$ system, the one-gluon exchange 
interaction also contributes a two-quark spin-flip operator.

\end{itemize}

\pagestyle{headings}
\chapter{Introduction}
\pagenumbering{arabic}
\setcounter{page}{2}

It has been widely accepted, since the beginning of the 20th century, that 
the visible matter in the universe is composed of protons and neutrons (or 
baryons), and electrons (or leptons). However, the discovery of the 
positron in 1933, predicted by Dirac a few years earlier, suggested that 
short-lived, transient particles may exist alongside the stable protons and 
electrons. This was confirmed in 1936, when a heavier, unstable 
electron-like particle, the muon ($\mu$), was discovered in cosmic ray 
experiments by Anderson and Neddermeyer. This discovery was followed up in 
1947, when the existence of the neutral ($\pi^0$) and charged ($\pi^\pm$) 
pions, predicted earlier by Yukawa to be the carriers of the strong 
nuclear force, was confirmed by a similar experiment. These particles were 
the first ones of a large number of short-lived, unstable baryons, mesons 
and leptons which were subsequently produced in copious numbers by 
accelerator experiments. In particular, the pions were shown to be the 
lightest members of a new family of particles known as mesons, to denote 
that they are intermediate in mass between the baryons and leptons.

\section{The quark model}

Around 1960, the number of short-lived baryons 
($\Delta$,\,$\Sigma$\,,$\Lambda$\,,$\Xi$\,...) and mesons 
($\pi$\,,$K$\,,$\rho$\,,$\eta$\,...) that had been discovered by 
accelerator experiments was overwhelming. This suggested that the hypothesis 
of Mendeleev could be extended to the baryons and mesons; Rather than 
being elementary, they might possess substructure and could perhaps be 
classified according to a "periodic table" of subatomic particles. This 
notion, originally put forward by Gell-Mann~\cite{Gellmann} and 
Ne'eman~\cite{Neeman} among others, became known as the "Quark Model" and
attempts to explain the observed properties (spin, isospin, electric 
charge, parity) of the mesons and baryons by arranging them into 
multiplets according to the symmetry group $SU(3)$. It was found that 
three quark flavors, "up" ($u$), "down" ($d$) and "strange" ($s$) with 
spin 1/2 and fractional electric charges were required to accommodate all 
of the mesons and baryons known at that time. 

\noindent
The experimental discovery~\cite{Omega} of the $\Omega$ baryon, which 
was predicted by the quark model because of a gap in the "periodic table" 
of the baryons, soon provided dramatic confirmation of the quark 
hypothesis. Although the quarks were at first only thought of as a useful 
theoretical tool, their actual existence inside the proton was 
confirmed~\cite{struct} by deep inelastic $e^-p$ scattering (DIS) 
experiments at high energies. However, in spite of these remarkable 
successes, the quark model soon ran into a difficulty of symmetry. The 
spin-parity quantum numbers of the $\Delta$ resonance were empirically 
found to be consistent with a combined spin-flavor and configuration space 
wavefunction which is symmetric. This is inconsistent with Fermi-Dirac 
statistics, which requires that the total baryon wavefunction should be 
antisymmetric. 

\noindent
This critical problem was finally circumvented by the introduction of a 
new property for the quarks, "color", which allows the wavefunction to be 
made antisymmetric by means of three color quantum numbers. In order to 
avoid an undesirable proliferation of unobserved states, a further 
constraint was placed, namely that the quarks only combine into colorless 
states (or singlet representations of the color $SU(3)_C$ group). This 
restricts the possible ways of combining quarks and antiquarks to hadrons, 
the simplest being $q\bar q$ (mesons), $qqq$ (baryons) and $\bar q\bar 
q\bar q$ (antibaryons). Together with the proposal~\cite{gauge} that eight 
spin 1 gauge fields, "gluons", should be associated with the new symmetry 
group $SU(3)_C$, these notions were eventually developed into the theory 
of strong interactions, called Quantum Chromodynamics (QCD)~\cite{QCD}.

\noindent
It was also realized that a fourth quark is required in the theory of weak 
interactions to explain e.g. the observed rate for the decay 
$K^0\rightarrow \mu^+\mu^-$. The fourth quark was eventually discovered in 
the form of narrow resonances~\cite{Jpsi} in November 1974 at 
center-of-mass energies of 3.1 GeV and 3.7 GeV in $e^+e^-$ annihilation 
and, independently, in proton-proton collisions. These resonances, named 
$J/\psi$ and $\psi'$, were interpreted as 
mesonic bound states of the new "charm" quark and its antiquark, $c\bar 
c$. The charm quark turned out to have a mass of $\sim 1500$~MeV, and 
is thus much more massive than the $\sim 5$~MeV $u,d$ quarks and the $\sim 
100$~MeV $s$ quark. Later, as higher collision energies became available, 
an unexpected "bottom" ($b$) quark with a mass of $\sim 4800$~MeV was 
similarly discovered in the form of $b\bar b$ or $\Upsilon$ mesons. This 
again raised the question of a possible partner for the $b$ quark, and 
indeed an extraordinarily heavy "top" ($t$) quark was finally 
detected~\cite{toprefs} in 1995, by the proton-antiproton collider 
experiments at Fermilab. The $t$ quark turned out to have a mass 
of 175 GeV, which makes it the most massive elementary particle known, and 
it is too short-lived for mesonic $t\bar t$ bound states to form. 

\section{Quantum Chromodynamics}

In the theory of Quantum Chromodynamics (QCD), the interactions between 
quarks are mediated by eight massless vector bosons called gluons. 
However, a number of complications effectively prevent the 
properties of hadrons to be predicted from QCD; First of all, the theory 
is nonlinear due to gluon self-interactions, and it describes systems 
that interact strongly enough so that perturbative methods are 
inapplicable. Only at the very highest energy scales, where quarks become 
asymptotically free and the coupling between them small, can the 
predictions of perturbative QCD be compared with experimental results. At 
low energies, the quarks interact strongly, are confined into hadronic 
bound states and acquire effective masses. These constituent quark masses 
are for the light $u,d$ quarks of the order $\sim 400$ MeV.

\noindent
At present, the only way to analyze QCD at a fundamental level is the 
method of "lattice QCD" simulations, where the properties of hadrons are 
probed by means of numerical Monte Carlo algorithms. Although much 
progress is being made in the development of more efficient algorithms and 
the inclusion of dynamical fermions (unquenched lattice QCD) into the 
simulations, the applicability of such methods is still limited by the 
huge demands on computing power. In such a situation, it is natural to 
attempt to understand the properties of hadrons by means of effective 
theories and phenomenological, QCD-motivated models. The physical 
motivation of such an approach is that the fundamental degrees of freedom 
of QCD are quarks and gluons, whereas low-energy experiments observe 
hadrons, which at least at long range interact by Yukawa-type meson 
exchange. It is, therefore, a reasonable expectation that the low-energy 
properties of QCD can be described in terms of an effective theory. In the 
limit of vanishing quark masses, QCD exhibits an invariance under 
chiral transformations that involve left- and right-handed quark fields 
separately. This symmetry is only approximate for quarks with a nonzero 
mass. The absence of parity doublets in the low-energy region of the 
hadron spectra suggests that this chiral symmetry is spontaneously broken 
at low energies~\cite{Glozpap}.

\section{Heavy flavor mesons}

Mesons that contain either two heavy quarks ($c\bar c$, $b\bar b$, $c\bar 
b$) or one heavy quark and one light ($c\bar q$, $c\bar s$, $b\bar q$, 
$b\bar s$) are special, since their masses lie in a region 
which is intermediate between the high-energy perturbative regime of QCD 
and the low-energy regime where the dynamics are governed by chiral 
symmetry breaking. Thus these heavy flavor mesons are likely to share 
features that are encountered in these two limits of QCD. One task at hand 
is then to determine phenomenologically, or from lattice QCD~\cite{Green}, 
the functional form, strength and Lorentz structure of the $Q\bar q$ and 
$Q\bar Q$ interaction. 

\noindent
Although the 
nonrelativistic Schr\"odinger framework~\cite{nrpap} can be 
applied to $Q\bar Q$ systems with some success, a realistic treatment of 
the $Q\bar q$ system has {\it a priori} to be relativistic, as the 
velocity of the confined light constitutent quark is close to that of 
light. The papers presented in this thesis employ the covariant 
Blankenbecler-Sugar (BSLT) reduction~\cite{BSLT} of the Bethe-Salpeter 
equation, which has the advantage of formal similarity to the 
Schr\"odinger framework. An alternate approach is provided by the Gross 
quasipotential reduction~\cite{Grosseq}, which has been 
shown~\cite{Qqcomp} to yield comparable results for the spectra of $Q\bar 
Q$ and $Q\bar q$ mesons. 

\noindent
However, as the mass spectra of the $Q\bar Q$ and $Q\bar q$ mesons are 
well described~\cite{phenmod} by a large number of phenomenological and 
QCD-motivated models, the spectrum alone cannot discriminate between 
different assumptions for the $Q\bar Q$ and $Q\bar q$ interaction. 
Fortunately, as will be shown in this thesis, the observed rates for 
$\gamma$ and $\pi$ transitions in heavy flavor mesons may provide useful 
and constraining information on the quark-antiquark interaction, the 
quarkonium wave functions, and in particular, on the Lorentz structure of 
the effective confining interaction. As the negative energy components of 
the Bethe-Salpeter equation are eliminated in the BSLT (or Schr\"odinger) 
quasipotential reduction, two-quark transition operators that depend 
explicitly on the Lorentz structure of the $Q\bar Q$ interaction appear as 
a consequence~\cite{Coester}. In particular, it will be demonstrated in 
this thesis that a pure scalar confining interaction compares favorably 
with the current empirical knowledge of M1 transitions in the charmonium 
system. It is noteworthy, that similar results have been obtained within 
the instantaneous approximation to the Bethe-Salpeter 
equation~\cite{Snellman}, which treats the negative energy components 
explicitly, i.e. without two-quark currents.

\newpage

\section{Transitions in heavy flavor mesons}

The transitions considered in this thesis include the radiative E1 and M1 
transitions in the $Q\bar Q$ systems, the M1 transitions in the heavy-light 
charm~($D$) and strange charm~($D_s$) mesons, and the $\pi$ and $\pi\pi$ 
transitions in the $Q\bar Q$ and $Q\bar q$ systems. It is shown in 
papers~{\bf I} and~{\bf VI} that a possible solution to the long-standing 
overprediction~\cite{Smilga} by a factor \mbox{$\sim 3$} of the width for 
the M1 transition $J/\psi \rightarrow \eta_c\gamma$ emerges, if the 
two-quark exchange current operator associated with a scalar confining 
interaction is included along with a relativistic treatment of the single 
quark spin-flip operator. 

\noindent
On the other hand, the exchange charge 
contributions~\cite{Helminen} to the E1 transition rates are shown in 
paper~{\bf VI} to be highly suppressed by the large masses of the charm 
and bottom constituent quarks~\cite{E1lahde}. Similarly, the 
nonrelativistic predictions for the spin-flip M1 widths of the $Q\bar q$ 
mesons are shown in paper~{\bf V} to be unrealistic, as the confined light 
constituent quark requires a relativistic treatment. It is also shown that 
accidental cancellations in the single quark spin-flip operators render 
the M1 widths very sensitive to two-quark exchange current contributions. 
However, as the form of the $Q\bar q$ interaction is uncertain, the 
results are suggestive rather than definite, quantitative predictions. 

\noindent
In the heavy-light $D$ mesons, the excited states decay to the ground 
state predominantly through pion emission. In this thesis, the pionic 
transitions in the $D$ mesons are described in terms of the chiral 
pseudovector Lagrangian which couples the pions to the light constituent 
quarks. It is shown in paper~{\bf II} that the $D^*\rightarrow D\pi$ 
transitions can provide useful and constraining information on the 
pion-quark axial coupling $g_A^q$. Also, the 
axial charge component of the amplitude for pion emission is shown to be 
highly affected by two-quark axial exchange charge contributions 
associated with the $Q\bar q$ interaction. The pionic transitions which 
are driven by the axial charge operator may, therefore, provide information 
on the Lorentz structure of the $Q\bar q$ interaction. In particular, it 
is shown that a scalar confining interaction has the effect of reducing 
the widths for such transitions.

\noindent
The chiral Lagrangian may, when augmented with a Weinberg-Tomozawa term for 
dipion emission, describe the $\pi\pi$ widths of the excited 
$L=1$ $D$ mesons. In this thesis the $\pi\pi$ widths of the $D$ mesons are 
shown to be of significant magnitude compared to the widths for single pion 
emission. This is known to be the case for the strange $K_2^*$ meson, 
where the empirical $\pi\pi$ width is $\sim 1/2$ of the $\pi$ width. 
This model for pseudoscalar emission has also been applied to the $D_s^* 
\rightarrow D_s\pi^0$ transition, which can then be used to extract the 
coupling of $\eta$ mesons to quarks and baryons from the empirical 
branching ratios for those transitions, once an estimate for the 
$\pi^0-\eta$ mixing angle is available. The value for the $\eta$-nucleon 
pseudovector coupling constant $f_{\eta NN}$ so obtained, is shown to be 
much smaller than that suggested by naive $SU(3)$ symmetry, but consistent 
with other recent phenomenological analyses of e.g. photoproduction of 
$\eta$ mesons on the nucleon.

\noindent
Whereas the dipion transitions in the $D$ mesons may be modeled in terms 
of the chiral Lagrangian, the $\pi\pi$ transitions between $S$-wave $c\bar 
c$ or $b\bar b$ states are likely to involve a broad 
$\sigma$ meson or a glueball. It is shown, within a model where the 
coupling of dipions to heavy quarks is mediated by a broad and 
heavy scalar meson, that the empirical $\pi\pi$ energy spectra constrain 
the $\sigma$ meson mass to $\sim 500$ MeV. A possible explanation for the 
anomalous double-peaked $\pi\pi$ spectrum of 
the $\Upsilon(3S)\rightarrow \Upsilon(1S) \pi\pi$ transition is obtained, 
if the $\pi\pi$ emission is described in terms of a heavier~($\sim 1500$ 
MeV) scalar meson.

\newpage

\section{Notation and layout}

Throughout this thesis, the natural units with $\hbar c = 1$ and the 
$\delta_{\mu\nu}$ metric have been employed. The Euclidean 
$\delta_{\mu\nu}$ or Pauli metric assigns imaginary time components to 
four-vectors. The momentum 
four-vector $k$ is thus of the form $k = (\vectr{$k$}, ik_0)$, where the 
three-vector has been denoted by bold-faced type. However, for typesetting 
reasons, three-vectors in exponents have been denoted with an arrow, 
according to $\vec k$. Also, in obvious cases the modulus $|\vectr{$k$}|$ 
has been denoted simply by $k$. In the Pauli metric the square of a 
four-vector is of the form
\begin{equation}
k^2 = k_\mu k_\mu = \vectr{$k$}^2 - k_0\,^2 = -m^2,
\end{equation}
and the Dirac $\gamma_\mu$ matrices are all hermitian with square equal to 
one. The explicit form of these matrices in the Pauli metric is then 
$\gamma_\mu$ = (\vectr{$\gamma$},$\gamma_4$) and $\gamma_5 = 
\gamma_1\gamma_2\gamma_3\gamma_4$, where
\begin{equation}
\vectr{$\gamma$} = \left( \begin{array}{cc} 
	0 & -i\vectr{$\sigma$} \\
	i\vectr{$\sigma$} & 0
       \end{array}\right), \quad
\gamma_4 = \left( \begin{array}{rr} 
        {\bf 1} & 0 \\
        0	& -{\bf 1}
       \end{array}\right), \quad
\gamma_5 = \left( \begin{array}{rr} 
        0 & -{\bf 1} \\
        -{\bf 1} & 0
       \end{array}\right).
\end{equation}
Factors of $i$ are also introduced into the Dirac current 
and charge density operators to make them real-valued, and for Lagrangians 
which include a $\gamma_5$, in order to assure hermiticity. 

\noindent
A number of abbreviations that are frequently used in this thesis are OGE 
(one-gluon exchange), BSLT (Blankenbecler-Sugar-Logunov-Tavkhelidze), NRIA 
(non-relativistic impulse approximation) and RIA (relativistic impulse 
approximation). Excited states in the heavy quarkonium systems have been 
denoted either by the $\psi(nJ)$ or the primed notation, where the $n$th 
excited state is denoted by $n$ primes, e.g. $\psi(3S) \equiv \psi''$. 
Note that in the primed notation, the primes refer to radial excitations 
only.

\noindent
This thesis contains a summary which comprises six chapters, and reprints 
of selected research papers that have been signed by the author. Chapter 2 
of the summary presents the Blankenbecler-Sugar quasipotential reduction, 
the $Q\bar Q$ and $Q\bar q$ Hamiltonian models and the numerical results 
for the spectra of the heavy flavor mesons. Chapter 3 discusses the 
calculations of the electromagnetic E1 and M1 widths of papers~{\bf 
I},~{\bf V} and~{\bf VI}, while chapter 4 presents the calculation of the 
pionic transitions in the $D$ mesons of paper~{\bf II} and the estimation 
of the $\eta$-nucleon pseudovector coupling $f_{\eta NN}$ from paper~{\bf V}. 
Chapter 5 deals with the $\pi\pi$ transitions in the $D$ mesons 
(paper~{\bf III}) and the model for the $\pi\pi$ transitions in the $Q\bar 
Q$ mesons from paper~{\bf IV}. Chapter 6 contains a concluding discussion.

\newpage

\chapter{Models for the Spectra of $Q\bar Q$ and $Q\bar q$ Mesons}

Although several phenomenological models that employ a nonrelativistic 
treatment of the heavy quarkonia~\cite{nrpap} have succeeded in describing 
many features of the $c\bar c$ and $b\bar b$ systems, a realistic 
treatment of the heavy-light mesons has {\it a priori} to be relativistic 
as the velocity of the confined light constituent quark is close to that 
of light. Also in the case of charmonium and bottomonium, the compact size 
of the $Q\bar Q$ system causes the charm and bottom quarks to move with 
relativistic velocities, in spite of their large masses. The reason for 
this is the effective confining interaction, which has a string tension of 
$\sim 1$~GeV/fm and confines the constituent quarks to a region of 
radius~$<0.5$~fm. In this situation, a quasipotential reduction of the 
relativistic Bethe-Salpeter equation suggests itself as a natural 
framework for a covariant description of the heavy quarkonium systems. 

\section{The BSLT quasipotential reduction}

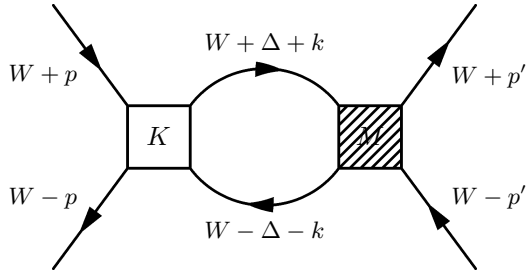
\begin{figure}[h!]
\begin{center}
\begin{fmffile}{Bethe2}
\begin{fmfgraph*}(200,100) \fmfpen{thin}
\fmfleftn{l}{2}\fmfrightn{r}{2}
\fmfrpolyn{empty,label=$K$}{G}{4}
\fmfpolyn{shaded,label=$M$}{K}{4}
\fmf{fermion,label=\small{$W - p$},label.side=right}{G1,l1}
\fmf{fermion,label=\small{$W + p$},label.side=right}{l2,G2}
\fmf{fermion,label=\small{$W - p'$},label.side=right}{r1,K1}
\fmf{fermion,label=\small{$W + p'$},label.side=right}{K2,r2}
\fmf{fermion,label=\small{$W + \Delta + k$},left=.5,tension=.5}{G3,K3}
\fmf{fermion,label=\small{$W - \Delta - k$},left=.5,tension=.5}{K4,G4}
\end{fmfgraph*}
\end{fmffile}
\caption{Illustration of the choice of frame and variables for the 
derivation of the Blankenbecler-Sugar (BSLT) reduction of the 
Bethe-Salpeter scattering equation for unequal quark masses. The upper and 
lower quark lines are taken to have masses $m_Q$ and $m_{\bar q}$, 
respectively. The four-momenta are defined as $W = (0,iP_0/2)$, $\Delta = 
(0,i[m_Q^{\,2} - m_{\bar q}^{\,2}]/4W_0)$, $p = (\vectr{$p$},ip_0$) and $k 
= (\vectr{$k$},ik_0$).}
\label{LN-diag}
\end{center}
\end{figure}

The field-theoretical scattering matrix $S$ may be written in the 
form
\begin{equation}
S_{fi}\:\:=\:\:\delta_{fi} - i\,(2\pi)^4\delta(P_f - P_i)\:{\cal M}_{fi},
\label{smatr}
\end{equation}
where the second term on the r.h.s. has been defined, for notational 
convenience, with a minus sign. The scattering amplitude $M$ is then 
defined as
\begin{equation}
{\cal M}_{fi} = \bar u(\vectr{$p$}_Q')\bar u(\vectr{$p$}_{\bar q}')\,M\,
u(\vectr{$p$}_Q) u(\vectr{$p$}_{\bar q}),
\label{mmatr}
\end{equation}
where $p_i$ and $p_i'$ denote the initial and final four-momenta of the 
quarks. Note that the antiquark will be described throughout by positive 
energy spinors. The Bethe-Salpeter equation for the scattering amplitude 
$M$ can then be written (schematically) in the form
\begin{equation}
M = K + K\,G\,M,
\label{BSeq}
\end{equation}
or explicitly, for an arbitrary frame, as
\begin{equation}
M(p',p,P) = K(p',p,P) + i\int\frac{d^4k}{(2\pi)^4}\:
K(p',k,P)\:G(k,P)\:M(k,p,P),
\label{BSeq2}
\end{equation}
where $P$ is the total four-momentum of the $Q\bar q$ system, and $p,k$ 
and $p'$ denote the initial, intermediate, and final relative four-momenta 
of the constituent quarks. In eq.~(\ref{BSeq2}), $K$ denotes the 
interaction kernel of the Bethe-Salpeter equation, which in the 
nonrelativistic limit corresponds to the potential defined for the 
Schr\"odinger equation. This can be seen by comparison of eq.~(\ref{BSeq}) 
and eq.~(\ref{smatr}) in the Born approximation. Also, $G$ denotes the 
Green's function of the Bethe-Salpeter equation, which is here taken to be 
the free fermion propagator. When bound states are considered, the 
inhomogeneous term in eq.~(\ref{BSeq2}) is dropped. The second term of the 
Bethe-Salpeter scattering equation is illustrated, along with the choice 
of momentum variables for the Blankenbecler-Sugar quasipotential 
reduction, by Fig.~\ref{LN-diag}.

%\begin{figure}[h!]
%\begin{center}
%\begin{fmffile}{Bethe}
%\begin{fmfgraph*}(200,100) \fmfpen{thin}
%\fmfleftn{l}{2}\fmfrightn{r}{1}
%\fmfrpolyn{empty,label=$K$}{G}{4}
%\fmfpolyn{shaded,label=$\Gamma$,tension=.6,smooth,pull=1.2}{K}{3}
%\fmf{fermion,label=\small{$\frac{1}{2}P - p$},label.side=right}{G1,l1}
%\fmf{fermion,label=\small{$\frac{1}{2}P + p$},label.side=right}{l2,G2}
%\fmf{dbl_plain,label=\small{$P$},label.side=right}{r1,K1}
%\fmf{fermion,label=\small{$\frac{1}{2}P + k$},left=.5,tension=.5}{G3,K2}
%\fmf{fermion,label=\small{$\frac{1}{2}P - k$},left=.5,tension=.5}{K3,G4}
%\end{fmfgraph*}
%\end{fmffile}
%\caption{Illustration of the Bethe-Salpeter vertex function $\Gamma(p,P)$ 
%(the second term on the r.h.s. of eq.~(\ref{BSeq2})) for an arbitrary frame 
%and arbitrary quark masses, in terms of total and relative four-momentum 
%variables.}
%\label{BS-diag}
%\end{center}
%\end{figure}

\vspace{.5cm}

\noindent
It is instructive, in order to perform the BSLT quasipotential reduction, 
to introduce the variables presented in Fig.~\ref{LN-diag} and write the 
Bethe-Salpeter equation, schematically, as two coupled integral equations, 
\begin{eqnarray}
M &=& U + U\,g\,M \label{eq1} \\
U &=& K + K\,(G - g)\,U. \label{eq2}
\end{eqnarray}
Here the quasipotential $U$ is defined by eq.~(\ref{eq2}) in terms of the 
Bethe-Salpeter propagator $G$ and a three-dimensional propagator 
$g$. The propagator $g$ is then constructed so that it has an identical 
elastic unitarity cut (right hand cut) as $G$ in the physical region. The 
approximation $U \simeq K$ will be employed here, in order to arrive at a 
major simplification of the Bethe-Salpeter problem. The propagators $G$ 
and $g$ will have identical discontinuities across the right hand cut if 
$\mathrm{Disc}\:G = 2i\,\mathrm{Im}\:g$. By means of the Cutkosky rules, 
$\mathrm{Im}\:g$ may then be obtained as
\begin{eqnarray}
\mathrm{Im}\:g &=& -\frac{2\pi^2}{(2\pi)^4}
\left[\,\gamma^Q\,(W + k + \Delta) + im_Q\,\right]
\left[\,\gamma^{\bar q}\,(W - k - \Delta) + im_{\bar q}\,\right] \nonumber \\
&& \quad\quad \delta^{(+)}\left[\,(W + k + \Delta)^2 + m_Q^{\,2}\,\right]\:
\delta^{(+)}\left[\,(W - k - \Delta)^2 + m_{\bar q}^{\,2}\,\right], 
\label{Img}
\end{eqnarray}
where it has been indicated that only the positive energy roots of the 
arguments in the delta functions are to be included. The complete 
propagator $g$ is then reconstructed by means of the dispersion integral
\begin{equation}
g\:\:\:=\:\:\: \frac{1}{\pi}\int_{0}^{\infty} 
\frac{d\vectr{$q$}^2}{\vectr{$q$}^2 - 
\vectr{$p$}^2 - i\epsilon}\:\:\mathrm{Im}\:g\,(W',k,\Delta'),
\label{disp}
\end{equation}
where $W'$ is defined as $q_0/2$ with $q_0$ on shell. Evaluation of the 
above integral yields the following form for the BSLT propagator $g$:
\begin{equation}
g\:\:\:=\:\:\:-\,\frac{1}{2}\:\frac{\delta(k_0)}{(2\pi)^3}\:\frac{
[\gamma_0^QE_Q(\vectr{$k$}) - \vectr{$\gamma$}^Q\cdot\vectr{$k$} - im_Q]\:
[\gamma_0^{\bar q}E_{\bar q}(\vectr{$k$}) + 
\vectr{$\gamma$}^{\bar q}\cdot\vectr{$k$} - im_{\bar q}]}
{(E_Q(\vectr{$k$}) + E_{\bar q}(\vectr{$k$}))\:(\vectr{$k$}^2 - 
\vectr{$p$}^2 - i\epsilon)},
\end{equation}
where the delta function ensures the condition $k_0 = 0$ in the resulting 
three-dimensional integral equation. Note that the (in principle 
arbitrary) variable $\Delta$ was chosen so that this condition is realized 
also in the case of unequal constituent quark masses. By introduction of 
the positive energy projection operators $\Lambda_+^i$, the above 
propagator can be written in the form  
\begin{equation}
g\:\:\:=\:\:\:-\,\frac{\delta(k_0)}{(2\pi)^3}\:
\frac{2m_Qm_{\bar q}}{E_Q(\vectr{$k$}) + E_{\bar q}(\vectr{$k$})}\:
\frac{\Lambda_+^Q(\vectr{$k$})\,\Lambda_+^{\bar q}(-\vectr{$k$})}
{\vectr{$k$}^2 - \vectr{$p$}^2 - i\epsilon}.
\label{BSLTprop}
\end{equation}
This form is convenient when matrix elements are taken between 
positive-energy spinors according to
\begin{equation}
{\cal M},{\cal V}(\vectr{$p$}',\vectr{$p$})\:\: = \:\: 
\bar u_Q(\vectr{$p$}')\bar u_{\bar q}(-\vectr{$p$}')
\:M,U(\vectr{$p$}',\vectr{$p$})\: u_Q(\vectr{$p$}) u_{\bar q}(-\vectr{$p$}),
\end{equation}
which, together with eq.~(\ref{eq1}), yields the three-dimensional BSLT 
scattering equation
\begin{equation}
{\cal M}(\vectr{$p$}',\vectr{$p$}) = {\cal V}(\vectr{$p$}',\vectr{$p$}) 
- \int\frac{d^3k}{(2\pi)^3}\:
{\cal V}(\vectr{$p$}',\vectr{$k$})\left(\frac{2m_Qm_{\bar q}}
{E_Q(\vectr{$k$}) + E_{\bar q}(\vectr{$k$})}\right)\frac{1}{\vectr{$k$}^2 
- \vectr{$p$}^2 - i\epsilon}
\:{\cal M}(\vectr{$k$},\vectr{$p$}),
\label{BSLTeq}
\end{equation}
where ${\cal V}$ denotes the nonlocal interaction operator as obtained from 
the Feynman rules for $S_{fi}$ using eq.~(\ref{smatr}) in the Born 
approximation. The above extension of the original BSLT equation to the 
case of unequal masses is similar to that of ref.~\cite{Lomon}, which has 
been employed in ref.~\cite{LNscat} for the case of $\Lambda N$ scattering.

\noindent
The elimination of the negative energy components of the Bethe-Salpeter 
equation in the derivation of eq.~(\ref{BSLTeq}) has been 
shown~\cite{Coester} to give rise to two-quark exchange current operators 
that depend explicitly on the Lorentz structure of the quark-antiquark 
interaction. These may then contribute significantly to the strong and 
electromagnetic transition rates in the $Q\bar q$ and $Q\bar Q$ systems. In 
particular, the exchange current operator associated with the scalar 
confining interaction has been shown to be of decisive importance for the 
M1 transitions of heavy quarkonia~\cite{old-new}. It should be noted that 
the appearance of such two-quark operators depends on the type of 
quasipotential reduction.
\pagebreak

\noindent
Although eq.~(\ref{BSLTeq}) is a widely used quasipotential reduction of 
the type discussed in this thesis, there is in principle an infinite 
number of different ways to reduce the Bethe-Salpeter equation to 
a 3-dimensional form. Another commonly used reduction is the Thompson 
equation~\cite{Thompson}, which differs from the BSLT equation by the 
choice of the dispersion integral~(\ref{disp}). These have been shown to 
produce results that are very close to the full Bethe-Salpeter equation in 
ref.~\cite{Fleischer}. There exists also a large 
variety of quasipotential reductions that differ in the choice of the 
propagator~(\ref{BSLTprop}), which attempt to include the effects of 
intermediate negative energy states by various combinations of the 
negative energy projection operators~\cite{diffprop}.

\noindent
It is also noteworthy that the Bethe-Salpeter equation in the ladder 
approximation has been shown~\cite{GrossBS}, not to reduce to the desired 
one-body (Dirac) equation when one of the quarks becomes much heavier than 
the other. However, a large number of quasipotential reductions (e.g. 
Gross) are known to be closely related to the Dirac equation. This 
suggests that such reductions are more appropriate for two-quark systems 
with a large difference between the constituent masses, while the BSLT 
equation is ideal for quarkonia such as $c\bar c$ and $b\bar b$. As the 
light constituent quarks in $Q\bar q$ mesons have masses that are lighter 
than those of the heavy quarks by factors of $3-10$, then the Gross and 
BSLT reductions are expected to give results of similar quality, which 
indeed appears to be the case~\cite{Qqcomp}.

\section{The BSLT and Lippmann-Schwinger equations}

As eq.~(\ref{BSLTeq}) is similar to the nonrelativistic 
Lippmann-Schwinger equation, except for the factor in parentheses, then it 
can be transformed into such an equation by means of the "minimal 
relativity" ansatz~\cite{minrel}
\begin{eqnarray}
T(\vectr{$p$}',\vectr{$p$}) &=&
\left(\frac{m_Q+m_{\bar q}}{E_Q(\vectr{$p$}') + 
E_{\bar q}(\vectr{$p$}')}\right)^{\frac{1}{2}}
{\cal M}(\vectr{$p$}',\vectr{$p$})
\left(\frac{m_Q+m_{\bar q}}{E_Q(\vectr{$p$}) + 
E_{\bar q}(\vectr{$p$})}\right)^{\frac{1}{2}},\\
V(\vectr{$p$}',\vectr{$p$}) &=&
\left(\frac{m_Q+m_{\bar q}}{E_Q(\vectr{$p$}') + 
E_{\bar q}(\vectr{$p$}')}\right)^{\frac{1}{2}}
\:{\cal V}(\vectr{$p$}',\vectr{$p$})\:\,
\left(\frac{m_Q+m_{\bar q}}{E_Q(\vectr{$p$}) + 
E_{\bar q}(\vectr{$p$})}\right)^{\frac{1}{2}}.
\label{modpot}
\end{eqnarray}
This yields the equation
\begin{equation}
T(\vectr{$p$}',\vectr{$p$}) = V(\vectr{$p$}',\vectr{$p$}) 
- \int\frac{d^3k}{(2\pi)^3}\:V(\vectr{$p$}',\vectr{$k$})\:
\frac{2\mu}{\vectr{$k$}^2 - \vectr{$p$}^2 - i\epsilon}
\:T(\vectr{$k$},\vectr{$p$}),
\label{LSequiv}
\end{equation}
which is formally identical to the Lippmann-Schwinger equation. Here $\mu$ 
stands for the usual reduced mass of the two-quark system. The advantage 
of eq.~(\ref{LSequiv}) is that it can be transformed to a 
Schr\"odinger-type differential equation where the potential is given by 
eq.~(\ref{modpot}). This transformation gives the differential equation
\begin{equation}
\left(H_0 - \frac{\vectr{$p$}^2}{2\mu}\right)\:
\psi_{\mathrm {nlm}} (\vectr{$r$}) = - V\:\psi_{\mathrm {nlm}} (\vectr{$r$}),
\end{equation}
where $H_0$ is the kinetic energy operator of the nonrelativistic 
Schr\"odinger equation. The factor $\vectr{$p$}^2$ can be expressed in 
terms of the total energy of the $Q\bar q$ state and the constituent quark 
masses $m_Q$ and $m_{\bar q}$. 
\pagebreak

\noindent
The eigenvalue $\varepsilon$ of the BSLT equation is obtained as
\begin{equation}
\varepsilon\:\: = \:\: \frac{\vectr{$p$}^2}{2\mu} = \frac{\left[ E^2 - 
(m_Q+m_{\bar q})^2 \right ] \left[ E^2 - (m_Q-m_{\bar q})^2 \right ]}{8\mu 
E^2},
\label{EBSLT}
\end{equation}
where $E$ is the mass of the $Q\bar q$ state. The BSLT equation can thus be 
expressed as an eigenvalue equation of the form
\begin{equation}
(H_0 + H_{\mathrm {int}})\:\psi_{\mathrm {nlm}} (\vectr{$r$}) =
\varepsilon\:\psi_{\mathrm {nlm}} (\vectr{$r$}),
\label{blank}
\end{equation}
where the interaction Hamiltonian $H_{\mathrm {int}}$ is given in terms of 
the potential defined in eq.~(\ref{modpot}). The introduction of the 
quadratic mass operator~(\ref{EBSLT}) leads to an effective weakening of 
the repulsive kinetic energy operator, which means that higher excited 
states will have lower masses in the BSLT equation than they would in the 
Schr\"odinger framework. The BSLT eigenvalue $\varepsilon$, expressed 
in terms of the Schr\"odinger excitation energy $E_{\mathrm {ex}} = E - 
(m_1 + m_2)$, is of the form
\begin{equation}
\varepsilon\:\:=\:\:\frac{E_{\mathrm {ex}}}{8\mu}\:
\left[\frac{(E_{\mathrm {ex}} + 2(m_Q+m_{\bar q}))\:
(E_{\mathrm {ex}}^2 + 2E_{\mathrm {ex}}(m_Q+m_{\bar q}) + 4m_Qm_{\bar q})}
{E_{\mathrm {ex}}^2 + 2 E_{\mathrm {ex}} (m_Q+m_{\bar q}) + 
(m_Q+m_{\bar q})^2}\right],
\end{equation}
where the expression in parentheses tends toward $8\mu$ when $m_Q,m_{\bar 
q} \rightarrow \infty$. This demonstrates that in the limit of heavy quark 
masses, or when the quark masses become large compared to the excitation 
energy $E_{\mathrm {ex}}$, the BSLT equation reduces to the nonrelativistic 
Schr\"odinger equation.

\noindent
Although the role of the BSLT potential $V$ as given by eq.~(\ref{modpot}) 
is equivalent to that of the nonrelativistic, static potential in the 
Schr\"odinger framework, the multiplication of the full non-local 
interaction (in momentum space) by the minimal relativity square root 
factors is shown in the next section to have important consequences, not 
only for the numerical treatment of eq.~(\ref{blank}) but also for the 
modeling of the strong and electromagnetic transitions between $Q\bar q$ 
and $Q\bar Q$ states. In particular, the well-known problem of too singular 
and thus illegal hyperfine operators in the Schr\"odinger equation is shown 
to disappear in the BSLT framework.

\section{The $Q\bar Q$ interaction in the BSLT framework}

The interaction between heavy quarks and heavy or light antiquarks is 
dominated by the (presumably linearly) rising confining interaction. The 
observed spectra of $Q\bar Q$ mesons also require the presence of a 
short-range hyperfine interaction that gives rise to e.g. the $J/\psi - 
\eta_c$ splitting. The one-gluon exchange (OGE) interaction~\cite{Rujula} 
of perturbative QCD is a natural candidate for the $Q\bar Q$ systems, 
whereas the origin of the hyperfine interaction in the $Q\bar q$ systems 
is less obvious. A recently suggested possibility is the pointlike 
instanton induced interaction proposed by ref.~\cite{instpap}. The 
interaction Hamiltonians used in this thesis in conjunction with the 
covariant BSLT equation are, therefore, of the form
\begin{equation}
H_{\mathrm {int}} = V_{\mathrm {conf}} + V_{\mathrm {OGE}} + V_{\mathrm
{inst}}, \label{ham}
\end{equation}
with confining, OGE, and instanton induced components, respectively. The 
effective confining interaction is taken to have scalar Lorentz 
structure, while the OGE interaction has vector coupling structure.  
In the nonrelativistic approximation, the effective linear confining 
interaction has the form (in the $LS$-coupling scheme):
\begin{eqnarray}
V_{\mathrm {conf}} &=& cr\left[1 - 
\frac{3}{2}\frac{\vectr{$P$}^2}{m_Q^{\,2} m_{\bar q}^{\,2}}
\left(\frac{m_Q^{\,2} + m_{\bar q}^{\,2} + m_Q m_{\bar 
q}}{3}\right)\right] + \frac{c}{4 m_Q m_{\bar q} r} \nonumber \\
&&-\frac{c}{r}\frac{m_Q^{\,2} + m_{\bar q}^{\,2}}
{4 m_Q^{\,2} m_{\bar q}^{\,2}}\:\vectr{$S$}\cdot\vectr{$L$}
+\frac{c}{r}\frac{m_Q^{\,2} - m_{\bar q}^{\,2}}
{8 m_Q^{\,2} m_{\bar q}^{\,2}}\:(\vectr{$\sigma$}_Q - 
\vectr{$\sigma$}_{\bar q})\cdot\vectr{$L$},
\label{statconf}
\end{eqnarray} 
where the string tension $c$~\cite{phenmod} is of the order 
$\sim 1$ GeV/fm. The above form contains also the momentum dependent terms 
from eq.~(\ref{modpot}) up to second order in $v^2/c^2$. In the 
Schr\"odinger framework (i.e. without the minimal relativity factors), the 
numerical factor $3/2$ in front of the $\vectr{$P$}^2$ term would be $1$. 
Likewise, the Darwin-Foldy term in eq.~(\ref{statconf}) would vanish. In 
addition to the familiar Thomas-precession term, an antisymmetric 
spin-orbit interaction also appears for unequal quark masses, which mixes 
the states with $L=1$ and $J=1$.

\noindent
The interaction components associated with the perturbative OGE 
interaction are, to order $v^2/c^2$ in the nonrelativistic approximation, 
of the form
\begin{eqnarray}
V_{\mathrm {OGE}} &=& -\frac{4}{3}\,\alpha_s\,\left[\frac{1}{r} - 
\frac{3\pi}{2}\left(\frac{m_Q^{\,2} + m_{\bar q}^{\,2} + m_Q m_{\bar q}}
{3\:m_Q^{\,2} m_{\bar q}^{\,2}}\right)
\delta(\vectr{$r$}) + \frac{1}{2}\,
\frac{\vectr{$P$}^2}{m_Q m_{\bar q}\:r}\right]
\nonumber \\
&& + \frac{2}{3}\frac{\alpha_s}{r^3}
\left(\frac{m_Q^{\,2}+m_{\bar q}^{\,2}}{2 m_Q^{\,2} m_{\bar q}^{\,2}}
+\frac{2}{m_Q m_{\bar q}}\right) \vectr{$S$}\cdot\vectr{$L$}
+\frac{\alpha_s}{6r^3} \frac{m_Q^{\,2} - m_{\bar q}^{\,2}}{m_Q^{\,2} 
m_{\bar q}^{\,2}}\:(\vectr{$\sigma$}_Q - \vectr{$\sigma$}_{\bar 
q})\cdot\vectr{$L$} \nonumber \\
&& + \frac{8\pi}{9}\frac{\alpha_s}{m_Q m_{\bar q}}\:\delta(\vectr{$r$}) 
\:\vectr{$\sigma$}_Q\cdot\vectr{$\sigma$}_{\bar q}
\:+ \frac{\alpha_s}{3 m_Q m_{\bar q} r^3}\:S_{12},
\label{OGEstat}
\end{eqnarray}
where $\alpha_s$ denotes the strong coupling of perturbative
QCD, and $S_{12}$ is the tensor operator $S_{12} = 
3(\vectr{$\sigma$}_Q\cdot\vectr{$\hat r$})(\vectr{$\sigma$}_{\bar q}
\cdot \vectr{$\hat r$})-\vectr{$\sigma$}_Q\cdot\vectr{$\sigma$}_{\bar q}$. 
In the Schr\"odinger framework, the coefficients for the contact and 
$\vectr{$P$}^2$ terms would be $-\pi$ and $1$, respectively.

\noindent
The instanton induced interaction, considered by 
ref.~\cite{instpap} for systems with heavy quarks, consists of a
spin-independent term as well as a $\vectr{$\sigma$}_Q\cdot
\vectr{$\sigma$}_{\bar q}$ term which contributes to the 
pseudoscalar-vector splittings in heavy quarkonia. The effective instanton 
interaction derived in ref.~\cite{instpap} is of the form
\begin{eqnarray}
V_{\mathrm{inst}} &=& -\frac{\Delta M_Q\,\Delta M_{\bar q}}{4n}
\:\delta(\vectr{$r$}) + \frac{\Delta M_Q^{\mathrm{spin}}\,\Delta M_{\bar 
q}}{4n}\:\delta(\vectr{$r$)}\:\vectr{$\sigma$}_Q\cdot
\vectr{$\sigma$}_{\bar q},
\label{instpot}
\end{eqnarray}
where the factors $\Delta M_Q$ and $\Delta M_{\bar q}$ denote the mass 
shifts of the heavy and light constituent quarks due to the instanton 
induced interaction. These shifts are, for light constituent quarks, of 
the order of the constituent quark mass~($\sim 400$~MeV), and smaller 
($\sim 100$ MeV) for the charm quark. The parameter 
$M_Q^{\mathrm{spin}}$ controls the strength of the spin-spin interaction, 
which has the same sign as that from the perturbative OGE interaction. The 
parameter $n$ represents the instanton density, which is typically 
assigned values around $\sim 1\:\mathrm{fm}^{-4}$. The spin-independent 
term has scalar coupling for the light constituent quark line and a mixed 
scalar-$\gamma_0$ vertex for the heavy quark.

\section{Relativistic $Q\bar Q$ potentials}

For systems that contain light quarks, the above static interaction 
Hamiltonians have but qualitative value because of the slow convergence of 
the asymptotic expansion in $v/c$. Even for systems composed of heavy 
quarks only, the compact size of the wave functions lead to very large 
matrix elements in first order perturbation theory for the $\vectr{$P$}^2$ 
terms in eqs.~(\ref{statconf}) and~(\ref{OGEstat}). Therefore, it was 
chosen in ref.~\cite{E1lahde} to employ a local interaction model for the 
heavy quarkonium systems which takes into account the minimal relativity 
factors~(\ref{modpot}), as well as the relativistic effects due to the 
quark spinors and the running coupling of QCD. The central, 
spin-independent part of the OGE interaction is thus modified to
\begin{eqnarray}
V_{\mathrm {OGE}}^0\,(r) &=& -\frac{4}{3}\:\frac{2}{\pi}
\int_{0}^{\infty}dk\:j_0(kr)\:W_{Q\bar q}\:\frac{m_Qm_{\bar q}}{e_Qe_{\bar 
q}}\: \alpha_s(\vectr{$k$}^2),
\label{OGEpot}
\end{eqnarray}
where the following notation has been introduced for convenience:
\begin{equation}
e_Q=\sqrt{m_Q^2+\frac{\vectr{$k$}^2}{4}},\quad
e_{\bar q}=\sqrt{m_{\bar q}^2+\frac{\vectr{$k$}^2}{4}},\quad
W_{Q\bar q}=\left(\frac{m_Q+m_{\bar q}}{e_Q+e_{\bar q}}\right).
\end{equation}
For the running QCD coupling $\alpha_s(\vectr{$k$}^2)$, the
parameterization of ref.~\cite{alphas}:
\begin{equation}
\alpha_s(\vectr{$k$}^2)\:=\:\frac{12\pi}{27}\:
\ln^{-1}\left[\frac{\vectr{$k$}^2 + 4m_g^2}
{\Lambda_{\mathrm{QCD}}^2}\right].
\label{zzz}
\end{equation}
has been employed. Here the QCD scale parameter $\Lambda_{\mathrm{QCD}}$ 
and the dynamical gluon mass $m_g$, which determines the low-momentum 
cutoff of the inverse logarithmic behavior of $\alpha_s$ have been 
determined by a fit to the experimental spectra of the $Q\bar Q$ and 
$Q\bar q$ systems.

\noindent
In general, the relativistic effects in eq.~(\ref{OGEpot}) lead to a 
strong suppression of the short-range coulombic potential. On the other 
hand, the running coupling $\alpha_s$, when employed according to 
eq.~(\ref{zzz}), increases the strength of the OGE interaction for large 
distances. The end result is, that the OGE interaction, when calculated 
using eqs.~(\ref{OGEpot}) and~(\ref{zzz}) bears little or no resemblance 
to a coulombic potential, even for the heavy $c\bar c$ system. This can 
potentially have serious phenomenological consequences since models that 
employ a short-range coulombic interaction have, in general, provided good 
descriptions of the $c\bar c$ and $b\bar b$ spectra. However, the 
spin-independent part of the instanton induced interaction~(\ref{instpot}) 
has been shown in paper~{\bf VI} to provide the necessary short-range 
attraction, even if the OGE interaction becomes weak. In principle, the 
effective confining interaction is also subject to similar relativistic 
effects, but in view of its long-range nature, their effect will be very 
small.

\noindent
The hyperfine components of the $Q\bar Q$ interaction, as given by 
eqs.~(\ref{statconf}) and~(\ref{OGEstat}) are usually treated as first 
order perturbations since their behavior for small $r$ is too 
singular to allow for direct numerical treatment. Modification of those 
hyperfine components according to eq.~(\ref{modpot}) leads to expressions, 
which are weaker and more well-behaved, and may consequently be fully 
taken into account. The employment of individual wave functions for each 
member in a given hyperfine multiplet was shown, in paper~{\bf VI}, to be 
important for a realistic description of several electromagnetic E1 and M1 
transitions in the heavy quarkonium systems.

\noindent
The expressions for the local hyperfine components of the $Q\bar q$ 
interaction that take into account the minimal relativity factors and the 
running QCD coupling are, in configuration space, of the form
\begin{eqnarray}
V_{\mathrm {OGE}}^{\mathrm {LS}} &=& \frac{4}{3 \pi r}\vectr{$S$}\cdot 
\vectr{$L$} \int_0^\infty dk\:k\:j_1(kr)
\frac{W_{Q\bar q}}{e_Qe_{\bar q}}\:\left[2 + \frac{m_Q}{e_{\bar q} + m_q}
+ \frac{m_{\bar q}}{e_{\bar Q} + m_Q}\right] \alpha_s(\vectr{$k$}^2),
\quad\quad\label{OGEls} \\
V_{\mathrm {conf}}^{\mathrm {LS}} &=& - \frac{2}{\pi}\,\frac{c}{r}
\vectr{$S$}\cdot\vectr{$L$}\int_0^{\infty} dk\:\frac{j_1(kr)}{k}
\frac{W_{Q\bar q}}{e_Qe_{\bar q}}\left[\frac{e_{\bar q}}{e_Q+m_Q}
+\frac{e_Q}{e_{\bar q}+m_{\bar q}}\right],
\label{Confls} \\
V_{\mathrm {OGE}}^{\mathrm {SS}} &=&
\frac{4}{9\pi}\,\vectr{$\sigma$}_Q\cdot\vectr{$\sigma$}_{\bar q}\,
\int_0^\infty dk\:k^2\:j_0(kr)
\:\frac{W_{Q\bar q}}{e_Q e_{\bar q}}\:\alpha_s(\vectr{$k$}^2),
\label{OGEss} \\
V_{\mathrm {OGE}}^{\mathrm {T}} &=&
\frac{2}{9 \pi} S_{12}\int_0^\infty dk\:k^2\:j_2(kr)
\:\frac{W_{Q\bar q}}{e_Q e_{\bar q}}\:\alpha_s(\vectr{$k$}^2),
\label{OGEt}
\end{eqnarray}
for the spin-orbit, spin-spin and tensor components of the OGE 
interaction, and the spin-orbit (Thomas precession) term from the 
effective scalar confining interaction. Note that the 
expression~(\ref{Confls}) for the spin-orbit term associated 
with the linear scalar confining interaction is obtained by means of the 
representation $-\,8\pi c/\vectr{$k$}^4$ in momentum space. This can be 
understood as the Fourier transform of a modified linear potential 
$cr\,e^{-\lambda r}$ in the limit $\lambda\rightarrow 0$. The 
integral~(\ref{Confls}) is convergent even if that limit is taken 
analytically. The above expressions are also free of singularities that 
require a perturbative treatment. If the QCD coupling $\alpha_s$ is taken 
to be constant, then the hyperfine components, as given by 
eqs.~(\ref{OGEls})-(\ref{OGEt}), reduce to the static expressions of 
eqs.~(\ref{statconf}) and~(\ref{OGEstat}) for large distances.

\noindent
As the instanton induced interaction for $Q\bar q$ systems, as given by 
ref.~\cite{instpap}, consists of delta functions, it has to be treated as a 
first order perturbation. Such a treatment is very unfortunate here since 
the repulsive kinetic energy as given by the BSLT quadratic mass 
operator~(\ref{EBSLT}) is very sensitive to the ground state energy 
relative to the sum of the quark masses. A perturbative treatment of a 
strong attractive interaction component would thus effectively lead to 
unrealistically small level spacings between the higher excited states. In 
view of this, the delta function of eq.~(\ref{instpot}) has been treated 
according to
\begin{equation}
V_{\mathrm{inst}} = - \frac{\Delta M_Q\Delta M_{\bar q}}{4n} \:\:
\int_0^\infty dk\:k^2\:j_0(kr)\:W_{Q\bar q}\:
\frac{m_Q m_{\bar q}}{e_Qe_{\bar q}},
\label{instpot2}
\end{equation}
which effectively leads to a smeared-out form of the instanton induced 
interaction. While the presence of $W_{Q\bar q}$ is naturally suggested by 
the BSLT minimal relativity factors, the $m_Q/e_Q$ factors are entirely 
phenomenological, and have been inserted to allow for better convergence 
of the above integral. In the limit of very large constituent masses (the 
static limit), the above equation reduces to the form~(\ref{instpot}). The 
spin-spin component of the instanton induced interaction was found in 
ref.~\cite{instpap} to be significant for the heavy-light $Q\bar q$ 
systems, but very weak for the heavy-heavy $Q\bar Q$ mesons. Because of 
this, the spectra shown in Table~\ref{stat} and Fig.~\ref{spektrfig} do 
not include that interaction. The calculated $Q\bar q$ spectra employed in 
paper~{\bf II}, shown for the $D$ meson in Fig.~\ref{Dfig}, do not 
include the instanton induced interaction, since sufficient attraction was 
provided there by the OGE interaction, although at the price of an 
unrealistically large value for the QCD scale parameter 
$\Lambda_{\mathrm{QCD}}$.

\section{Spectra of heavy flavor mesons}

\begin{figure}[h!]
\parbox{0.495\textwidth}{\epsfig{file = 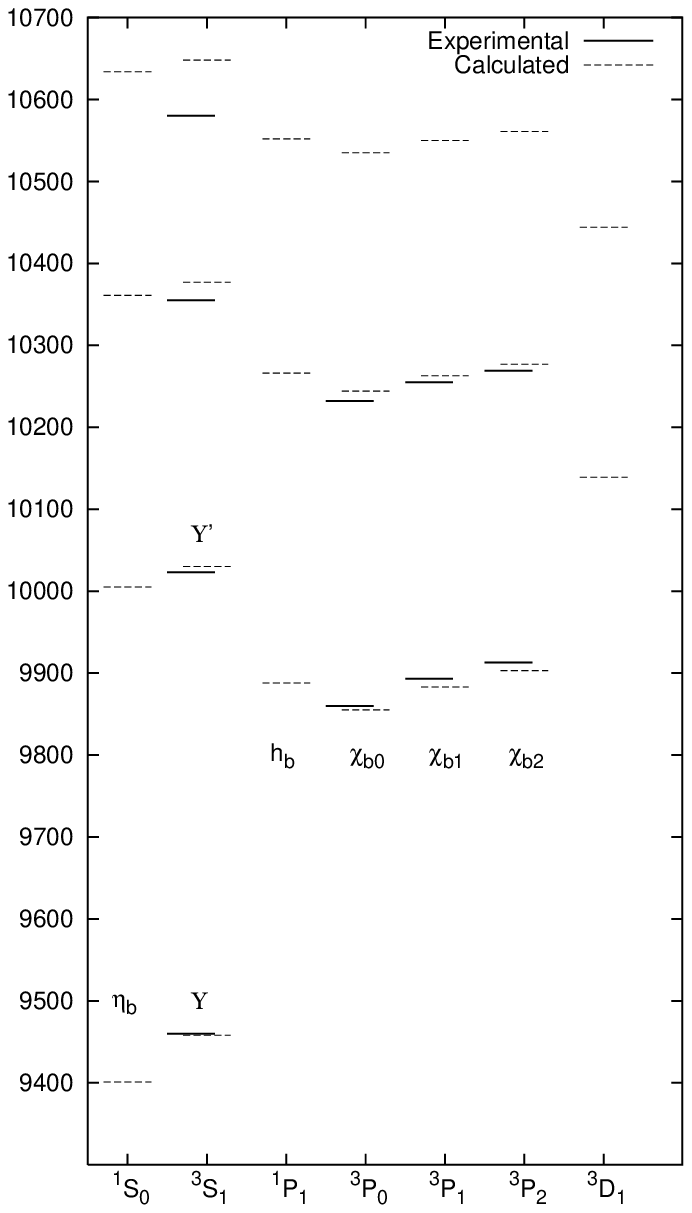,width=6.7cm}}
\parbox{0.495\textwidth}{\epsfig{file = 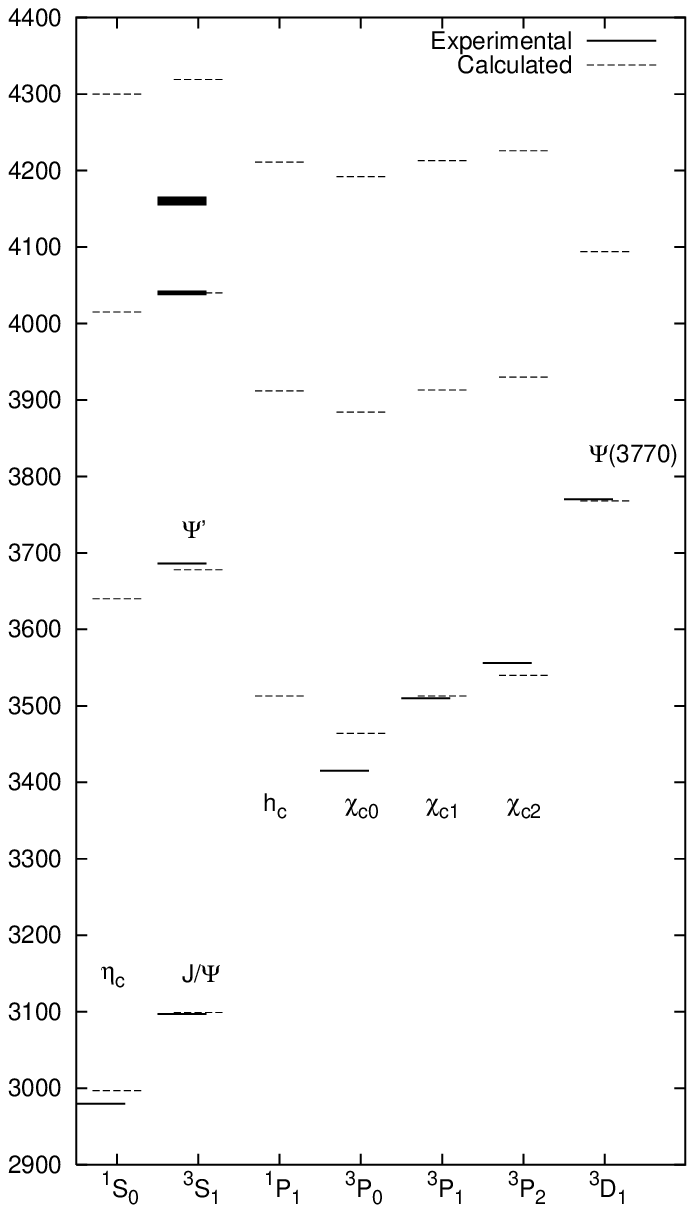,width=6.7cm}}
\caption{Calculated and experimental $b\bar b$ and $c\bar c$ spectra. All 
states are given in MeV, and correspond to the data in 
Table~\ref{stat}. The thickness of the lines denoting the 
experimentally determined states indicates the uncertainty in the mass of 
the state. Note that the identification of the $4\,^3S_1$ state in 
charmonium is uncertain, and may actually be a $2\,^3D_1$ state, or a 
mixture of the two. The threshold for $D\bar D$ decay is at $\sim 3750$ 
MeV, and for $B\bar B$ decay at $\sim 10500$ MeV.}
\label{spektrfig}  
\end{figure}

The $c\bar c$, $b\bar b$ and $B_c^\pm$ spectra that are shown in 
Table~\ref{stat} and Fig.~\ref{spektrfig} have been obtained by solution 
of the BSLT equation for a linear scalar confining interaction, and OGE + 
instanton components modeled according to the expressions given in this 
section. The hyperfine components have been taken fully into account, so 
that all the states given in Table~\ref{stat} are represented by different 
radial wave functions. This model has been employed for the calculations 
of the electromagnetic transitions in paper~{\bf VI} as well as the dipion 
transitions in paper~{\bf IV}.

\begin{table}[h!]
\centering{
\caption{Calculated and experimental $c\bar c$, $b\bar b$ and 
$B_c^\pm$ states rounded to the nearest MeV, as obtained in papers~{\bf 
IV} and~{\bf VI}. The states are classified according to excitation number 
$n$, total spin $S$, total orbital angular momentum $L$ and total angular 
momentum $J$. The experimental values are from ref.~\cite{PDG}, except for 
the recently observed~\cite{Belle} $\eta_c(2S)$.}
\vspace{.5cm}
\begin{tabular}{c||c|c|c|c|c}
$n\,^{2S+1}L_J$ & $b\bar b$ & Exp($b\bar b$) & $c\bar c$ & Exp($c\bar c$)
& $c\bar b$ \\ \hline\hline
&&&&&\\
$1\,^1S_0$ & 9401  &  --  & 2997 & $2980\pm 1.8$ & 6308 \\
$2\,^1S_0$ & 10005 &  --  & 3640 & $3654\pm 6$~\cite{Belle}   & 6888 \\
$3\,^1S_0$ & 10361 &  --  & 4015 &  -- & 7229 \\
$4\,^1S_0$ & 10634 &  --  & 4300 &  -- & 7488 \\
\vspace{-.22cm}
&&&&&\\
$1\,^3S_1$ & 9458  & 9460  & 3099 & 3097 & 6361 \\
$2\,^3S_1$ & 10030 & 10023 & 3678 & 3686 & 6910 \\
$3\,^3S_1$ & 10377 & 10355 & 4040 & $4040\pm 10$ & 7244 \\
$4\,^3S_1$ & 10648 & 10580 & 4319 & $4159\pm 20$ ? & 7500 \\
\vspace{-.22cm}
&&&&&\\
$1\,^1P_1$ & 9888  &  --  & 3513 &  --  & 6754 \\
$2\,^1P_1$ & 10266 &  --  & 3912 &  --  & 7126 \\
$3\,^1P_1$ & 10552 &  --  & 4211 &  --  & 7401 \\
\vspace{-.22cm}
&&&&&\\
$1\,^3P_0$ & 9855  & 9860  & 3464 & 3415 & 6723 \\
$2\,^3P_0$ & 10244 & 10232 & 3884 &  -- & 7107 \\
$3\,^3P_0$ & 10535 &  --   & 4192 &  -- & 7387 \\
\vspace{-.22cm}
&&&&&\\ 
$1\,^3P_1$ & 9883  & 9893  & 3513 & 3511 & 6751 \\
$2\,^3P_1$ & 10263 & 10255 & 3913 &  -- & 7125 \\
$3\,^3P_1$ & 10550 &  --   & 4213 &  -- & 7400 \\
\vspace{-.22cm}
&&&&&\\
$1\,^3P_2$ & 9903  & 9913  & 3540 & 3556 & 6770 \\
$2\,^3P_2$ & 10277 & 10269 & 3930 &  -- & 7136 \\
$3\,^3P_2$ & 10561 &  --   & 4226 &  -- & 7410 \\
\vspace{-.22cm}
&&&&&\\
$1\,^3D_3$ & 10158 &  --  & 3790 &  --  & 7009 \\
$1\,^3D_2$ & 10149 &  --  & 3784 &  --  & 7006 \\
$1\,^3D_1$ & 10139 &  --  & 3768 & $3770\pm 2.5$ & 6998 \\
\end{tabular}
\label{stat}}
\end{table}

\noindent
Although preliminary, the measured mass of the $B_c^\pm$ was reported in 
ref.~\cite{CDF} as $6.40\pm 0.39$ GeV, which is about $\sim 100$ MeV 
higher than the predicted 6308 MeV, and most other models~\cite{Bcmod} 
give even lower masses for the $B_c^\pm$ ground state. However, the 
predicted $B_c^\pm$ spectrum agrees very well with the QCD-inspired model 
of~\mbox{ref.~\cite{Chen}.}

\noindent
The quality of the calculated $Q\bar Q$ spectra in 
Table~\ref{stat} is generally quite satisfactory, as both the 
$\psi' - J/\psi$ and $J/\psi - \eta_c$ splittings are given realistically. 
In particular, the $\eta_c(2S)$ state has recently been reported 
by the BELLE collaboration~\cite{Belle} with a mass of about 3650~MeV. 
This suggests that the spin-spin splitting is much smaller for the $2S$ 
states than for the $J/\psi$ and the $\eta_c$, a feature which is well 
described by the present model. The main difficulty is the prediction 
of the hyperfine splittings in the $L=1$ multiplet of charmonium. 
Table~\ref{stat} and Fig.~\ref{spektrfig} indicate that the splittings are 
underpredicted for $c\bar c$ but in reasonable agreement with experiment 
for $b\bar b$. This problem can be traced, in part, to the weakness of the 
OGE tensor interaction as given by eq.~(\ref{OGEt}). 

\begin{table}[h!]
\parbox{0.55\textwidth}{
\begin{tabular}{c||c|c}
& Paper {\bf VI} & Other models \\ \hline\hline &&\\
$M_b$   & 4885 MeV      & 4870 MeV~\cite{Qqcomp}       \\
$M_c$   & 1500 MeV      & 1530 MeV~\cite{Qqcomp}       \\ &&\\
$\Lambda_{\mathrm{QCD}}$ & 260 MeV & 200-300 MeV~\cite{alphas} \\
$m_g$   & 290 MeV       & $m_g > \Lambda_{\mathrm{QCD}}$~\cite{alphas}\\
$c$     & 890 MeV/fm    & 912 MeV/fm~\cite{Qqcomp} \\  &&\\
$\frac{(\Delta M_c)^2}{4n}$ & 0.084 $\mathrm{fm}^2$ & $\sim 0.05$
$\mathrm{fm}^2$~\cite{instpap} \\
$\frac{(\Delta M_b)^2}{4n}$ & 0.004 $\mathrm{fm}^2$ & ? \\
\end{tabular}}
\parbox{0.44\textwidth}{
\caption{Quark masses and coupling constants used for the calculated 
spectra in Fig.~\ref{spektrfig}. The values should be considered as best 
fits within the BSLT model to the empirical $c\bar c$ and $b\bar b$ 
spectra.}
\label{partab}}
\end{table}

\noindent
The heavy quark masses are close to those obtained by Roberts {\it et al.} 
in ref.~\cite{Qqcomp} within the framework of the Gross equation. The 
values of $\Lambda_{\mathrm{QCD}}$ and $m_g$ are in line with those 
suggested in ref.~\cite{alphas}, while the string tension $c$ is somewhat 
smaller than that suggested by the lattice QCD calculations of 
ref.~\cite{Bali}. The strength of the instanton induced interaction in the 
$c\bar c$ system is comparable to the estimate given in 
ref.~\cite{instpap}. In spite of the generally satisfactory results, 
perfect agreement with experiment had to be sacrificed in the $b\bar b$ 
system in order to obtain an optimal description of the $c\bar c$ 
spectrum with the same set of parameters. As this nevertheless is a small 
effect, the results indicate that a flavor-independent confining 
interaction is a reasonable first approximation, in contrast to the 
instanton induced interaction, the strength of which depends explicitly on 
the quark flavors involved.

\begin{figure}[h!]
\parbox{0.66\textwidth}{\epsfig{file = 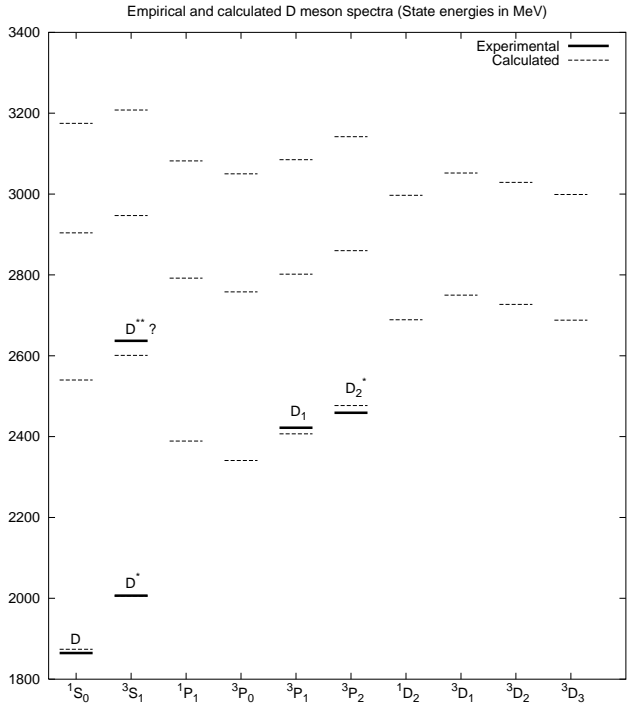, width = 0.65\textwidth}}
\parbox{0.33\textwidth}{
\caption{Experimental and calculated $D$ meson \mbox{spectra} from 
ref.~\cite{Lahdeout}. It should be noted that the excited $2S$ state 
$D^{**}$ at $\sim 2630$ MeV, which was reported by ref.~\cite{Qqs-exp} was 
not detected by a subsequent search~\cite{Qqs-exp2} and may therefore not 
exist at that energy.}
\label{Dfig}}
\end{figure}

\noindent
The spectra of the heavy-light $Q\bar q$ mesons that were obtained in 
ref.~\cite{Lahdeout} within the framework of the BSLT equation have been 
used here for calculation of the M1 and pion widths of the $Q\bar q$ 
states. The $D$ meson spectrum so obtained is shown in Fig.~\ref{Dfig}. 
Although quite satisfactory agreement with the empirical $Q\bar q$ 
spectra was achieved, this was only at the price of a very strong OGE 
interaction and the introduction of a negative constant into the scalar 
confining interaction, which was treated as a free parameter. The hyperfine 
components of the OGE and scalar confining interactions were treated as 
perturbations, while the instanton induced interaction was dropped since the 
OGE interaction was found to give sufficient attraction to account for the 
empirical spectra. However, later (unpublished) calculations of the $Q\bar 
q$ spectra have indicated that a nonperturbative treatment of the OGE 
spin-spin interaction strongly favors the inclusion of the instanton induced 
spin-spin interaction proposed by ref.~\cite{instpap}, in which case the 
spin-independent term of that interaction may also lead to a more realistic 
strength and low-momentum behavior of $\alpha_s$.
\pagebreak

\noindent
The small number of empirically known $Q\bar q$ states makes a determination 
of the quality of a given model difficult. However, the most significant 
unsolved problem in the case of the $Q\bar q$ spectra is the ordering of 
the $L=1$ multiplet, of which only two resonances have been detected so far. 
These probably correspond to spin-triplet states with $J=2$ and $J=1$ ($j_q 
= 3/2$ in the heavy quark limit), and are denoted $D_2^*$ and $D_1$, 
respectively (note the notational confusion). The empirical fact that the 
$D_2^*$ is higher in mass by $\sim 40$ MeV then suggests that the ordering 
of the $L=1$ states in the $D$ mesons is similar to that observed for the 
$c\bar c$ and $b\bar b$ systems. It is reassuring that this result is 
consistent with lattice QCD calculations of the spin-orbit splittings in 
heavy-light mesons~\cite{Qqlatt} although the problem of poor convergence 
of such calculations is still not solved to satisfaction. In view of these 
results, it appears that the possibility~\cite{QqIsgur} of spin-orbit 
inversion in heavy-light mesons is not realized. It should be noted, 
however, that the spin-orbit splittings as calculated from static 
expressions like eqs.~(\ref{statconf}) and~(\ref{OGEstat}) are unrealistic 
because of the low masses of the light constituent quarks. An unphysical 
dominance of the Thomas precession associated with the scalar confining 
interaction may then suggest that the spin-orbit splittings of the $L=1$ 
$Q\bar q$ states should be inverted.
\pagebreak

\noindent
It is instructive to compare the parameters of ref.~\cite{Lahdeout} with 
the values suggested by the more realistic calculation (paper {\bf VI}) of 
the $Q\bar Q$ spectra in Table~\ref{partab}. The heavy quark masses $m_c = 
1580$ MeV and $m_b = 4885$ MeV of ref.~\cite{Lahdeout}, though slightly 
higher, agree quite well with those of Table~\ref{partab}. On the other hand, 
the masses of the light quarks were obtained as $m_{u,d} = 450$ MeV and $m_s 
= 560$ MeV, respectively. Although the light quark mass is higher than the 
usual phenomenological value of $\sim 350$ MeV, it is close to the value 
420 MeV which has been employed~\cite{instpap} for the instanton induced 
interaction. It has also been suggested~\cite{Shur} that a natural value 
of the constituent quark mass is one third of the $\Delta$ mass, rather 
than the nucleon mass. 

\noindent
The parameters $\Lambda_{\mathrm {QCD}}$ and $m_g$ were obtained as 
280~MeV and 240~MeV in ref.~\cite{Lahdeout} and lead to a much stronger 
coupling $\alpha_s$ than is necessary in paper~{\bf VI}. Most other 
analyses~\cite{alphas} suggest a much weaker form. The phenomenological 
consequences of a very strong OGE interaction for light constituent quarks 
are potentially serious~\cite{Glozpap}, since incorrect ordering of the 
positive and negative parity nucleon and $\Lambda$ states may result. 
Moreover, the relativistic damping of the short-range part of the 
attractive coulombic OGE potential in ref.~\cite{Lahdeout} is such that 
the OGE interaction alone cannot describe e.g. the $1S - 1P$ level 
splittings in the $Q\bar q$ spectra. It was therefore necessary to push 
the other parameters of that model to their limits.

\noindent
In this situation, the instanton induced interaction of ref.~\cite{instpap} 
suggests itself naturally as it provides both a strong attraction in the 
$S$-wave and contributes to the $D^* - D$ splitting. Also, the OGE 
potential of eq.~(\ref{OGEpot}) for a system composed of two light~($\sim 
400$ MeV) quarks, becomes depleted for distances equal to the meson 
radius. The central OGE component may therefore only play a minor role in 
the dynamics of light constituent quarks, while for the $Q\bar q$ mesons, 
the short-range attraction may turn out to be best described by a 
combination of OGE and instanton induced components.

\newpage

\chapter{Electromagnetic Transitions}

The radiative transitions in heavy quarkonium ($c\bar c$, $b\bar b$,
$c\bar b$) systems have drawn much theoretical interest~\cite{Radref,Radref2}, 
as they can provide direct information on both the heavy quarkonium 
wave functions and the $Q\bar Q$ interaction. As reasonably reliable 
empirical data now exists for a number of transitions in both the $c\bar c$ 
and $b\bar b$ systems~\cite{NewPDG}, a fair assessment of the quality of 
theoretical models is already possible. The measured $\gamma$ transitions 
in the charmonium~($c\bar c$) system include the E1 transitions $\chi_{cJ}
\rightarrow J/\psi\,\gamma$ and $\psi\,'\rightarrow \chi_{cJ}\,\gamma$, as
well as the spin-flip M1 transitions $J/\psi \rightarrow \eta_c\gamma$ and
$\psi\,'\rightarrow \eta_c\gamma$. The situation concerning the
analogous transitions in the bottomonium~($b\bar b$) system is, however, 
less satisfactory as the total widths of the $\chi_{bJ}$ states are not 
known, and none of the spin-flip M1 transitions observed.

\noindent
Previous calculations of the E1 widths of heavy quarkonia have 
demonstrated that the E1 approximation leads to overpredictions of most 
transition rates, and that this overprediction can be significantly 
reduced, if not entirely eliminated, by the consideration of relativistic 
effects. On the other hand, theoretical predictions for the M1 transitions 
have remained unsatisfactory for a long time~\cite{Smilga} as the width for 
$J/\psi \rightarrow \eta_c\gamma$ has typically been overpredicted by a 
factor~$\sim 3$. However, calculations of M1 widths using the 
nonrelativistic Schr\"odinger equation in ref.~\cite{Schrcalc} and 
paper~{\bf I} have demonstrated that the M1 transitions in charmonium are 
sensitive both to the relativistic aspects of the spin-flip operator as 
well as the Lorentz structure of the $Q\bar Q$ interaction. The results 
obtained in papers~{\bf I} and~{\bf VI} suggest that a scalar confining 
interaction may explain the observed width of $\sim 1$~keV for 
$J/\psi\rightarrow \eta_c\gamma$, provided that an unapproximated single 
quark spin-flip operator is used. This conclusion is supported by the 
calculation of ref.~\cite{Snellman}, which is based on the instantaneous 
approximation to the Bethe-Salpeter equation, even though a quantitative 
understanding of the radiative transitions in charmonium was not achieved. 

\noindent
The situation concerning the M1 transitions in the heavy-light $Q\bar q$ 
mesons is more uncertain because of the scarcity of reliable empirical 
data. Only recently has a first measurement of the width of the $D^*$ state 
been published, which allows a determination of the partial width for the 
M1 transition $D^{\pm *}\rightarrow D^\pm \gamma$ from its known branching 
fraction. Even though the total width of the $D^{0*}$ is not known, the 
partial width for $D^{0*}\rightarrow D^0\gamma$ can be inferred from the 
measured width of the $D^{\pm *}$ and model calculations of the pionic 
widths of the $D^*$ mesons, as the relative branching fractions for $\pi$ 
and $\gamma$ emission are known. 

\noindent
As the velocity of the light constituent quark in the $Q\bar q$ systems is 
close to that of light, the relativistic corrections to both single quark 
and two-quark operators will {\it a priori} be large. It is shown in 
paper~{\bf V} that the large relativistic corrections to the single quark 
spin-flip operators, that yield unfavorable results for the M1 widths of 
$Q\bar q$ mesons, are in general counteracted by the two-quark operators 
associated with an interaction Hamiltonian that consists of OGE + scalar 
confinement components, thus allowing for better agreement with experiment. 
It is also suggested that the instanton induced interaction for $Q\bar q$ 
systems of ref.~\cite{instpap} may have a favorable effect on the 
predictions for the M1 widths of the $D^*$ mesons.

\section{Charge density and electric dipole operators}

\begin{figure}[ht!]
\begin{center}
\begin{tabular}{c c}
\begin{fmffile}{ex1}
\begin{fmfgraph*}(160,130) \fmfpen{thin}
\fmfcmd{%
 vardef port (expr t, p) =
  (direction t of p rotated 90)
   / abs (direction t of p)
 enddef;}
\fmfcmd{%
 vardef portpath (expr a, b, p) =
  save l; numeric l; l = length p;
  for t=0 step 0.1 until l+0.05:
   if t>0: .. fi point t of p
    shifted ((a+b*sind(180t/l))*port(t,p))
  endfor
  if cycle p: .. cycle fi
 enddef;}
\fmfcmd{%
 style_def brown_muck expr p =
  shadedraw(portpath(thick/2,2thick,p)
   ..reverse(portpath(-thick/2,-2thick,p))
   ..cycle)
 enddef;}
\fmfleft{i2,i1}
\fmfright{o2,o1}
\fmftop{o3}
\fmf{fermion,label=$p_1$}{i1,v3}
\fmf{fermion,label=$p_a$}{v3,v1}
\fmf{fermion,label=$p_1'$}{v1,o1}
\fmf{fermion,label=$p_2'$}{o2,v2}
\fmf{fermion,label=$p_2$}{v2,i2}
\fmf{photon,label=$q$,label.side=right}{v3,o3}
\fmf{brown_muck,lab.s=right,lab.d=4thick,lab=$V_{Q\bar Q}(k_2)$,label.side=right}{v1,v2}
\fmfdot{v1,v2,v3}
\fmfforce{(.1w,.85h)}{i1}
\fmfforce{(.9w,.85h)}{o1}
\fmfforce{(.1w,.15h)}{i2}
\fmfforce{(.9w,.15h)}{o2}
\fmfforce{(.2w,.h)}{o3}
\fmfforce{(.5w,.70h)}{v1}
\fmfforce{(.5w,.30h)}{v2}
\fmfforce{(.3w,.80h)}{v3}
\end{fmfgraph*}
\end{fmffile}
&
\begin{fmffile}{ex2}
\begin{fmfgraph*}(160,130) \fmfpen{thin}
\fmfcmd{%
 vardef port (expr t, p) =
  (direction t of p rotated 90)
   / abs (direction t of p)
 enddef;}
\fmfcmd{%
 vardef portpath (expr a, b, p) =
  save l; numeric l; l = length p;
  for t=0 step 0.1 until l+0.05:
   if t>0: .. fi point t of p
    shifted ((a+b*sind(180t/l))*port(t,p))
  endfor
  if cycle p: .. cycle fi
 enddef;}
\fmfcmd{%
 style_def brown_muck expr p =
  shadedraw(portpath(thick/2,2thick,p)
   ..reverse(portpath(-thick/2,-2thick,p))
   ..cycle)
 enddef;}
\fmfleft{i2,i1} 
\fmfright{o2,o1}
\fmftop{o3}
\fmf{fermion,label=$p_1$}{i1,v1}
\fmf{fermion,label=$p_b$}{v1,v3}
\fmf{fermion,label=$p_1'$}{v3,o1}
\fmf{fermion,label=$p_2'$}{o2,v2}
\fmf{fermion,label=$p_2$}{v2,i2}
\fmf{photon,label=$q$,label.side=right}{v3,o3}
\fmf{brown_muck,lab.s=right,lab.d=4thick,lab=$V_{Q\bar Q}(k_2)$,label.side=right}{v1,v2}
\fmfdot{v1,v2,v3}
\fmfforce{(.1w,.85h)}{i1}
\fmfforce{(.9w,.85h)}{o1}
\fmfforce{(.1w,.15h)}{i2}
\fmfforce{(.9w,.15h)}{o2}
\fmfforce{(.6w,.h)}{o3}
\fmfforce{(.5w,.70h)}{v1}
\fmfforce{(.5w,.30h)}{v2}
\fmfforce{(.7w,.80h)}{v3}
\end{fmfgraph*}
\end{fmffile}
\\ & \\
\begin{fmffile}{ex3}
\begin{fmfgraph*}(160,130) \fmfpen{thin}
\fmfcmd{%
 vardef port (expr t, p) =
  (direction t of p rotated 90)
   / abs (direction t of p)
 enddef;}
\fmfcmd{%
 vardef portpath (expr a, b, p) =
  save l; numeric l; l = length p;
  for t=0 step 0.1 until l+0.05:
   if t>0: .. fi point t of p
    shifted ((a+b*sind(180t/l))*port(t,p))
  endfor
  if cycle p: .. cycle fi
 enddef;}
\fmfcmd{%
 style_def brown_muck expr p =
  shadedraw(portpath(thick/2,2thick,p)
   ..reverse(portpath(-thick/2,-2thick,p))
   ..cycle)
 enddef;}
\fmfleft{i2,i1}
\fmfright{o2,o1}
\fmftop{o3}
\fmf{fermion,label=$p_1$}{i1,v3}
\fmf{fermion,label=$p_a$}{v3,v1} 
\fmf{fermion,label=$p_1'$}{v1,o1} 
\fmf{fermion,label=$p_2'$}{o2,v2}
\fmf{fermion,label=$p_2$}{v2,i2}
\fmf{photon,label=$q$,label.side=right}{v3,o3}
\fmf{brown_muck,lab.s=right,lab.d=4thick,lab=$V_{Q\bar Q}(k_2)$,label.side=right}{v1,v2}
\fmfdot{v1,v2,v3}
\fmfforce{(.1w,.85h)}{i1}
\fmfforce{(.9w,.75h)}{o1}
\fmfforce{(.1w,.05h)}{i2}
\fmfforce{(.9w,.05h)}{o2}
\fmfforce{(.5w,.60h)}{v1}
\fmfforce{(.5w,.20h)}{v2}
\fmfforce{(.6w,.80h)}{v3}
\fmfforce{(.5w,.h)}{o3}
\end{fmfgraph*}
\end{fmffile}
&
\begin{fmffile}{ex4}
\begin{fmfgraph*}(160,130) \fmfpen{thin}
\fmfcmd{%
 vardef port (expr t, p) =
  (direction t of p rotated 90)
   / abs (direction t of p)
 enddef;}
\fmfcmd{%
 vardef portpath (expr a, b, p) =
  save l; numeric l; l = length p;
  for t=0 step 0.1 until l+0.05:
   if t>0: .. fi point t of p
    shifted ((a+b*sind(180t/l))*port(t,p))
  endfor
  if cycle p: .. cycle fi
 enddef;}
\fmfcmd{%
 style_def brown_muck expr p =
  shadedraw(portpath(thick/2,2thick,p)
   ..reverse(portpath(-thick/2,-2thick,p))
   ..cycle)
 enddef;}
\fmfleft{i2,i1}
\fmfright{o2,o1}
\fmftop{o3}
\fmf{fermion,label=$p_1$}{i1,v1}
\fmf{fermion,label=$p_b$,label.side=right}{v1,v3}
\fmf{fermion,label=$p_1'$}{v3,o1}
\fmf{fermion,label=$p_2'$}{o2,v2}
\fmf{fermion,label=$p_2$}{v2,i2}
\fmf{photon,label=$q$,label.side=right}{v3,o3}
\fmf{brown_muck,lab.s=right,lab.d=4thick,lab=$V_{Q\bar Q}(k_2)$,label.side=right}{v1,v2}
\fmfdot{v1,v2,v3}
\fmfforce{(.1w,.75h)}{i1}
\fmfforce{(.9w,.85h)}{o1}
\fmfforce{(.1w,.05h)}{i2}
\fmfforce{(.9w,.05h)}{o2}
\fmfforce{(.5w,.60h)}{v1}
\fmfforce{(.5w,.20h)}{v2}
\fmfforce{(.4w,.80h)}{v3}
\fmfforce{(.3w,.h)}{o3}
\end{fmfgraph*}
\end{fmffile} \\
\end{tabular}
\caption{Two-quark contributions to the $Q\bar Q$ current and charge 
density operators. In the decomposition of the $Q\bar Q\gamma$ vertex, the 
irreducible two-quark contributions give rise to exchange current 
operators, the most important of which are illustrated by the upper Born 
diagrams. In order to obtain the correct two-quark contribution, the 
positive energy part of the intermediate propagator is subtracted in the 
lower Born diagrams, since that part is already accounted for by the impulse 
approximation. Note that similar diagrams describe photon emission by the 
heavy antiquark. In the case of the $Q\bar Q$ interaction, the scalar 
confining and vector OGE components have been taken into account. The 
contributions from the instanton induced interaction have not been 
considered but are in any case small for the $Q\bar Q$ systems.}
\label{feyn}
\end{center}
\end{figure}
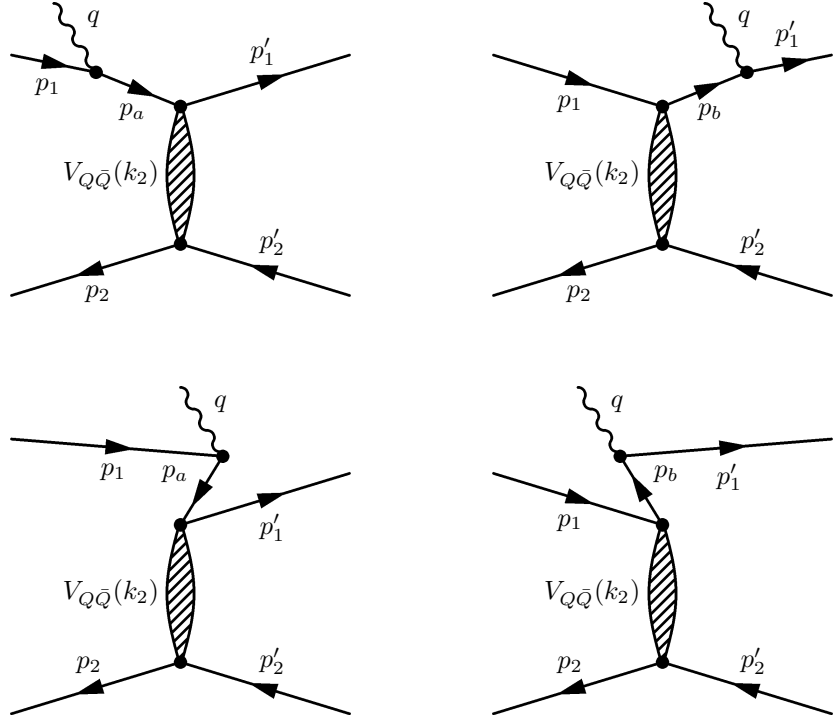

The electromagnetic transition amplitude for a two-quark system, in the 
impulse approximation, is of the form
\begin{equation}
T_{fi}\:=\:-\int d^3r_1 d^3r_2\:
\varphi_f^*(\vectr{$r$}_1,\vectr{$r$}_2)\:
\vectr{$\hat\varepsilon$}\cdot\left[e^{i\vec q\cdot\vec r_1}\,\vectr{$
\jmath$}_1(\vectr{$q$}) + e^{i\vec q\cdot\vec r_2}\,\vectr{$\jmath$}_2
(\vectr{$q$})\right] \varphi_i(\vectr{$r$}_1,\vectr{$r$}_2), \label{matr}
\end{equation}
where \vectr{$q$} and \vectr{$\hat\varepsilon$} denote the
momentum and polarization of the emitted photon, respectively, while
$\varphi_i$ and $\varphi_f$ denote the orbital wave functions of the
initial and final heavy quarkonium states. In the above equation, 
$\vectr{$\jmath$}_1$ and $\vectr{$\jmath$}_2$ denote the
single quark current operators of quarks 1 and 2, respectively. By Fourier 
transformation, the current operators may be rewritten as
\begin{eqnarray}
\vectr{$\jmath$}\,(\vectr{$q$}) &=& \int d^3r' e^{i\vec q\cdot\vec r\,'}
\vectr{$\jmath$}\,(\vectr{$r$}') \label{Four} \\
&=& - \int d^3r' \vectr{$r$}' (\vectr{$\jmath$}\cdot\nabla)\:e^{i\vec 
q\cdot\vec r\,'} - \int d^3r' e^{i\vec q\cdot\vec r\,'} \vectr{$r$}'
(\nabla\cdot\vectr{$\jmath$}\,), \label{Jackson}
\end{eqnarray}
from which the E1 approximation is obtained if the exponentials in
eq.~(\ref{Jackson}) are dropped (i.e. $\vectr{$q$}\rightarrow 0$). 
Application of the continuity equation then gives 
$\nabla\cdot\vectr{$\jmath$} = i\omega\rho$. For nonzero \vectr{$q$}, the 
second term in eq.~(\ref{Jackson}) has to be retained without 
approximation. Note that $\vectr{$\jmath$}\,(\vectr{$q$})$ is taken to 
contain the quantity in square brackets in eq.~(\ref{matr}). Application 
of eq.~(\ref{Jackson}) together with eq.~(\ref{matr}) then leads to the 
following form for the amplitude of a $\gamma$ transition,
\begin{equation}   
T_{fi}\:=\: i\,|\vectr{$q$}|\,\int d^3r_1 d^3r_2 \:
\varphi_f^*(\vectr{$r$}_1,\vectr{$r$}_2)\:
\vectr{$\hat\varepsilon$}\cdot\vectr{$d$}\,(\vectr{$r$}_1,\vectr{$r$}_2)
\,\varphi_i(\vectr{$r$}_1,\vectr{$r$}_2).\label{matr2}
\end{equation}
The dipole operator $\vectr{$d$}(\vectr{$r$}_1,\vectr{$r$}_2)$,
\begin{equation}
\vectr{$d$}\,(\vectr{$r$}_1,\vectr{$r$}_2) = \int d^3r' e^{i\vec q\cdot\vec 
r\,'}\vectr{$r$}'\, \rho(\vectr{$r$}',\vectr{$r$}_1,\vectr{$r$}_2),
\end{equation}
reduces to the E1 approximation in the limit $\vectr{$q$}\rightarrow 0$. 
In general, the charge density operator $\rho(\vectr{$r$}')$ contains, in 
addition to the single quark contribution $\rho_{\mathrm{sq}}$, an 
exchange part $\rho_{\mathrm{ex}}$, which arises from the two-quark 
currents that are illustrated by the diagrams in Fig.~\ref{feyn}. A 
necessary constraint is that two-quark contributions to the charge density 
must have vanishing volume integrals. The dipole operator that corresponds 
to the single quark charge operator 
$\rho_{\mathrm{sq}}(\vectr {$r$}',\vectr{$r$}) = 
\rho_1(\vectr{$r$}',\vectr{$r$}_1) + \rho_2(\vectr 
{$r$}',\vectr{$r$}_2)$ may be expressed as
\begin{equation}
\vectr{$d$}\,_{\mathrm{sq}}(\vectr{$r$}_1,\vectr{$r$}_2) =
\int \frac{d^3q}{(2\pi)^3}\,d^3r'\,e^{i\vec q_f\cdot\vec r\,'}
\left[\rho_1(\vectr{$q$})\,\vectr{$r$}'\,e^{i\vec q\cdot(\vec r_1 - \vec 
r\,')}\:+\:\rho_2(\vectr{$q$})\,\vectr{$r$}'\,e^{i\vec q\cdot(\vec r_2 - 
\vec r\,')}\right],  
\end{equation}
\newpage

\noindent
which upon evaluation yields
\begin{equation}
\vectr{$d$}\,_{\mathrm{sq}}(\vectr{$r$}_1,\vectr{$r$}_2) = 
\lim_{q\rightarrow q_f}
\left[\vectr{$r$}_1\:e^{i\vec q\cdot\vec r_1\,}\rho_1(\vectr{$q$}) + 
e^{i\vec q\cdot\vec r_1\,}\: i\nabla_{\vec q}\,\rho_1(\vectr{$q$})\right]
+ (1\rightarrow 2),
\label{sqop}
\end{equation}
where $q_f$ refers to the physical photon momentum of each transition. 
The above form reduces to the E1 expression by the substitution 
$q_f\rightarrow 0$. The nonrelativistic single quark dipole operator is 
therefore of the form
\begin{equation}
\vectr{$d$}\,_{\mathrm{sq}}(\vectr{$r$}_1,\vectr{$r$}_2) = 
Q_1\,\vectr{$r$}_1\:
e^{i\vec q_f\cdot\vec r_1\,} +  Q_2\,\vectr{$r$}_2\:e^{i\vec q_f\cdot\vec r_2\,},
\label{dip}
\end{equation}
where $Q_1$ is the charge of the heavy quark, while $Q_2$ denotes that of 
the heavy antiquark. Insertion of eq.~(\ref{dip}) into eq.~(\ref{matr2}) 
yields the single quark dipole operator
\begin{equation}
\vectr{$d$}\,_{\mathrm{sq}}(\vectr{$r$}) = \left[\frac{Q_1m_2 -
Q_2m_1}{m_1+m_2}\right]\:\vectr{$r$}\:e^{i\vec q_f\cdot\vec r/2\,}.
\label{cmdip}
\end{equation}

\noindent
The charge density operator in eq.~(\ref{sqop}) is, to second order in 
$v/c$, of the form~\cite{Helminen}
\begin{equation}
\rho_{\mathrm{sq}} \simeq Q_1\left[1 - \frac{\vectr{$q$}^2}{8m^2} + 
\frac{i\vectr{$\sigma$}_1\cdot\vectr{$p$}'_1\times\vectr{$p$}_1}{4m^2}
\:\right] + (1\rightarrow 2), \label{rsq}
\end{equation}  
where the second term on the r.h.s. is the relativistic Darwin-Foldy term. 
The effect of this term is very small because of the large masses of the 
heavy constituent quarks. The justification of this expansion lies in the 
small coefficient of that term; It has been shown e.g. in papers {\bf I} and 
{\bf VI} that such an expansion cannot be used for the magnetic moment 
operator. 

\noindent
If the two-quark exchange charge operators from the Born
diagrams in Fig.~\ref{feyn} are decomposed as 
$\rho_{\mathrm{ex}}(\vectr{$r$}',\vectr{$r$}_1,\vectr{$r$}_2) 
= \rho_{\mathrm{ex}1}(\vectr{$r$}',\vectr{$r$}_1) + 
\rho_{\mathrm{ex}2}(\vectr{$r$}',\vectr{$r$}_2)$, then the contribution 
from quark~1 may be expressed as   
\begin{equation}
\rho_{\mathrm{ex}1}(\vectr{$r$}',\vectr{$r$}_1) = \int\frac{d^3q}{(2\pi)^3}\,
e^{i\vec q\cdot(\vec r_1 - \vec r\,')}\, \int\frac{d^3k_2}{(2\pi)^3}\,
e^{-i\vec k_2\cdot\vec r}\:\rho_{\mathrm{ex}1}(\vectr{$q$},\vectr{$k$}_2).
\end{equation}
Here $\vectr{$k$}_2$ is the momentum transfered to the heavy antiquark 
and~\vectr{$r$} is defined as $\vectr{$r$}_1 - \vectr{$r$}_2$. The exchange 
charge contribution to the two-quark dipole 
operator 
\begin{equation}
\vectr{$d$}\,_{\mathrm{ex}}\,(\vectr{$r$}_1,\vectr{$r$}_2) = \int d^3r' 
e^{i\vec q_f\cdot\vec 
r\,'}\vectr{$r$}'\,\rho_{\mathrm{ex}}(\vectr{$r$}',\vectr{$ 
r$}_1,\vectr{$r$}_2)
\end{equation}
from quark~1 may then be expressed as
\begin{eqnarray}
\vectr{$d$}\,_{\mathrm{ex1}}\,(\vectr{$r$}_1)\:=\:\vectr{$r$}_1\,e^{i\vec
q_f\cdot\vec r_1}&&\!\!\!\!\!\!\!\!\!\!\int\frac{d^3k_2}{(2\pi)^3}\,
e^{-i\vec k_2\cdot\vec r}\,\rho_{\mathrm{ex}1}(\vectr{$q$}_f,\vectr{$k$}_2)
\nonumber \\
&&\!\!\!\!\!\!\!\!\!\!
-\lim_{q\rightarrow q_f}\left[e^{i\vec q\cdot\vec r_1}\,i\nabla_{\vec q}  
\!\int\frac{d^3k_2}{(2\pi)^3}\,e^{-i\vec k_2\cdot\vec r}\,
\rho_{\mathrm{ex}1}(\vectr{$q$},\vectr{$k$}_2)\right], \label{exop}
\end{eqnarray}
which again reduces to the E1 approximation by the substitution $q_f  
\rightarrow 0$.

\noindent
The exchange charge density operators that are associated with the $Q\bar
Q$ interaction have been extracted in \mbox{ref.~\cite{Helminen},} for 
different Lorentz invariants for equal-mass systems. When generalized to 
unequal quark masses, the required operators are obtained as   
\begin{eqnarray}
&&\rho_{\mathrm{ex}}^{\,\mathrm c} = 
\frac{Q_1}{4m_1^{\,3}}\,q^2\,V_c(\vectr{$k$}_2\,) 
+ \frac{Q_2}{4m_2^{\,3}}\,q^2\,V_c(\vectr{$k$}_1\,),
\label{rconf} \\
&&\rho_{\mathrm{ex}}^{\,\mathrm g} =
\frac{Q_1}{4m_1^{\,2}}
\left[\frac{\vectr{$q$}\cdot\vectr{$k$}_2}{m_1} + \frac{2}{3}\,
\frac{\vectr{$q$}\cdot\vectr{$k$}_2\:\vectr{$\sigma$}_1\cdot\vectr{$\sigma$}
_2}{m_2}\right] 
V_g(\vectr{$k$}_2\,) +
\frac{Q_2}{4m_2^{\,2}}
\left[\frac{\vectr{$q$}\cdot\vectr{$k$}_1}{m_2} \right. \\
&& \hspace{7cm} \left.
+ \frac{2}{3}\,\frac{\vectr{$q$}\cdot\vectr{$ 
k$}_1\:\vectr{$\sigma$}_1\cdot\vectr{$\sigma$}_2}{m_1}\right]
V_g(\vectr{$k$}_1\,). \nonumber
\label{roge}
\end{eqnarray}
In the above expressions, $V_c$ and $V_g$ denote the Fourier transforms of 
the confining and OGE interactions, respectively. Evaluation of 
eq.~(\ref{exop}) thus yields the dipole operators
\begin{eqnarray}
\vectr{$d$}\,_{\mathrm{ex}}^{\mathrm{Conf}}(\vectr{$r$})\!\!\! &=& 
\!\!\!q_f^2\,
\left[\frac{Q_1}{4m_1^{\,3}}\,\frac{m_2}{m_1+m_2} -
\frac{Q_2}{4m_2^{\,3}}\,\frac{m_1}{m_1+m_2}\right]\:
\vectr{$r$}\:\:V_c(r)\:\:e^{i\vec q_f\cdot\vec r/2\,}, \label{cdip} \\
\vectr{$d$}\,_{\mathrm{ex}}^{\mathrm{Oge}}(\vectr{$r$}) \!\!\! &=& \!\!\!
\left[\frac{Q_1}{4m_1^{\,2}}\left(\frac{1}{m_1} +
\frac{2}{3}\frac{\vectr{$\sigma$}_1\!\cdot\vectr{$\sigma$}_2}{m_2}\right)
-\frac{Q_2}{4m_2^{\,2}}\left(\frac{1}{m_2} +
\frac{2}{3}\frac{\vectr{$\sigma$}_1\!\cdot\vectr{$\sigma$}_2}{m_1}\right)
\right]\:\vectr{$r$}\:\:e^{i\vec q_f\cdot\vec r/2\,}\:\:
\frac{\partial V_g(r)}{r\,\partial r}. \nonumber
\end{eqnarray}
Here $V_c(r)$ is the linear confining interaction, while $V_g(r)$ denotes 
the form of the OGE interaction in configuration space, which is taken to 
include the effects of the running coupling of QCD. 

\section{Current density and magnetic moment operators}

In the impulse approximation, the spin-flip magnetic moment operator for 
M1 transitions between $S$-wave heavy quarkonium states have been derived 
from the amplitude
\begin{equation}
T_{fi}\:=\:-(2\pi)^3\,\delta^3(P_f - P_i - q_f)
\int d^3r\: \varphi_f^*(\vectr{$r$})
\:\vectr{$\hat\varepsilon$}\cdot\left[e^{i\vec q\cdot\vec r/2}\,\vectr{$
\jmath$}_1(\vectr{$q$}) + e^{-i\vec q\cdot\vec r/2}\,\vectr{$
\jmath$}_2(\vectr{$q$})\right] \varphi_i(\vectr{$r$}),   
\label{matrM1}
\end{equation}
where $\vectr{$r$} = \vectr{$r$}_1 - \vectr{$r$}_2$. Expansion of the 
exponential in eq.~(\ref{Four}) according to $\simeq 1 + 
i\vectr{$q$}\cdot\vectr{$r$}'$ then yields the M1 and E2 amplitudes for 
photon emission. Upon isolation of the M1 contribution, the matrix element 
for $J/\psi\rightarrow \eta_c\gamma$ and $\psi'\rightarrow \eta_c\gamma$ may 
be written in the form
\begin{equation}
{\cal M}_{fi} =\,i\int d^3r \: \varphi_f^*(\vectr{$r$})\:\:\vectr{$q$}\times
\vectr{$\hat\varepsilon$}\cdot\vectr{$\mu$}_{\mathrm{sf}}
\:\:\varphi_i(\vectr{$r$}),
\label{matr3}
\end{equation}  
where $\vectr{$\mu$}_{\mathrm{sf}}$ denotes the spin-flip part of the 
magnetic moment operator
\begin{equation}
\vectr{$\mu$} = \frac{1}{2}\int d^3r'\: \vectr{$r$}'\times 
\vectr{$\jmath$}\,(\vectr{$r$}'). 
\label{mueq}
\end{equation}
In eq.~(\ref{mueq}), the current operator consists of a single
quark contribution $\vectr{$\jmath$}_{\mathrm{sq}}$ and a two-quark 
contribution $\vectr{$\jmath$}_{\mathrm{ex}}$, which arises from the pair 
terms given in Fig.~\ref{feyn}. The corresponding single quark magnetic 
moment operator may be expressed as
\begin{equation}
\vectr{$\mu$}_{\mathrm{sq}}\:=\:\frac{1}{2} \int
\frac{d^3q}{(2\pi)^3}\,d^3r'\,\left[\,\vectr{$r$}'\times
\vectr{$\jmath$}_1(\vectr{$q$})\,e^{i\vec q\cdot(\vec r/2 - \vec r\,')}\:   
+\vectr{$r$}'\times \vectr{$\jmath$}_2(\vectr{$q$})\,e^{-i\vec q\cdot(\vec 
r/2 + \vec r\,')}\right],
\end{equation}  
which yields
\begin{equation}
\vectr{$\mu$}_{\mathrm{sq}}\:=\:\lim_{\vec q \rightarrow
0}\left[-\frac{i}{2}\:\nabla_q\times
\left(e^{i\vec q\cdot\vec r/2}\,\vectr{$\jmath$}_1(\vectr{$q$}) +
e^{-i\vec q\cdot\vec r/2}\,\vectr{$\jmath$}_2(\vectr{$q$})\right)\:\right].
\label{musq}
\end{equation}

\noindent
The magnetic moment operator is given by eq.~(\ref{musq}) in the
nonrelativistic impulse approximation. However, previous work has
demonstrated that the static magnetic moment operators of the baryons 
receive large corrections from the canonical boosts of the constituent quark
spinors~\cite{Barpap}. Furthermore, it has been shown in paper~{\bf I} that 
the nonrelativistic impulse approximation does not provide a satisfactory 
description of the spin-flip magnetic moment operators for $Q\bar Q$ 
systems, even though the masses of the charm and bottom constituent quarks 
are large. The matrix element that corresponds to eq.~(\ref{matr3}) in the 
relativistic impulse approximation is of the form
\begin{equation}
{\cal M}_{fi}^{\mathrm{Rel}} =\,i\int\frac{d^3P}{(2\pi)^3}\,d^3r\,d^3r'
\:e^{i\vec P\cdot (\vec r\,' - \vec r\,)}\:
\varphi_f^*(\vectr{$r$}')\:\:\vectr{$q$}\times \vectr{$\hat\varepsilon$}\cdot
\vectr{$\mu$}\,_{\mathrm{sq}}^{\mathrm{Rel}}(\vectr{$P$})
\:\:\varphi_i(\vectr{$r$}),\label{matr4}
\end{equation}
where the final and initial state coordinates $\vectr{$r$}'$ and 
\vectr{$r$} are defined as $\vectr{$r$}'_1 - \vectr{$r$}'_2$ and 
$\vectr{$r$}_1 - \vectr{$r$}_2$ respectively. In eq.~(\ref{matr4}), the 
momentum variable \vectr{$P$} is
defined as $\vectr{$P$} = (\vectr{$p$}' + \vectr{$p$})/2$, where 
$\vectr{$p$}'$ and \vectr{$p$} are the relative momenta in the 
representation $\vectr{$p$}_1 = \vectr{  
$P$}_i/2 + \vectr{$p$}$, $\vectr{$p$}_2 = \vectr{$P$}_i/2 - \vectr{$p$}$ 
and $\vectr{$p$}'_1 = \vectr{$P$}_f/2 + \vectr{$p$}'$, $\vectr{$p$}'_2 = 
\vectr{$P$}_f/2 - \vectr{$p$}'$. The relativistic single quark magnetic 
moment operator that appears in the matrix element~(\ref{matr4}) is of the 
form
\begin{equation}
\vectr{$\mu$}\,_{\mathrm{sq}}^{\mathrm{Rel}}\:=\:\lim_{\vec q \rightarrow 0}   
\left[-\frac{i}{2}\:\nabla_q\times\left[
e^{i\vec q\cdot(\vec r\,'+\vec r\,)/4}
\,\left(\vectr{$\jmath$}_1(\vectr{$q$},\vectr{$P$}) +
\vectr{$\jmath$}_2(\vectr{$q$},\vectr{$P$})\right)\right]\:\right],
\label{musqrel} 
\end{equation}
where the single quark current operators 
$\vectr{$\jmath$}_i\,(\vectr{$q$},\vectr{$P$})$ are now treated without 
approximation. 
\pagebreak

\noindent
In the nonrelativistic approximation, the spin-dependent part of the 
single quark current operator is given by
\begin{equation}
\vectr{$\jmath$}\,_{\mathrm{sq}}^{\mathrm{spin}} = 
\frac{ie}{2}\,(\vectr{$\sigma$}_1 + \vectr{$\sigma$}_2)\times\vectr{$q$}\,
\left[\frac{Q_1}{2m_1} + \frac{Q_2}{2m_2}\right]
+ \frac{ie}{2}\,(\vectr{$\sigma$}_1 - \vectr{$\sigma$}_2)
\times\vectr{$q$}\,\left[\frac{Q_1}{2m_1} - \frac{Q_2}{2m_2}\right],
\label{nrscurr}
\end{equation}
where the first term describes the
magnetic moment of the two-quark system whereas the second term is the
spin-flip operator for an M1 transition in the nonrelativistic impulse
approximation (NRIA). In order to obtain the relativistic single quark 
current operator to be used with eq.~(\ref{musqrel}), the nonrelativistic 
current operator for quark~1 should be replaced according to
\begin{eqnarray}
\vectr{$\jmath$}_1 &=& \frac{e\,Q_1}{2m_1}\,\left[\,\vectr{$p$}_1 + 
\vectr{$p$}'_1 + i\vectr{$\sigma$}_1\times(\vectr{$p$}'_1 - 
\vectr{$p$}_1)\,\right] \label{relcurr} \\
&\longrightarrow &
e\,Q_1\:\sqrt{\frac{(E_1'+m_1)(E_1+m_1)}{4E_1'E_1}}\left[\frac{\vectr{ 
$p$}_1}{E_1+m_1} + \frac{\vectr{$p$}'_1}{E_1'+m_1} \right.\nonumber \\ 
&& \left. \hspace{4.5cm} + \:\: 
i\vectr{$\sigma$}_1\times\left(\frac{\vectr{$p$}'_1}{E_1'+m_1}
- \frac{\vectr{$p$}_1}{E_1+m_1}\right)\,\right], \nonumber
\end{eqnarray}
and similarly for quark~2. In the above equation, the energy factors of the
quarks are defined as $E_1 = \sqrt{\vectr{$p$}_1^{\,2} + m_1^{\,2}}$ and 
$E_1' = \sqrt{\vectr{$p$}_1^{'2} + m_1^{\,2}}$. The spin-flip magnetic 
moment operator in the non-relativistic impulse 
approximation~(NRIA) may be obtained by insertion of eq.~(\ref{nrscurr}) 
into eq.~(\ref{musq}), giving
\begin{equation}
\vectr{$\mu$}\,_{\mathrm{sq}}\:=\:\frac{e}{2}\left[
\frac{Q_1}{2m_1} - \frac{Q_2}{2m_2}\right]\:(\vectr{$\sigma$}_1 - 
\vectr{$\sigma$}_2).
\label{NRIA}
\end{equation}
The corresponding operator in the relativistic impulse approximation (RIA)
has been considered in refs.~\cite{Lahdeout,Barpap}, and may for
transitions between $S$-wave states be expressed as
\begin{equation}
\vectr{$\mu$}\,_{\mathrm{sq}}^{\mathrm{Rel}}\:=\:\frac{e}{2}\left[
\frac{Q_1}{2m_1}f_1^\gamma -
\frac{Q_2}{2m_2}f_2^\gamma\right]\:(\vectr{$\sigma$}_1 - 
\vectr{$\sigma$}_2),
\label{RIA}
\end{equation}
where the relativistic factors $f_i^\gamma$ are defined as
\begin{equation}
f_i^\gamma = \frac{m_i}{3E_i}\left[2+\frac{m_i}{E_i}\right],
\label{RIA2}  
\end{equation}
where $E_i$ denotes the energy factor $E_i = \sqrt{\vectr{$P$}^2 + 
m_i^{\,2}}$. This shows that a relativistic treatment will lead to an 
effective weakening of the NRIA spin-flip operator.

\noindent
In addition to the above single quark current operators, the pair terms in 
Fig.~\ref{feyn} also give large contributions to the magnetic moment 
operators of mesons and baryons~\cite{Tsushima}. However, in the case of the 
magnetic moments of the baryons, additional complications are known to arise 
from flavor dependent meson exchange interactions which also contribute 
significant exchange current operators~\cite{Barpap}. If the exchange 
current operators of Fig.~\ref{feyn} are decomposed as
$\vectr{$\jmath$}_{\mathrm{ex}}(\vectr{$q$},\vectr{$k$}_1,\vectr{$k$}_2) = 
\vectr{$\jmath$}_{\mathrm{ex1}}(\vectr{$q$},\vectr{$k$}_2)
+\vectr{$\jmath$}_{\mathrm{ex2}}(\vectr{$q$},\vectr{$k$}_1)$, then the 
contribution to the two-quark magnetic moment operator may be written in the 
form
\begin{equation}
\vectr{$\mu$}_{\mathrm{ex}}
\:=\:\frac{1}{2}\int\frac{d^3q\,d^3k_2}{(2\pi)^6}\,d^3r'\,
\left[e^{i\vec q\cdot(\vec r/2 - \vec r\,')}\,e^{i\vec k_2\cdot\vec
r}\,\vectr{$r$}'\times \vectr{$\jmath$}_{\mathrm{ex1}}(\vectr{$q$},
\vectr{$k$}_2)\:\:+\:\: (1\rightarrow 2)\right],
\end{equation}
where it is again understood that $\vectr{$r$}\rightarrow -\vectr{$r$}$ in 
the contribution from quark~2. Evaluation of the above equation leads to an  
expression analogous to eq.~(\ref{musq}),
\begin{eqnarray}
\vectr{$\mu$}_{\mathrm{ex}} &=& \lim_{\vec q \rightarrow
0}\left[-\frac{i}{2}\:\nabla_q\times
\left(e^{i\vec q\cdot\vec r/2}\!\int \frac{d^3k_2}{(2\pi)^3}\, 
e^{-i\vec k_2\cdot\vec 
r}\,\vectr{$\jmath$}_{\mathrm{ex1}}(\vectr{$q$},\vectr{$k$}_2) 
\right.\right.\nonumber \\
&& \left.\left. \hspace{3.5cm} + \:\: e^{-i\vec q\cdot\vec r/2}\!\int 
\frac{d^3k_1}{(2\pi)^3}\, e^{i\vec k_1\cdot\vec 
r}\,\vectr{$\jmath$}_{\mathrm{ex2}}(\vectr{$q$},\vectr{$k$}_1)
\right)\:\right].
\label{muex}
\end{eqnarray}

\noindent
As the exchange current operators for most Lorentz invariants do not depend
explicitly on the photon momentum \vectr{$q$}, one notable exception being
that for the scalar invariant~\cite{Tsushima}, then the exchange magnetic
moment operators turn out to be difficult to calculate directly from
eq.~(\ref{muex}). It has therefore been shown in paper~{\bf VI} that a 
convenient way to extract the two-quark magnetic moment operators 
results, if eq.~(\ref{muex}) is cast in the form
\begin{equation}
\vectr{$\mu$}\,_{\mathrm{ex}}\:=\:\lim_{\vec q \rightarrow 0}
\left[\,-\frac{i}{2}\,\int\!\frac{d^3k}{(2\pi)^3}\,e^{-i\vec k\cdot\vec
r}\:\nabla_q\times\left\{
\vectr{$\jmath$}_{\mathrm{ex1}}\left(\frac{\vectr{$q$}}{2} + 
\vectr{$k$}\right) +
\vectr{$\jmath$}_{\mathrm{ex2}}\left(\frac{\vectr{$q$}}{2} - 
\vectr{$k$}\right) \right\}\right], \label{muex2}
\end{equation}
which is similar to that obtained in ref.~\cite{Tsushima}. By means of 
eq.~(\ref{muex2}), it is now possible to consider the
two-quark current operators for the scalar confining and vector OGE
interactions, as calculated from the diagrams in Fig.~\ref{feyn} in 
paper~{\bf V} and ref.~\cite{Tsushima}. The two-quark current operator 
associated with the scalar confining interaction is then of the form
\begin{eqnarray}
\vectr{$\jmath$}\,_{\mathrm{ex}}^{\:\:\mathrm{c}}(\vectr{$q$},\vectr{$k$}_1,
\vectr{$k$}_2)\:\:=\:\: - e\,\left(\frac{Q_1^*\vectr{$P$}_1}{m_1^{\,2}} + 
\frac{Q_2^*\vectr{$P$}_2}{m_2^{\,2}}\right.\!\!\!\!\!\!\!\!\!\!\!
&&\left.+\:\:\frac{i}{2}(\vectr{$\sigma$}_1+\vectr{$\sigma$}_2)\times\vectr{ 
$q$}\left[\frac{Q_1^*}{2m_1^{\,2}} + \frac{Q_2^*}{2m_2^{\,2}}\right]\right.
\quad\quad\quad \nonumber \\
&&\left.+\:\:\frac{i}{2}(\vectr{$\sigma$}_1-\vectr{$\sigma$}_2)\times\vectr{ 
$q$}\left[ \frac{Q_1^*}{2m_1^{\,2}} - \frac{Q_2^*}{2m_2^{\,2}}\right]\right), 
\label{2qcurr}
\end{eqnarray}
where the variables $Q_1^*$ and $Q_2^*$ are defined as $Q_1^* =
V_c(\vectr{$k$}_2)Q_1$ and $Q_2^* = V_c(\vectr{$k$}_1)Q_2$, respectively. 
The corresponding current operator for the OGE interaction may be expressed 
as
\begin{eqnarray}
\vectr{$\jmath$}\,_{\mathrm{ex}}^{\:\:\mathrm{g}}
(\vectr{$q$},\vectr{$k$}_1,\vectr{$k$}_2)
\!\!&=&\!\! -e\,\left(Q_1^*\left[ \frac{i\vectr{$\sigma$}_1\times\vectr{$
k$}_2}{2m_1^{\,2}} + \frac{2\vectr{$P$}_2 + i\vectr{$\sigma$}_2\times\vectr{$
k$}_2}{2m_1m_2}\right] + Q_2^*\left[
\frac{i\vectr{$\sigma$}_2\times\vectr{$k$}_1}{2m_2^{\,2}} \right.\right. 
\quad\quad\quad\quad \\
&& \hspace{7cm} \left.\left. 
+\:\frac{2\vectr{$P$}_1 + i\vectr{$\sigma$}_1\times
\vectr{$k$}_1}{2m_1m_2}\right]\right), \nonumber
\label{Ogecurr}
\end{eqnarray}
with $Q_1^* = V_g(\vectr{$k$}_2)Q_1$ and $Q_2^* = V_g(\vectr{$k$}_1)Q_2$. As 
the above equation depends only on $\vectr{$k$}_1$ and $\vectr{$k$}_2$, the 
OGE magnetic moment operator is most conveniently calculated using 
eq.~(\ref{muex2}). The corresponding spin-flip operators for transitions 
between $S$-wave quarkonium states have been obtained in paper~{\bf V} as
\begin{equation}
\vectr{$\mu$}_{\mathrm{ex}}^{\:\mathrm{Conf}} =
-\frac{eV_c(r)}{4}\left\{\left[\frac{Q_1}{m_1^{\,2}} -
\frac{Q_2}{m_2^{\,2}}\right]\:(\vectr{$\sigma$}_1 - \vectr{$\sigma$}_2)
+\left[\frac{Q_1}{m_1^{\,2}} + 
\frac{Q_2}{m_2^{\,2}}\right]\:(\vectr{$\sigma$}_1 + 
\vectr{$\sigma$}_2)\right\}  
\label{muc}
\end{equation}
for the scalar confining interaction, and
\begin{eqnarray}
\vectr{$\mu$}_{\mathrm{ex}}^{\:\mathrm{Oge}}
&=& -\frac{eV_g(r)}{8}\left\{\left[\frac{Q_1}{m_1^{\,2}} - \frac{Q_2}{m_2^{\,2}}
- \frac{Q_1-Q_2}{m_1m_2}\right]\:(\vectr{$\sigma$}_1 - \vectr{$\sigma$}_2) 
\right. \nonumber \\
&& \hspace{3cm} \left. 
+\left[\frac{Q_1}{m_1^{\,2}} + \frac{Q_2}{m_2^{\,2}}
+\frac{Q_1+Q_2}{m_1m_2}\right]\:(\vectr{$\sigma$}_1 + 
\vectr{$\sigma$}_2)\right\}
\label{mug}
\end{eqnarray}
for the OGE interaction. 
\pagebreak

\noindent
For equal constituent quark masses,
eqs.~(\ref{muc}) and~(\ref{mug}) reduce to the expressions given in
ref.~\cite{Tsushima}. Note that the presence of a spin-flip term in the
OGE operator~(\ref{mug}) is solely a consequence of the difference in mass
between the constituent quarks, and will thus not contribute to the M1 
widths of the charmonium and bottomonium states. Similarly, the 
terms that are symmetric in the quark spins vanish for equal mass
quarkonia. However, in the case of the $B_c^\pm$ system, these terms will
contribute to the magnetic moment of the $c\bar b$ system. Also the
spin-flip M1 transitions in the $B_c^\pm$ system will receive a 
contribution from the OGE operator.

\section{Widths for radiative decay}

The widths for E1 dominated transitions of the type
$\chi_{cJ}\rightarrow J/\psi\,\gamma$ or $\psi'\rightarrow
\chi_{cJ}\,\gamma$ have in paper~{\bf VI} been calculated according to
\begin{equation}
\Gamma\:\:=\:\:{\cal S}_{fi}
\,\frac{2J_f\!+\!1}{3}\:q^3\alpha\:\frac{M_f}{M_i}
\left[\:\frac{4}{9}\:|{\cal M}_0|^2 + \frac{8}{9}\:|{\cal M}_2|^2\right],  
\label{dyndec}
\end{equation}
where $J_f$ is the total angular momentum of the final quarkonium state,   
and $q$ is the momentum of the emitted photon. The widths for 
$\psi'\rightarrow \chi_{cJ}\,\gamma$ with $J = 0,1,2$ are then expected to  
scale as $1\,:\,3\,:\,5$ respectively, but that result is highly modified by the 
large hyperfine splittings of the $Q\bar Q$ with $L=1$. The statistical 
factor ${\cal S}_{fi}$ is defined as in ref.~\cite{Quigg} and assumes the 
values ${\cal S}_{fi} = 1$ for a triplet-triplet transition and ${\cal 
S}_{fi} = 3$ for a singlet-singlet transition. On the other hand, in 
paper~{\bf VI} the widths for transitions between $D$- and $P$-wave states 
were obtained from
\begin{equation}
\Gamma\:\:=\:\:4\,{\cal S}_{fi}
\,\frac{2J_f\!+\!1}{27}\:q^3\alpha\:\frac{M_f}{M_i}\:|{\cal M}_0|^2,
\quad\quad {\cal S}_{fi}\:\:=\:\: 18\,
\left\{\begin{array}{ccc}
2   & 1 & J_d \\
J_p & 1 & 1   \\
\end{array}
\right \}^2, \label{Ddec}
\end{equation}
where $J_d$ and $J_p$ are the total angular momenta of the $D$- and
$P$-wave states, respectively. The values of ${\cal S}_{fi}$ are then given 
by the above Wigner~$6j$ symbol. Note that the triangularity of the $6j$
symbol requires that $|J_d-J_p|$ = 1 or 0. Transitions that
change the value of $J$ by more than one unit are thus forbidden. In
eqs.~(\ref{dyndec}) and~(\ref{Ddec}), ${\cal M}_0$ and ${\cal M}_2$ denote 
radial matrix elements for $S$- and $D$-wave photon emission, respectively.
The radial matrix element for $S$-wave emission receives contributions not   
only from the impulse approximation, eq.~(\ref{cmdip}), but also from the
confinement and OGE operators in eq.~(\ref{cdip}). That matrix
element may thus be expressed as
\begin{equation}
{\cal M}_0\:\:=\:\:\int_0^\infty dr\:r\,u_f(r)\,u_i(r)\:
j_0\left(\frac{qr}{2}\right)\left[\:\left<Q\right>_{\mathrm{IA}}
+ q^2\,V_c(r)\,\left<Q\right>_c +
\left(\frac{\partial V_g(r)}{r\,\partial r}\right)
\left<Q\right>_g\right],
\label{smatr2}
\end{equation}
where $u_i$ and $u_f$ are the reduced radial wave functions for the initial
and final heavy quarkonium states. Similarly, the matrix element for
$D$-wave emission, which vanishes in the E1 approximation, is of the form
\begin{equation}
{\cal M}_2\:\:=\:\:\left<Q\right>_{\mathrm{ID}}
\int_0^\infty dr\:r\,u_f(r)\,u_i(r)\:
j_2\left(\frac{qr}{2}\right).
\label{dmatr}
\end{equation}
The contribution from this matrix element is usually very small as the 
product $qr\ll 1$ for typical values of the photon momenta and quarkonium 
radii. Therefore, that matrix element has not been included in 
eq.~(\ref{Ddec}).

\noindent
The impulse approximation charge factor
$\left<Q\right>_{\mathrm{IA}}$, and the exchange charge factors
$\left<Q\right>_c$ for the scalar confining interaction and
$\left<Q\right>_g$ for the OGE interaction that appear in
eqs.~(\ref{smatr2}) and~(\ref{dmatr}) are of the form
\begin{equation}
\left<Q\right>_{\mathrm{IA}}\:\:=\:\:
\left[\,Q_1\left(1-\frac{q^2}{8m_1^{\,2}}\right)\,\frac{m_2}{m_1+m_2}\:-\:
Q_2\left(1-\frac{q^2}{8m_2^{\,2}}\right)\,\frac{m_1}{m_1+m_2}\,\right]
\end{equation}
for the impulse approximation, where the quark charge operators have been
multiplied with the Darwin-Foldy terms from eq.~(\ref{rsq}), and
\begin{eqnarray}
\left<Q\right>_c &=&
\left[\frac{Q_1}{4m_1^{\,3}}\,\frac{m_2}{m_1+m_2} -
\frac{Q_2}{4m_2^{\,3}}\,\frac{m_1}{m_1+m_2}\right], \label{OGEQ} \\
\left<Q\right>_g &=&
\left[\frac{Q_1}{4m_1^{\,2}}\left(\frac{1}{m_1} +
\frac{2}{3}\frac{\left<S_f|\vectr{$\sigma$}_1\!\cdot\vectr{$\sigma$}_2|S_i\right>}
{m_2}\right) -\frac{Q_2}{4m_2^{\,2}}\left(\frac{1}{m_2} +
\frac{2}{3}\frac{\left<S_f|\vectr{$\sigma$}_1\!\cdot\vectr{$\sigma$}_2|S_i\right>}
{m_1}\right)\right], \nonumber
\end{eqnarray}
for the charge factors that are associated with the scalar confining and OGE 
interactions, respectively. In the spin dependent terms of eqs.~(\ref{OGEQ}), 
$S_i$ and $S_f$ denote the total spins of the initial and final quarkonium 
states. For triplet-triplet and singlet-singlet transitions,
$\left<S_f|\vectr{$\sigma$}_1\!\cdot\vectr{$\sigma$}_2|S_i\right> = +1$ and 
$-3$, respectively. The charge factor $\left<Q\right>_{\mathrm{ID}}$ in 
eq.~(\ref{dmatr}) is defined as $\left<Q\right>_{\mathrm{ID}} = 
\lim_{q\rightarrow 0}\left<Q\right>_{\mathrm{IA}}$. This is permissible 
since the Darwin-Foldy and exchange charge terms are very small compared to 
the dominant dipole contribution, which in itself is already insignificant 
because of the suppression by the $j_2$ function in the matrix element. 

\noindent
The expression for the width of a spin-flip M1 transition between
$S$-wave heavy quarkonium states can be written in the form
\begin{equation}
\Gamma_{\mathrm{M1}}\:=\:\:\frac{16}{2S_i\!+\!1}\:q^3\alpha\:\frac{M_f}{M_i}
\:|{\cal M}_{\gamma}|^2,
\label{M1dec}
\end{equation}
where ${\cal M}_{\gamma}$ denotes the radial matrix element for an M1 
transition and $S_i$ is the total spin of the initial state. That matrix 
element consists of the relativistic impulse approximation, scalar 
confining and OGE components, according to
\begin{equation}
{\cal M}_{\gamma}\:\:=\:\:{\cal M}_{\gamma}^{\mathrm{RIA}} + {\cal
M}_{\gamma}^{\mathrm{Conf}} + {\cal M}_{\gamma}^{\mathrm{Oge}},
\label{M1matr}
\end{equation}
where the matrix element in the relativistic impulse approximation is 
defined according to
\begin{equation}
{\cal M}_{\gamma}^{\mathrm{RIA}}\:\:=\:\:
\frac{2}{\pi} \int_0^\infty dr\,dr'\,r\,r'u_f(r')\,u_i(r) 
\int_0^\infty dP\,P^2 \: \frac{1}{4}\left[\frac{Q_1}{m_1}f_1^\gamma -
\frac{Q_2}{m_2}f_2^\gamma\right]\: j_0\left(r'P\right)j_0\left(rP\right),
\label{RIAmatr}
\end{equation}  
where the factors $f_i^\gamma$ are given by eq.~(\ref{RIA2}). The matrix
elements associated with the scalar confining and vector OGE interactions, 
which have been shown to be large in papers~{\bf I},{\bf V} and~{\bf VI} 
are, in the nonrelativistic approximation, of the form
\begin{eqnarray}
{\cal M}_\gamma^{\mathrm{Conf}} &=& -\int_0^\infty
dr\,u_f(r)\,u_i(r)\,\frac{V_c(r)}{4}\left[\frac{Q_1}{m_1^{\,2}} -  
\frac{Q_2}{m_2^{\,2}}\right], \label{NRConf} \\
{\cal M}_\gamma^{\mathrm{Oge}}  &=& -\int_0^\infty
dr\,u_f(r)\,u_i(r)\,\frac{V_g(r)}{8}\left[\frac{Q_1}{m_1^{\,2}} -
\frac{Q_2}{m_2^{\,2}} - \frac{Q_1-Q_2}{m_1 m_2}\right].\label{NROge}
\end{eqnarray}
In particular, eq.~(\ref{NRConf}) is shown, in the next section, to 
provide an explanation for the experimental width of the M1 transition 
$J/\psi\rightarrow \eta_c\gamma$.

\section{E1 and M1 transitions in heavy quarkonia}

A detailed comparison of the numerical results obtained in papers~{\bf V} 
and~{\bf VI} with experimental results and other theoretical calculations is 
instructive, as there are issues with several of the E1 and M1 transitions 
that are not readily apparent by casual inspection of the large amount of 
numerical data presented in those papers. This is even more important as the 
branching fractions for various transitions are typically better known than 
the total width of the initial state. With this in mind, the most important 
ones of the M1 and E1 transitions given in Tables~\ref{M1tab} and~\ref{ccE1} 
are discussed below.

\subsection{The M1 transition $J/\psi\rightarrow \eta_c\,\gamma$}

The major importance of this M1 transition, from both experimental and
theoretical points of view, has been stated e.g. in the review of 
ref.~\cite{Smilga}. The experimental width of $1.14\,\pm\,0.39$ keV has been 
difficult to explain theoretically, since nonrelativistic calculations 
overestimate this width by a factor~$\sim 3$. A possible solution for this 
overprediction, which was already hinted at in ref.~\cite{Schrcalc}, is 
presented in Table~\ref{M1tab}, where the exchange current contribution from 
the scalar confining interaction brings the width down to the desired level. 
The importance of such negative energy components for the transition 
$J/\psi\rightarrow \eta_c\,\gamma$ has also been demonstrated within the 
framework of the instantaneous approximation to the Bethe-Salpeter equation 
in ref.~\cite{Snellman} and within the Schr\"odinger approach in paper~{\bf I}. 

\noindent
If the entire $Q\bar Q$ potential had effective vector coupling structure, 
which has often been suggested in the literature~\cite{Vectref}, then no 
exchange current contributions would arise, as a vector interaction 
contributes a significant spin-flip operator only if the quark and antiquark 
masses are unequal, and agreement with experiment would thus be excluded. 
Furthermore, a large family of effective vector confining interactions 
have been shown to be inconsistent with QCD by ref.~\cite{Gromes}. 
However, it has also been shown in paper~{\bf I} that an expansion of 
the RIA spin-flip operator to order $v^2/c^2$ overestimates the relativistic 
correction to the static (NRIA) result, which originally led to an opposite 
conclusion~\cite{Schrcalc} concerning the usefulness of a scalar two-quark 
spin-flip operator. It was also suggested that the charm quark might possess 
a large anomalous magnetic moment, but that possibility has apparently not 
been substantiated.

\subsection{The M1 transition $\psi\,'\rightarrow \eta_c\,\gamma$}

This nonrelativistically forbidden M1 transition has also proved challenging
to explain theoretically, since the (near) orthogonality of the quarkonium
wave functions renders the results hypersensitive to small effects. In the 
recent calculation by ref.~\cite{Snellman}, where good agreement with 
experiment was found for $J/\psi\rightarrow \eta_c\,\gamma$, the width for 
$\psi\,'\rightarrow \eta_c\,\gamma$ was however overpredicted by almost an 
order of magnitude. It is shown in Table~\ref{M1tab} that the M1 model 
employed in paper~{\bf VI} gives a width of $\sim 1.1$ keV for that 
transition, which is close to the upper uncertainty limit of the current
empirical result $0.84\,\pm\,0.24$ keV~\cite{NewPDG}. That such a favorable 
result is obtained depends on several factors in the present work, such as 
the employment of $\psi\,'$ and $\eta_c$ wave functions that model the 
spin-spin interaction in the $S$-wave. 

\noindent
The choice of approximation for the M1 matrix element is also important in 
this respect. The amplitude~(\ref{matrM1}) has the additional advantage of 
allowing the use of a realistic photon momentum in the 
expression~(\ref{M1dec}) for the M1 width. Also, this treatment yields the 
same spin-flip operators as in the calculation of the exchange magnetic 
moment operators in ref.~\cite{Tsushima}, where the rigorous M1 
approximation was used. Furthermore, the M1 approximation has been taken 
to affect the entire factor in brackets in eq.~(\ref{matrM1}), which leads to 
the elimination of the photon momentum~\vectr{$q$} from the RIA matrix 
element~(\ref{RIAmatr}). If the exponentials were
separated from the current operators in eq.~(\ref{matrM1}), then the width
for $\psi\,'\rightarrow \eta_c\,\gamma$ would be overpredicted by a factor  
$\sim 4$. However, if spin-averaged wave functions were employed, as in 
paper~{\bf I}, then the conclusion would be exactly the opposite; In  
that case the present treatment would lead to unfavorable results. As seen
from Table~\ref{M1tab}, the exchange current operator associated with the 
scalar confining interaction gives the main contribution to the width for 
$\psi\,'\rightarrow \eta_c\,\gamma$ within this calculation. 

\noindent
The present treatment of the M1 approximation may be regarded as consistent 
since it leads to the correct spin-flip operators and simultaneously allows 
a realistic photon momentum to be used. However, the large photon momentum 
introduces an additional uncertainty, which involves boosts on the $Q\bar 
Q$ wavefunction in the final state, an effect which is yet to be 
considered.

\subsection{Other M1 transitions}

In principle, the width for $\Upsilon\rightarrow \eta_b\,\gamma$ could be 
predicted with much better accuracy than the corresponding one in the $c\bar 
c$ system, because of the large mass of the bottom quark. In particular, the 
exchange current contribution from the scalar confining interaction is much 
smaller than for $c\bar c$. The largest uncertainty is introduced by the 
unknown photon momentum for the $\Upsilon\rightarrow \eta_b\,\gamma$ 
transition, as the mass of the $\eta_b$ state is not known empirically. As 
realistic models of the spin-spin splittings for $S$-wave quarkonia give an 
$\eta_b$ mass around 9400 MeV, then the width for $\Upsilon\rightarrow 
\eta_b\,\gamma$ is likely to be less than 10 eV, as given in 
Table~\ref{M1tab}. 

\noindent
In addition to the M1 transitions discussed above, predictions have also
been given in paper~{\bf VI} for M1 transitions between $Q\bar Q$ states 
below the thresholds for fragmentation. Among these is the transition 
$\psi\,'\rightarrow \eta_c'\,\gamma$, which is similar to 
$J/\psi\,\rightarrow \eta_c \,\gamma$. 
As recent experimental results indicate that the mass of the $\eta_c'$ is 
much higher than previously thought~\cite{Belle}, then the amount of phase 
space available for $\psi\,'\rightarrow \eta_c'\,\gamma$ is also smaller. 
The predicted width for that transition is thus significantly smaller than 
the values suggested by previous work~\cite{Snellman}. The width for 
$\eta_c'\rightarrow J/\psi\,\gamma$ is also sensitive to particulars of 
the model because of cancellations in the matrix element and the large 
photon momentum involved. 
The results in Table~\ref{M1tab} suggest that the width for this transition 
should be around 2 keV. As the experimental situation concerning the 
$\eta_c'$ continues to improve, then the width for $\eta_c'\rightarrow
J/\psi\,\gamma$ may possibly be measured in the near future.

\noindent
In the case of the $b\bar b$ system, the number of measurable M1 
transitions is larger since the $3S$ states of bottomonium lie below the 
threshold for $B\bar B$ fragmentation. These widths are difficult to 
predict and provide an important test for models of the M1 transitions. 
The results of paper~{\bf VI} suggest that the widths for transitions 
which do not change the principal quantum number of the quarkonium state 
should be highly suppressed, whereas the widths for transitions from 
excited $\eta_b$ states to the $\Upsilon$ ground state are predicted to 
have larger widths of about 100 eV.

\newpage

\begin{table}[h!]
\centering{
\caption{The M1 transitions between low-lying $S$-wave states in the
charmonium~($c\bar c$), bottomonium~($b\bar b$) and $B_c^\pm $~($c\bar 
b,\bar c b$) systems. Experimental data~\cite{NewPDG} is available only for 
$J/\psi\rightarrow \eta_c\gamma$ and $\psi\,'\rightarrow \eta_c\gamma$. 
The quoted photon momenta $q_\gamma$ have been obtained by combination of 
the empirical masses of the spin triplet states with splittings given by the 
Hamiltonian model of paper~{\bf VI} in Table~\ref{stat}. Note that the 
OGE interaction contributes only to M1 transitions between $B_c$ states.}
\vspace{1.0cm}
\footnotesize
\begin{tabular}{l||r|r|r||c|c|c}
\multicolumn{1}{c||}{Transition} &
\multicolumn{3}{c||}{Matrix element\,\,[fm]} & \multicolumn{3}{c}{Width} \\
&\multicolumn{3}{c||}{}&\multicolumn{3}{c}{} \\
& \multicolumn{1}{c|}{NRIA} & \multicolumn{1}{c|}{RIA} &
\multicolumn{1}{c||}{Exch} & \,NRIA\, & \,RIA\, & {\bf RIA+Exch} \\
\hline\hline
&&&&&& \\
$J/\psi\rightarrow \eta_c\gamma$ & $4.356\cdot 10^{-2}$ &
$3.762\cdot 10^{-2}$ & $-8.724\cdot 10^{-3}$ & 2.85 & 2.12 & {\bf 1.25 keV} 
\\
\vspace{-.2cm}
\scriptsize{$q_\gamma : 116$ MeV}&&&&&& \scriptsize{ex: $1.14\pm
0.39$} \\
&&&&&& \\
$\psi{\,'}\rightarrow \eta_c\gamma$  & $3.985\cdot 10^{-3}$ &
$-5.14\cdot 10^{-4}$ & $ 2.826\cdot 10^{-3}$ & 3.35 & 0.06 & {\bf 1.13 keV} 
\\
\vspace{-.2cm}
\scriptsize{$q_\gamma : 639$ MeV}&&&&&& \scriptsize{ex: $0.84\pm
0.24$} \\
&&&&&& \\
$\psi{\,'}\rightarrow \eta_c'\gamma$ & $4.344\cdot 10^{-2}$ &
$3.735\cdot 10^{-2}$ & $-1.870\cdot 10^{-2}$ & 0.18 & 0.13 & {\bf 0.03 keV} 
\\
\vspace{-.2cm}
\scriptsize{$q_\gamma : 46$ MeV}&&&&&& \\
&&&&&& \\
$\eta_c'\rightarrow J/\psi\,\gamma$ & $-4.271\cdot 10^{-3}$ &
$-7.584\cdot 10^{-3}$ & $5.206\cdot 10^{-3}$ & 5.89 & 18.6 & {\bf 1.83 keV} 
\\
\vspace{-.2cm}
\scriptsize{$q_\gamma : 502$ MeV}&&&&&& \\   
&&&&&& \\
\hline
&&&&&& \\   
$\Upsilon\rightarrow \eta_b\gamma$ & $-6.71\cdot 10^{-3}$ &
$-6.39\cdot 10^{-3}$ & $ 2.41\cdot 10^{-4}$ & 9.2 & 8.3 & {\bf 7.7 eV} \\
\vspace{-.2cm}
\scriptsize{$q_\gamma : 59$ MeV}&&&&&& \\
&&&&&& \\
$\Upsilon'\rightarrow \eta_b\gamma$ & $-3.94\cdot 10^{-4}$ &
$-1.44\cdot 10^{-4}$ & $ -8.76\cdot 10^{-5}$ & 31.8 & 4.3 & {\bf 11.0 eV} 
\\
\vspace{-.2cm}
\scriptsize{$q_\gamma : 603$ MeV}&&&&&& \\
&&&&&& \\
$\Upsilon'\rightarrow \eta_b'\,\gamma$ & $-6.70\cdot 10^{-3}$ &
$-6.39\cdot 10^{-3}$ & $ 5.45\cdot 10^{-4}$ & 0.70 & 0.64 & {\bf 0.53 eV} 
\\
\vspace{-.2cm}
\scriptsize{$q_\gamma : 25$ MeV}&&&&&& \\
&&&&&& \\
$\eta_b'\rightarrow \Upsilon\,\gamma$ & $4.18\cdot 10^{-4}$ &
$6.30\cdot 10^{-4}$ & $ -1.31\cdot 10^{-4}$ & 71.5 & 162 & {\bf 102 eV} \\
\vspace{-.2cm}
\scriptsize{$q_\gamma : 530$ MeV}&&&&&& \\
&&&&&& \\
$\Upsilon''\rightarrow \eta_b''\,\gamma$ & $-6.70\cdot 10^{-3}$ &
$-6.35\cdot 10^{-3}$ & $ 8.02\cdot 10^{-4}$ & 0.18 & 0.16 & {\bf 0.13 eV} 
\\
\vspace{-.2cm}
\scriptsize{$q_\gamma : 16$ MeV}&&&&&& \\
&&&&&& \\   
$\Upsilon''\rightarrow \eta_b'\,\gamma$ & $-3.59\cdot 10^{-4}$ &
$-1.11\cdot 10^{-4}$ & $ -1.55\cdot 10^{-4}$ & 5.3 & 0.5 & {\bf 2.9 eV} \\
\vspace{-.2cm}
\scriptsize{$q_\gamma : 350$ MeV}&&&&&& \\
&&&&&& \\
$\Upsilon''\rightarrow \eta_b\,\gamma$ & $-2.10\cdot 10^{-4}$ &
$-6.59\cdot 10^{-5}$ & $ -3.77\cdot 10^{-5}$ & 30.2 & 3.0 & {\bf 7.3 eV} \\ 
\vspace{-.2cm}
\scriptsize{$q_\gamma : 910$ MeV}&&&&&& \\
&&&&&& \\
$\eta_b''\rightarrow \Upsilon'\,\gamma$ & $3.96\cdot 10^{-4}$ & 
$6.05\cdot 10^{-4}$ & $ -2.25\cdot 10^{-4}$ & 13.7 & 32.0 & {\bf 12.6 eV} 
\\
\vspace{-.2cm}
\scriptsize{$q_\gamma : 311$ MeV}&&&&&& \\
&&&&&& \\
$\eta_b''\rightarrow \Upsilon\gamma$ & $2.05\cdot 10^{-4}$ &  
$3.02\cdot 10^{-4}$ & $ -4.68\cdot 10^{-5}$ & 68.7 & 149 & {\bf 106 eV}   
\\
\vspace{-.2cm}
\scriptsize{$q_\gamma : 842$ MeV}&&&&&& \\
&&&&&& \\
\hline
&&&&&& \\
$B_c^*\rightarrow B_c\gamma$ & $1.851\cdot 10^{-2}$ &
$1.496\cdot 10^{-2}$ & $0.326\cdot 10^{-3}$ & 50.0 & 32.6 & {\bf 34.0 eV}
\\
\vspace{-.2cm}
\scriptsize{$q_\gamma : 53$ MeV}&&&&&& \\
&&&&&& \\
${B_c^*}'\rightarrow B_c\gamma$  & $1.015\cdot 10^{-3}$ &
$-1.437\cdot 10^{-3}$ & $ 2.524\cdot 10^{-3}$ & 179 & 360 & {\bf 206 eV}
\\
\vspace{-.2cm}
\scriptsize{$q_\gamma : 576$ MeV}&&&&&& \\
&&&&&& \\
${B_c^*}'\rightarrow B_c'\gamma$ & $1.849\cdot 10^{-2}$ &
$1.480\cdot 10^{-2}$ & $-5.186\cdot 10^{-3}$ & 3.61 & 2.31 & {\bf 0.98 eV}
\\
\vspace{-.2cm}
\scriptsize{$q_\gamma : 22$ MeV}&&&&&& \\
&&&&&& \\
$B_c'\rightarrow B_c^*\,\gamma$ & $-1.067\cdot 10^{-3}$ &
$-3.089\cdot 10^{-3}$ & $1.028\cdot 10^{-2}$ & 411 & 3440 & {\bf 39.5 eV}
\\
\vspace{-.2cm}
\scriptsize{$q_\gamma : 507$ MeV}&&&&&& \\
\end{tabular}
\label{M1tab}}
\end{table}

\newpage

\subsection{The E1 transitions $\chi_{cJ}\rightarrow J/\psi\,\gamma$ and 
$\psi\,'\rightarrow \chi_{cJ}\,\gamma$}

The E1 transitions from the spin-triplet $P$-wave states are in principle 
the simplest to predict accurately, as the wave functions involved do not 
contain any nodes. Although the empirical data from ref.~\cite{PDG} has 
suggested that the E1 widths are generally overpredicted~\cite{Radref}, that 
discrepancy is apparently resolved by the new data presented in the 2002 
edition of the PDG~\cite{NewPDG}. However, the results in Table~\ref{ccE1} 
indicate that the rigorous E1 approximation, with $q = M_i - M_f$, 
overpredicts the widths by a factor $\sim 2$. If the E1 approximation is 
removed, the recoil of the $J/\psi$ can be accounted for, in which case 
that overprediction is eliminated.

\noindent
Prediction of the widths for $\gamma$ transitions from the $\psi'$ state 
has proved to be difficult, as the E1 approximation typically overpredicts 
the widths by at least a factor $\sim 2$. The present empirical 
data~\cite{NewPDG} on the $\psi'$ suggests that the widths for 
$\psi\,'\rightarrow \chi_{cJ}\,\gamma$ should be around 25 keV. As 
demonstrated in paper~{\bf VI}, the E1 approximation yields widths 
in excess of 40 keV. This is puzzling, since recoil effects 
are small and cannot explain this overprediction. Also, it is seen by 
inspection of Table~\ref{ccE1} that the predicted relative widths also do 
not agree well with experiment, although the experimental uncertainties are 
considerable. Not surprisingly, the matrix elements in Table~\ref{ccE1} 
reveal that these transitions are very sensitive to small hyperfine effects 
in the $Q\bar Q$ wave functions, which has also been demonstrated in 
ref.~\cite{Schrcalc}. It is therefore conceivable that small modifications 
of the $Q\bar Q$ wave functions may be sufficient to solve this 
overprediction. It should also be noted that significant reductions of the 
E1 widths were achieved in ref.~\cite{Eichtenqq} by consideration of closed 
$c\bar q - q\bar c$ fragmentation channels.

\subsection{The E1 transitions from the $\chi_{bJ}$ states} 

The calculated widths for the $\chi_{bJ}\rightarrow \Upsilon\gamma$ 
transitions agree rather well with those of the other models presented in 
Table~\ref{restabb}, although they appear to be slightly larger. If the 
calculated E1 widths are used to predict the total widths of the $\chi_{bJ}$ 
states, then it is found that the width of the $\chi_{b2}$ should be $164\pm 
22$ keV and that of the $\chi_{b1}$ about $93\pm 22$ keV. Similarly, the 
calculated E1 width of the $\chi_{b0}$ suggests that the total width of that 
state is at least $\sim 440$ keV. This situation is similar to that observed 
for $c\bar c$~\cite{PDG}, where the $\chi_{c2}$ is wider than the 
$\chi_{c1}$ by about a factor $\sim 2$. 

\noindent
The E1 transitions from the $\chi_{bJ}(2P)$ states in bottomonium provide a
useful test for theoretical models since experimental data is available on 
all six branching fractions~\cite{PDG}, even though the total widths of the
$\chi_{bJ}(2P)$ states are not known. These data indicate that the widths 
for transitions to the $\Upsilon$ should be about one half of those for 
transitions to the $\Upsilon(2S)$, even though much more phase space is 
available for the former. Indeed, it can be seen from Table~\ref{restabb} 
that spin-averaged wave functions do not provide a good description of the 
experimental branching fractions even though the hyperfine splittings
of the $\chi_{bJ}(2P)$ states are small. On the other hand, much better 
agreement with experiment is obtained if the hyperfine effects are 
accounted for by the $Q\bar Q$ wave functions. The calculated widths for
$\chi_{bJ}(2P)\rightarrow \Upsilon(2S)\,\gamma$ may be used to estimate the 
total widths of the $\chi_{bJ}(2P)$ states from the known branching 
fractions. The predicted width of the $\chi_{b2}(2P)$ state is then $100 \pm 
15$ keV, while that of the $\chi_{b1}(2P)$ is $72 \pm 14$ keV. The 
$\chi_{b0}(2P)$ state appears to be significantly broader, but because of 
the large errors in the reported E1 branching fractions, only a rough 
estimate of $267 \pm 140$ keV is possible.

\newpage

\begin{table}[h!]
\centering{
\footnotesize
\caption{The E1 dominated transitions between low-lying states in the
charmonium~($c\bar c$) system, together with the empirical data given by
ref.~\cite{PDG}. The column "IA" contains the matrix element~(\ref{smatr2})
in the impulse approximation, while in the column labeled
"DYN", the exchange charge contributions have been included. Note that a 
good experimental candidate~\cite{PDG} for the $^3D_1$ state is the 
$\psi\,(3770)$ resonance.}
\vspace{.4cm}
\begin{tabular}{l||r|r|r|c||c|c} 
\multicolumn{1}{c||}{Transition} &
\multicolumn{3}{c|}{${\cal M}_0$\,\,[fm]} &
\multicolumn{1}{c||}{${\cal M}_2$\,\,[fm]}
&\multicolumn{2}{c}{Width} \\
&\multicolumn{3}{c|}{}& &\multicolumn{2}{c}{} \\
&\multicolumn{1}{c|}{IA} & \multicolumn{1}{c|}{DYN} &
\multicolumn{1}{c|}{E1} & \,\, & \,E1\, & {\bf DYN} \\
\hline\hline
&&&&&& \\
$\chi_{c2}\rightarrow J/\psi\,\gamma$ & \,\,0.2389\,\, & \,\,0.2442\,\, &
\,\,0.2632\,\, & $~~7.145\cdot 10^{-3}$ & 558 keV & {\bf 343 keV} \\
\vspace{-.2cm}
\scriptsize{$q_\gamma : 429$ MeV}&&&&&&   
\scriptsize{ex: $389\pm 60$} \\  
&&&&&& \\
$\chi_{c1}\rightarrow J/\psi\,\gamma$ & \,\,0.2464\,\, & \,\,0.2519\,\, &
\,\,0.2673\,\, & $~~5.729\cdot 10^{-3}$ & 422 keV & {\bf 276 keV} \\
\vspace{-.2cm}
\scriptsize{$q_\gamma : 390$ MeV}&&&&&&
\scriptsize{ex: $290\pm 60$} \\
&&&&&& \\
$\chi_{c0}\rightarrow J/\psi\,\gamma$ & \,\,0.2556\,\, & \,\,0.2612\,\, &
\,\,0.2701\,\, & $~~3.345\cdot 10^{-3}$ & 196 keV & {\bf 144 keV} \\  
\vspace{-.2cm}
\scriptsize{$q_\gamma : 303$ MeV}&&&&&&
\scriptsize{exp: $165\pm 40$} \\
&&&&&& \\
$\psi\,'\rightarrow \chi_{c0}\,\gamma$ & \,\,$-0.2685$\,\, &
\,\,$-0.2686$\,\, &
\,\,$-0.2840$\,\, & $-6.106\cdot 10^{-3}$ & 44.6 keV & {\bf 33.1 keV} \\
\vspace{-.2cm}
\scriptsize{$q_\gamma : 261$ MeV}&&&&&&
\scriptsize{ex: $26.1\pm 4.5$} \\
&&&&&& \\
$\psi\,'\rightarrow \chi_{c1}\,\gamma$ & \,\,$-0.3126$\,\, &
\,\,$-0.3126$\,\, &
\,\,$-0.3202$\,\, & $-3.028\cdot 10^{-3}$ & 45.8 keV & {\bf 38.7 keV} \\
\vspace{-.2cm}
\scriptsize{$q_\gamma : 171$ MeV}&&&&&&   
\scriptsize{ex: $25.2\pm 4.5$} \\
&&&&&& \\
$\psi\,'\rightarrow \chi_{c2}\,\gamma$ & \,\,$-0.3440$\,\, &
\,\,$-0.3442$\,\, &
\,\,$-0.3489$\,\, & $-1.871\cdot 10^{-3}$ & 37.1 keV & {\bf 33.1 keV} \\
\vspace{-.2cm}
\scriptsize{$q_\gamma : 127$ MeV}&&&&&&
\scriptsize{ex: $20.4\pm 4.0$} \\
&&&&&& \\
$h_c\rightarrow \eta_c\,\gamma$ & \,\,0.2098\,\, & \,\,0.2091\,\,  &
\,\,0.2289\,\, & $~~7.377\cdot 10^{-3}$ & 661 keV & {\bf 370 keV}  \\
\vspace{-.2cm}
\scriptsize{$q_\gamma : 493$ MeV}&&&&&& \\
&&&&&& \\
$\eta_c'\rightarrow h_c\,\gamma$ & \,\,$-0.3420$\,\, & \,\,$-0.3424$\,\,
& \,\,$-0.3465$\,\, & $-1.618\cdot 10^{-3}$ & 61.5 keV & {\bf 55.0 keV} \\
\vspace{-.2cm}
\scriptsize{$q_\gamma : 125$ MeV}&&&&&& \\
&&&&&& \\
\hline   
&&&&&& \\
$\psi''\rightarrow \chi_{c0}\,\gamma$ & \,\,$-0.0456$\,\, & \,\,
$-0.0450$\,\, & \,\,$-0.0199$\,\, & $~~0.926\cdot 10^{-2}$ & 2.69 keV &   
{\bf 9.86 keV}  \\
\vspace{-.2cm}
\scriptsize{$q_\gamma : 577$ MeV}&&&&&& \\
&&&&&& \\
$\psi''\rightarrow \chi_{c1}\,\gamma$ & \,\,$-0.0306$\,\, & \,\,
$-0.0298$\,\, & \,\,$-0.0033$\,\, & $~~1.016\cdot 10^{-2}$ & 0.13 keV &
{\bf 9.57 keV} \\
\vspace{-.2cm}
\scriptsize{$q_\gamma : 494$ MeV}&&&&&& \\
&&&&&& \\
$\psi''\rightarrow \chi_{c2}\,\gamma$ & \,\,$-0.0168$\,\, & \,\,
$-0.0161$\,\, & \,\,0.0123\,\, & $~~1.099\cdot 10^{-2}$ & 2.38 keV &  
{\bf 5.75 keV} \\
\vspace{-.2cm}
\scriptsize{$q_\gamma : 455$ MeV}&&&&&& \\
&&&&&& \\
$\psi''\rightarrow \chi_{c0}'\,\gamma$ & \,\,$-0.4315$\,\, & \,\,  
$-0.4315$\,\, & \,\,$-0.4497$\,\, & $-7.344\cdot 10^{-3}$ & 21.3 keV &
{\bf 17.8 keV} \\
\vspace{-.2cm}
\scriptsize{$q_\gamma : 153$ MeV}&&&&&& \\
&&&&&& \\
$\psi''\rightarrow \chi_{c1}'\,\gamma$ & \,\,$-0.4860$\,\, & \,\,
$-0.4861$\,\, & \,\,$-0.4995$\,\, & $-5.399\cdot 10^{-3}$ & 42.6 keV &
{\bf 37.3 keV} \\
\vspace{-.2cm}
\scriptsize{$q_\gamma : 125$ MeV}&&&&&& \\
&&&&&& \\
$\psi''\rightarrow \chi_{c2}'\,\gamma$ & \,\,$-0.5280$\,\, & \,\,
$-0.5283$\,\, & \,\,$-0.5391$\,\, & $-4.367\cdot 10^{-3}$ & 53.7 keV &
{\bf 48.2 keV} \\
\vspace{-.2cm}   
\scriptsize{$q_\gamma : 109$ MeV}&&&&&& \\
&&&&&& \\
\hline   
&&&&&& \\
$^3D_3 \rightarrow \chi_{c2}\,\gamma$ & \,\,$0.4164$\,\, &
\,\,$0.4194$\,\, & \,\,$0.4353$\,\, & -- & 243 keV & {\bf 192 keV} \\
\vspace{-.22cm}
\scriptsize{$q_\gamma : 227$ MeV} &&&&& \\
&&&&& \\
$^3D_2 \rightarrow \chi_{c2}\,\gamma$ & \,\,$0.4188$\,\, &
\,\,$0.4219$\,\, & \,\,$0.4367$\,\, & -- & 56.5 keV & {\bf 45.2 keV} \\
\vspace{-.22cm}
\scriptsize{$q_\gamma : 221$ MeV} &&&&& \\
&&&&& \\
$^3D_2 \rightarrow \chi_{c1}\,\gamma$ & \,\,$0.3920$\,\, &
\,\,$0.3953$\,\, & \,\,$0.4145$\,\, & -- & 262 keV & {\bf 198 keV} \\
\vspace{-.22cm}
\scriptsize{$q_\gamma : 263$ MeV} &&&&& \\
&&&&& \\
$^3D_1 \rightarrow \chi_{c2}\,\gamma$ & \,\,$0.4216$\,\, &
\,\,$0.4246$\,\, & \,\,$0.4372$\,\, & -- & 5.06 keV & {\bf 4.13 keV} \\
\vspace{-.22cm}
\scriptsize{$q_\gamma : 206$ MeV} &&&&& \\
&&&&& \\
$^3D_1 \rightarrow \chi_{c1}\,\gamma$ & \,\,$0.3963$\,\, &
\,\,$0.3997$\,\, & \,\,$0.4164$\,\, & -- & 123 keV & {\bf 94.9 keV} \\
\vspace{-.22cm}
\scriptsize{$q_\gamma : 248$ MeV} &&&&& \\
&&&&& \\
$^3D_1 \rightarrow \chi_{c0}\,\gamma$ & \,\,$0.3578$\,\, & 
\,\,$0.3619$\,\, & \,\,$0.3889$\,\, & -- & 370 keV & {\bf 251 keV} \\
\vspace{-.22cm}
\scriptsize{$q_\gamma : 336$ MeV} &&&&& \\
\end{tabular}
\label{ccE1}} 
\end{table}

\newpage

\subsection{The E1 transitions from the $\Upsilon$ states}

The experimental situation concerning the $\Upsilon(2S)\rightarrow
\chi_{bJ}\,\gamma$ transitions has lately become more uncertain since the total
width of the $\Upsilon(2S)$ as reported by ref.~\cite{PDG}, originally  
given as $\sim 27$ keV, has increased over time and now stands at $44\pm 7$
keV. This situation is analogous to that for the $\psi'$, which has   
undergone a similar, albeit smaller, increase. This has made the model     
predictions in Table~\ref{restabb}, which originally fitted the
experimental data very well, much less satisfactory. It is therefore very
difficult to judge the quality of any given prediction until the 
experimental situation is stabilized. Still, it is noteworthy that the 
calculation of paper~{\bf VI} gives slightly better agreement with 
experiment than the other models in Table~\ref{restabb}.

\noindent
As the reported total width of the $\Upsilon(3S)$ state~\cite{PDG}, $26.3
\pm 3.5$ keV, is better known than that of the $\Upsilon(2S)$ state, then  
it is expected that systematic uncertainties in the reported experimental 
results for $\Upsilon(3S)\rightarrow \chi_{bJ}(2P)\,\gamma$ should be
smaller than for the analogous $\Upsilon(2S)\rightarrow \chi_{bJ}\,\gamma$ 
transitions. By inspection of Table~\ref{restabb}, it can be seen that the
$\Upsilon(3S)\rightarrow \chi_{bJ}(2P)\,\gamma$ transitions are generally
rather well described by a number of models, although the calculation of  
ref.~\cite{Schrcalc}, where spin-averaged wave functions were employed, 
apparently underpredicts the empirical widths. Also, the results of
ref.~\cite{GrE1bb} appear to compare slightly more favorably with
experiment than those of the calculation in paper~{\bf VI}.

\noindent
While the $\Upsilon(3S)\rightarrow \chi_{bJ}(2P)\,\gamma$ transitions are 
relatively well described by different models, the situation concerning the
$\Upsilon(3S)\rightarrow \chi_{bJ}\,\gamma$ transitions remains unsettled  
because of a strong cancellation in the E1 matrix element. However, 
experimental detection of these transitions may be a formidable task since 
the widths are an order of magnitude smaller than those of any previously
measured E1 transition in the $b\bar b$ system. Inspection of the results in 
paper~{\bf VI} reveals that within the dynamical model, the width for 
$\Upsilon(3S)\rightarrow \chi_{b0}\,\gamma$ should be the largest and that 
for $\Upsilon(3S)\rightarrow \chi_{b2}\,\gamma$ the smallest. It is 
encouraging that the same pattern is also predicted in Table~\ref{ccE1} for 
the analogous transitions in the $c\bar c$ system, where the widths are much
larger relative to the other E1 transitions. 

\subsection{Other E1 transitions}

The widths for E1 transitions from the $\psi(3S)$ state have also been given 
in Table~\ref{ccE1}. Those results suggest that the transitions
to the $(2P)$ states should have widths that are comparable to those for
the $\psi\,'\rightarrow \chi_{cJ}\,\gamma$ transitions. On the other hand,
the widths for the $\psi(3S)\rightarrow \chi_{cJ}\,\gamma$ transitions are
predicted to be smaller by factors $3-4$. The empirical detection of any of
these transitions will probably be difficult since the $\psi(3S)$ state lies 
above the $D\bar D$ threshold. However, since the spin
singlet $h_c$ state is well below threshold, then the photon produced in
the $h_c\rightarrow \eta_c\gamma$ transition may be detected in the
near future. The dynamical model yields a width of 370 keV for this 
transition, which is then the largest E1 width in the $c\bar c$ system, 
although the difference between the E1 approximation and the dynamical
model is large for that transition. 

\noindent
The E1 transitions from the $^3D_1$ state are also of particular interest, 
as that state probably corresponds to the empirical $\psi(3770)$ resonance. 
The predictions of paper~{\bf VI} suggest that the transitions to the 
$\chi_{c1}$ and $\chi_{c0}$ states should be detectable by experiment, 
whereas that to the $\chi_{c2}$ state is highly suppressed by 
the statistical factor ${\cal S}_{fi}$.

\newpage

\begin{table}[h!]   
\centering{   
\caption{Comparison of the predicted E1 widths in the
bottomonium ($b\bar b$) system with those of other models that
use a scalar confining interaction. All widths are given in keV. The
experimental widths have been extracted from the branching fractions and
total widths reported by ref.~\cite{PDG}.}
\vspace{.4cm}
\begin{tabular}{c||c|c|c|c}
& GS (ref.~\cite{Schrcalc}) & GZ (ref.~\cite{GrE1bb})
& paper~{\bf VI} & Exp (ref.~\cite{PDG}) \\
\vspace{-.3cm}
&&&& \\
\hline\hline
\vspace{-.2cm}
&&&& \\
$\chi_{b2}\rightarrow \Upsilon\,\gamma$ & 33.0 & 33.8 & {\bf 36.0} &
$22 \pm 3$\% \\
$\chi_{b1}\rightarrow \Upsilon\,\gamma$ & 29.8 & 30.4 & {\bf 32.5} &
$35 \pm 8$\% \\
$\chi_{b0}\rightarrow \Upsilon\,\gamma$ & 25.7 & 25.3 & {\bf 26.6} &
$< 6$ \% \\   
\vspace{-.2cm}
&&&& \\
\hline
\vspace{-.2cm}
&&&& \\
$\Upsilon'\rightarrow \chi_{b0}\,\gamma$ & 0.73 & 0.76 & {\bf 1.01} &
$1.7 \pm 0.5$ keV \\ 
$\Upsilon'\rightarrow \chi_{b1}\,\gamma$ & 1.62 & 1.37 & {\bf 1.80} &
$3.0 \pm 0.7$ keV \\
$\Upsilon'\rightarrow \chi_{b2}\,\gamma$ & 1.84 & 1.45 & {\bf 2.03} &
$3.1 \pm 0.7$ keV \\
\vspace{-.2cm}
&&&& \\
\hline
\vspace{-.2cm}
&&&& \\
$\chi_{b2}'\rightarrow \Upsilon'\,\gamma$ & 12.9 & 16.2 & {\bf 16.4}
& $16.4 \pm 2.4$ \% \\
$\chi_{b1}'\rightarrow \Upsilon'\,\gamma$ & 11.9 & 14.7 & {\bf 15.1}
& $21 \pm 4$ \% \\
$\chi_{b0}'\rightarrow \Upsilon'\,\gamma$ & 10.6 & 12.3 & {\bf 12.3}
& $4.6 \pm 2.1$ \% \\
\vspace{-.2cm}
&&&& \\
\hline
\vspace{-.2cm}
&&&& \\
$\chi_{b2}'\rightarrow \Upsilon\,\gamma$ & 18.2 & 10.4 & {\bf 11.4}
& $7.1 \pm 1.0$ \% \\ 
$\chi_{b1}'\rightarrow \Upsilon\,\gamma$ & 11.8 & 7.51 & {\bf 8.40}
& $8.5 \pm 1.3$ \% \\
$\chi_{b0}'\rightarrow \Upsilon\,\gamma$ & 6.50 & 3.57 & {\bf 4.93}  
& $0.9 \pm 0.6$ \% \\
\vspace{-.2cm}
&&&& \\
\hline
\vspace{-.2cm}  
&&&& \\
$\Upsilon''\rightarrow \chi_{b0}\,\gamma$ & 0.114 & 0.029 & {\bf 0.15}
& -- \\
$\Upsilon''\rightarrow \chi_{b1}\,\gamma$ & 0.003 & 0.095 & {\bf 0.11}
& -- \\
$\Upsilon''\rightarrow \chi_{b2}\,\gamma$ & 0.194 & 0.248 & {\bf 0.04}  
& -- \\
\vspace{-.2cm}
&&&& \\
\hline
\vspace{-.2cm}
&&&& \\
$\Upsilon''\rightarrow \chi_{b0}'\,\gamma$ & 1.09 & 1.30 & {\bf 1.14}
& $1.4 \pm 0.3$ keV \\
$\Upsilon''\rightarrow \chi_{b1}'\,\gamma$ & 2.15 & 2.34 & {\bf 2.12}
& $3.0 \pm 0.5$ keV \\
$\Upsilon''\rightarrow \chi_{b2}'\,\gamma$ & 2.29 & 2.71 & {\bf 2.50}
& $3.0 \pm 0.6$ keV \\
\end{tabular}
\label{restabb}}
\end{table}

\noindent
The determination of the photon momenta for transitions in the 
bottom-charm $B_c^\pm$ system has to rely on model predictions for the 
masses of the $c\bar b$ states. The uncertainty introduced 
by this is, however, rather small for the E1 transitions, as the model 
predictions for the major level splittings agree with each 
other to a large extent~\cite{Quigg}. Inspection of the results in 
paper~{\bf VI} reveals that the predicted widths are similar to those 
obtained by ref.~\cite{Quigg}, although differences exist for transitions 
like $B_c^*(2S)\rightarrow B_{c0}^*\,\gamma$ and $B_{c2}^*(2P)\rightarrow 
B_c^*(2S)\,\gamma$, where the widths are sensitive to the effects of the 
hyperfine components of the $Q\bar Q$ interaction. It is noteworthy that 
while the predicted widths for the $B_{cJ}^*\rightarrow B_c^*\,\gamma$ 
transitions agree rather well with those from ref.~\cite{Quigg}, there is a 
significant disagreement for $B_{c1}\rightarrow B_c\,\gamma$. When the 
somewhat different photon momenta are accounted for, this disagreement 
amounts to about a factor~$\sim 3$. 

\noindent
An issue not considered in paper~{\bf VI} is the spin mixing of the $L=1$ 
states with $J=1$, that is due to the antisymmetric spin-orbit 
interaction which was not included in the $Q\bar Q$ interaction 
Hamiltonian. This mixing, which was considered in ref.~\cite{Quigg}, has 
the effect of allowing "spin-flip" E1 transitions of the type $B_{c1}^* 
\rightarrow B_c\,\gamma$. 
However, the widths for such "forbidden" transitions were found in 
ref.~\cite{Quigg} to be typically suppressed by a factor $\sim 100$ relative 
to the "allowed" ones considered in paper~{\bf VI}.

\newpage

\section{M1 transitions in heavy-light mesons}

\noindent
The M1 widths of the spin-flip M1 transition between the vector and 
pseudoscalar states in the charm mesons have been calculated in paper~{\bf 
V}, where the $Q\bar q$ interaction was modeled as a scalar confining + OGE 
potential. A comparison of the results presented in Table~\ref{Dpmtab} is 
instructive since the total width of the $D^{\pm *}$ state has recently been 
measured by the CLEO collaboration~\cite{Dwidth}. The width is reported as 
$\Gamma(D^{\pm *}) = 96 \pm 4 \pm 22$ keV, where the latter error represents 
the systematic uncertainty. The reported~\cite{PDG} branching ratio
of $1.6\pm 0.4$\% for radiative decay then gives a width of $1.5\pm 0.6$ keV 
for the M1 transition $D^{\pm *}\rightarrow D^\pm \gamma$. Here most of the
uncertainty can be traced to the systematic errors of the experimental 
result. Such a comparison shows that the results of Table~\ref{Dpmtab} 
reproduce the empirical width fairly well for a range of values of the light 
constituent quark mass $m_q$. The value $m_q = 450$ MeV corresponds to the 
potential model of ref.~\cite{Lahdeout}, while the value $m_q = 420$ MeV has 
been suggested in ref.~\cite{instpap}. Indeed, for a light constituent quark 
mass of 420 MeV, a width for radiative M1 decay which is close to 1.5 keV is 
reproduced. Values close to that are also favored by the analysis within 
the framework of the Gross equation by ref.~\cite{Goity}.

\begin{table}[h!]
\parbox{0.63\textwidth}{\footnotesize\centering
\begin{tabular}{c|c|c|c}
$D^{0 *}\rightarrow D^0 \gamma$
& NRIA            & RIA           & {\bf RIA + Exch}      \\
\hline\hline
&&& \\
450 MeV & 21.1 keV      & 8.86 keV      & {\bf 8.95 keV}      \\
&&& \\
420 MeV & 23.5 keV      & 9.18 keV      & {\bf 9.89 keV}      \\
&&& \\
390 MeV & 26.4 keV      & 9.52 keV      & {\bf 11.1 keV}      \\
\vspace{.4cm}
\end{tabular}
\begin{tabular}{c|c|c|c}
$D^{\pm *} \rightarrow D^\pm \gamma$
& NRIA    & RIA                & {\bf RIA + Exch}         \\      
\hline\hline
&&& \\
450 MeV & 0.58 keV      & $9.4\cdot 10^{-3}$ keV & {\bf 1.09 keV}     \\   
&&& \\
420 MeV & 0.79 keV      & $1.5\cdot 10^{-2}$ keV & {\bf 1.43 keV}     \\
&&& \\
390 MeV & 1.07 keV      & $2.2\cdot 10^{-2}$ keV & {\bf 1.90 keV}     \\
\vspace{.4cm}
\end{tabular}
\begin{tabular}{c|c|c|c}
$D_s^{\pm *}\rightarrow D_s^\pm \gamma$
& NRIA    & RIA                & {\bf RIA + Exch}         \\
\hline\hline
&&& \\
560 MeV & 0.18 keV      & $2.6\cdot 10^{-4}$ keV & {\bf 0.38 keV}     \\
&&& \\
530 MeV & 0.26 keV      & $3.9\cdot 10^{-5}$ keV & {\bf 0.49 keV}     \\  
&&& \\
500 MeV & 0.36 keV      & $8.6\cdot 10^{-4}$ keV & {\bf 0.64 keV}     \\
\end{tabular}}
\hfill
\parbox{0.36\textwidth}{
\caption{Numerical results from paper~{\bf V} for the M1 transitions between 
ground state vector and pseudoscalar mesons in the $D$ and $D_s$ systems, 
for \mbox{different} values of the light and \mbox{strange} constituent 
quark masses. 
The charm quark mass of 1580 MeV corresponds to the potential model 
of ref.~\cite{Lahdeout}. In the right-hand column, the two-quark exchange 
current contributions from the scalar confining and OGE interactions have 
been added to the RIA result.}
\label{Dpmtab}}
\end{table}

\noindent
Even though the total width of the neutral $D^{0*}$ meson has not yet been
determined~\cite{PDG}, considerable information about the expected width
for $D^{0*}\rightarrow D^0 \gamma$ may be extracted from the reported
branching fraction of $38.1 \pm 2.9$\%~\cite{PDG}, since the corresponding
width for pion emission can be constrained by means of the empirically
determined width of the $D^{\pm *}$ and model calculations of the pionic 
transitions in $D$ mesons~\cite{Goity,EichtenD}. If one notes that the 
branching fraction of $D^{\pm *}\rightarrow D^0 \pi^\pm$ is reported as $67.7\pm 
0.5$\%~\cite{PDG} which implies a width for this transition of $\sim 65 \pm 
14$ keV, then it is found from the model calculation of paper~{\bf III} that 
this corresponds to a width of $\sim 40\pm 10$ keV for $D^{0*}\rightarrow 
D^0 \pi^0$. As the relative branching fractions for
$\pi^0$ and $\gamma$ emission by the $D^{0*}$ are well known~\cite{PDG}, the
best estimate for the width of $D^{0*}\rightarrow D^0 \gamma$ is $\sim 25$
keV, which is close to that preferred by ref.~\cite{Goity}. There remains, 
however, a considerable uncertainty of $\sim \pm 10$ keV from the systematic 
errors in the empirical measurement of $\Gamma(D^{\pm *})$.

\noindent
The results of paper~{\bf V} in Table~\ref{Dpmtab} underpredict this 
expectation by about a factor~$\sim 2$, which underlines a basic 
weakness of the present approach to the M1 transitions in the $D$ mesons.
Firstly, as the two-quark spin-flip operators of eqs.~(\ref{muc}) 
and~(\ref{mug}) are given for the nonrelativistic approximation, a 
relativistic treatment of those operators will, in general, lead to a 
weakening of the two-quark contributions to the matrix element for an M1 
transition. It is thus entirely possible that a fully relativistic 
treatment will render 
the two-quark contributions too weak to account for the experimental data on  
M1 decay of the $D^\pm$ meson as well. This conclusion is in line with that 
reached in ref.~\cite{Goity}, where the discrepancy between the model and 
experiment was parameterized in terms of a large anomalous magnetic moment 
for the light quark.

\noindent
Another interesting possibility discussed in paper~{\bf V}
suggests itself, in view of the problems of fitting the 
spectra of the $Q\bar q$ mesons within the framework of the BSLT equation 
using a OGE interaction alone, namely that the instanton induced 
interaction for $Q\bar q$ systems given in ref.~\cite{instpap} may also 
contribute a significant two-quark current. The interaction of 
ref.~\cite{instpap}, which was found to be short-ranged, attractive and 
with negative sign, has scalar coupling to the light constituent quark. It 
has been noted in paper~{\bf V} that such an interaction adds up 
constructively with the OGE contribution, and counteracts that from the 
scalar confining interaction. An overall favorable effect on the widths 
for M1 transitions may thus be obtained, which may be inferred from the 
matrix elements given in paper~{\bf V}.

\chapter{Pionic Transitions}

The pionic transitions in the heavy-light $D$ mesons are instructive, as they 
can provide direct information on the strength of the coupling between pions 
and light constituent quarks. Furthermore, as the charm quark in the $D$ 
mesons does not couple to pions, the decay mechanism is governed by the pion 
coupling to the light constitutent quark alone. As a first approximation, 
the pion-light constitutent quark coupling can be taken as being independent 
of the quark-antiquark interaction in the $D$ meson. It has been shown in 
paper~{\bf II}, where the pion emission was modeled in terms of the chiral 
pseudovector Lagrangian, that while this assumption is reasonable for the 
axial current term, it leads to a large overestimate of the axial charge 
contribution. In this case two-quark pair terms analogous to those that are 
required for a realistic description of the M1 transition $J/\psi\rightarrow 
\eta_c\gamma$ may reduce the axial charge amplitude to a realistic level. An 
analogous suppression of the $S$-wave pion transitions was achieved in 
ref.~\cite{Goity} within the framework of the Gross quasipotential 
reduction.

\noindent
The empirical information on the widths and branching fractions of excited 
charm mesons is still very limited. Absolute values, albeit with large 
uncertainties, are known for the $D_1(2420)$ and $D_2^*(2460)$ mesons, and 
recently the width of the $D^*(2010)^\pm$, for which only an upper bound of 
0.131~MeV~\cite{PDG} was available earlier, has now been reported as 
$96 \pm 4 \pm 22$ keV by the CLEO collaboration~\cite{Dwidth}. This result 
is shown to be 
consistent with values of the pion-quark axial coupling $g_A^q$ that are 
slightly smaller than 1. The observed widths of the $L=1$ $D_1(2420)$ and 
$D_2^*(2460)$ mesons are also shown to be fairly well described, although a 
slight underprediction is expected as it has been demonstrated in paper~{\bf 
III} that two-pion $\pi\pi$ decay may also contribute significantly to the 
total widths of those mesons. In particular, the analogy with the 
$K_2^*(1430)$ strange meson suggests that $\pi\pi$ transitions may account 
for a significant fraction of the observed total widths. 

\noindent
The flavor symmetry breaking decay mode $D_s^*\rightarrow D_s\pi^0$, which 
has been considered in paper~{\bf V} is mainly due to a small isoscalar $\eta$ 
meson component in the physical $\pi^0$ meson, as only the $\eta$ meson can 
couple to the strange quark in the $D_s$ meson. The known ratio of the 
branching fractions for $D_s^*\rightarrow D_s\pi^0$ and $D_s^*\rightarrow 
D_s\gamma$ may be used to extract the coupling of $\eta$ mesons to strange 
quarks, once the value of the $\pi^0\!-\!\eta$ mixing angle is known. 
Applied to the quark model for the baryons, an $\eta NN$ pseudovector 
coupling constant, small enough to be consistent with the phenomenological 
analysis of photoproduction of the $\eta$ on the nucleon and the reaction 
$pp\rightarrow pp\eta$, has been obtained in paper~{\bf V}.

\newpage

\section{The amplitude for pion emission}

\noindent
In paper~{\bf II}, the emission of pions from a $D$ meson was described 
by the pseudovector Lagrangian, which constitutes the lowest order chiral 
coupling for pions to light constituent quarks:
\begin{equation}
{\cal L}_{qq\pi}\:\:=\:\:i{g_A^q\over 
2f_\pi}\:\bar\psi_q\,\gamma_5\gamma_\mu
\,\partial_\mu\,\pi_a\tau_a\,\psi_q.
\label{lagr}
\end{equation} 
Here $g_A^q$ denotes the axial coupling constant of pions to light 
constituent quarks, and $f_\pi$ is the pion decay constant, the empirical 
value of which is 93 MeV. The axial coupling constant is conventionally 
taken to be equal to, or somewhat less than, unity~\cite{Weinpap}. The 
Lagrangian~(\ref{lagr}) yields the following transition amplitude for pion 
emission:
\begin{equation}
T_\pi\:=\: \xi \frac{g_A^q}{2f_\pi}\: \bar 
u_q(\vectr{$p$}')\gamma_5\gamma_\mu q_\mu\,u_q(\vectr{$p$}),
\end{equation}
where $\xi$ is an isospin factor, the value of which is $\sqrt{2}$ for 
$\pi^\pm$ and 1~for $\pi^0$ emission. Decomposition of the above amplitude 
into axial current and axial charge components yields the pion emission 
amplitude in the impulse approximation,
\begin{eqnarray}
T_\pi &=& -i\,\xi{g^q_A\over 2f_\pi}\:\sqrt{\frac{(E'_q + m_q)(E_q + 
m_q)}{4E'_q E_q}} \left[1-\frac{\vectr{$P$}^2 - \vectr{$q$}^2/4}{ 
3(E'_q + m_q)(E_q + m_q)}\right]\:\vectr{$\sigma$}_q\cdot\vectr{$
q$} \nonumber \\
&& +i\,\xi{g^q_A\over 2f_\pi}\:\frac{2m_q+E_q+E'_q}{
\sqrt{4E_qE'_q(E_q+m_q)(E'_q+m_q)}}\:\:\omega_\pi\,\vectr{$\sigma$}_q\cdot 
\vectr{$P$},
\label{piRIA}
\end{eqnarray}
where the first (axial current) term gives rise to the $P$-wave 
transitions $D^*\rightarrow D\pi$ and the $D$-wave transitions from the 
$L=1$ charm mesons. On the other hand, the axial charge term leads to 
$S$-wave transitions from states with $L=1$. 

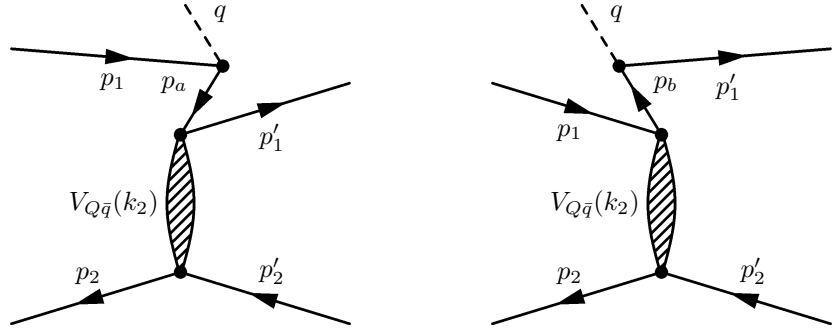
\begin{figure}[h!]
\centering{
\begin{tabular}{c c}
\begin{fmffile}{piex3}
\begin{fmfgraph*}(160,130) \fmfpen{thin}
\fmfcmd{%
 vardef port (expr t, p) =
  (direction t of p rotated 90)
   / abs (direction t of p)
 enddef;}
\fmfcmd{%
 vardef portpath (expr a, b, p) =
  save l; numeric l; l = length p;
  for t=0 step 0.1 until l+0.05:
   if t>0: .. fi point t of p
    shifted ((a+b*sind(180t/l))*port(t,p))
  endfor
  if cycle p: .. cycle fi
 enddef;}
\fmfcmd{%
 style_def brown_muck expr p =
  shadedraw(portpath(thick/2,2thick,p)
   ..reverse(portpath(-thick/2,-2thick,p))
   ..cycle)
 enddef;}
\fmfleft{i2,i1}
\fmfright{o2,o1}
\fmftop{o3}
\fmf{fermion,label=$p_1$}{i1,v3}
\fmf{fermion,label=$p_a$}{v3,v1} 
\fmf{fermion,label=$p_1'$}{v1,o1} 
\fmf{fermion,label=$p_2'$}{o2,v2}
\fmf{fermion,label=$p_2$}{v2,i2}
\fmf{dashes,label=$q$,label.side=right}{v3,o3}
\fmf{brown_muck,lab.s=right,lab.d=4thick,lab=$V_{Q\bar q}(k_2)$,label.side=right}{v1,v2}
\fmfdot{v1,v2,v3}
\fmfforce{(.1w,.85h)}{i1}
\fmfforce{(.9w,.75h)}{o1}
\fmfforce{(.1w,.05h)}{i2}
\fmfforce{(.9w,.05h)}{o2}
\fmfforce{(.5w,.60h)}{v1}
\fmfforce{(.5w,.20h)}{v2}
\fmfforce{(.6w,.80h)}{v3}
\fmfforce{(.5w,.h)}{o3}
\end{fmfgraph*}
\end{fmffile}
&
\begin{fmffile}{piex4}
\begin{fmfgraph*}(160,130) \fmfpen{thin}
\fmfcmd{%
 vardef port (expr t, p) =
  (direction t of p rotated 90)
   / abs (direction t of p)
 enddef;}
\fmfcmd{%
 vardef portpath (expr a, b, p) =
  save l; numeric l; l = length p;
  for t=0 step 0.1 until l+0.05:
   if t>0: .. fi point t of p
    shifted ((a+b*sind(180t/l))*port(t,p))
  endfor
  if cycle p: .. cycle fi
 enddef;}
\fmfcmd{%
 style_def brown_muck expr p =
  shadedraw(portpath(thick/2,2thick,p)
   ..reverse(portpath(-thick/2,-2thick,p))
   ..cycle)
 enddef;}
\fmfleft{i2,i1}
\fmfright{o2,o1}
\fmftop{o3}
\fmf{fermion,label=$p_1$}{i1,v1}
\fmf{fermion,label=$p_b$,label.side=right}{v1,v3}
\fmf{fermion,label=$p_1'$}{v3,o1}
\fmf{fermion,label=$p_2'$}{o2,v2}
\fmf{fermion,label=$p_2$}{v2,i2}
\fmf{dashes,label=$q$,label.side=right}{v3,o3}
\fmf{brown_muck,lab.s=right,lab.d=4thick,lab=$V_{Q\bar q}(k_2)$,label.side=right}{v1,v2}
\fmfdot{v1,v2,v3}
\fmfforce{(.1w,.75h)}{i1}
\fmfforce{(.9w,.85h)}{o1}
\fmfforce{(.1w,.05h)}{i2}
\fmfforce{(.9w,.05h)}{o2}
\fmfforce{(.5w,.60h)}{v1}
\fmfforce{(.5w,.20h)}{v2}
\fmfforce{(.4w,.80h)}{v3}
\fmfforce{(.3w,.h)}{o3}
\end{fmfgraph*}
\end{fmffile} \\
\end{tabular}
\label{pidiag}
\caption{Irreducible two-quark contributions associated with the $Q\bar q$ 
interaction to the axial current and axial charge operators, with 
four-momentum variables defined as for Fig.~\ref{feyn}. The pion always 
interacts with the light constituent quark, as the charm quark does not 
couple to pions.}}
\end{figure}
\pagebreak

\noindent
As demonstrated in paper~{\bf II}, two-quark axial exchange current 
operators illustrated by Fig.~\ref{pidiag} give large contributions to 
several $\pi$ transitions, in particular to those which involve the axial 
charge amplitude. In that case the axial exchange charge operator is of 
equal magnitude as the single quark operator, while the axial exchange 
current operators typically represent $\sim 10$\% corrections to the 
impulse approximation result. If, in the static approximation, the axial 
current $A_{\mu a} = (\vectr{$A$}_a, iA_{0a})$ of the light constituent 
quark is expressed as
\begin{equation}
\vectr{$A$}_a = -g_A^q\vectr{$\sigma$}_q \tau_a,
\label{axial}
\end{equation}
then the contribution to the axial current from a scalar confining 
interaction is, to lowest order in $v/c$, of the form~\cite{Ax-Tsushima}
\begin{equation}
\vectr{$A$}_a^{\mathrm{Conf}} = -\frac{g_A^q}{4 m_q^{\,3}}
\:V_c(\vectr{$k$}_2)
\:\left[3\,\vectr{$\sigma$}_q\vectr{$P$}^2 - 
\frac{1}{4}\vectr{$\sigma$}_q\vectr{$k$}_2\hspace{.1pt}^2 - 
4\,\vectr{$P$}\,\vectr{$\sigma$}_q\cdot\vectr{$P$} + 
2i\,\vectr{$P$}\times\vectr{$k$}_2\right]\:\tau_a,
\label{confaxial} 
\end{equation}
where $V_c(\vectr{$k$}_2)$ is the Fourier transform of the scalar confining 
interaction. The above expression does not include the corrections from 
the canonical boost factors on the single quark spinors that are included 
in the single quark operator, eq.~(\ref{piRIA}). Moreover, a factor 
$m_q^{-2}$ in the axial exchange current operator~(\ref{confaxial}) arises 
as the static approximation to the propagator of the intermediate negative 
energy quark. 

\noindent
Hence a more realistic evaluation requires that those factors are taken 
into account. For simplicity, the same spinor factors as for the single 
quark operator were used in paper~{\bf II}. The so obtained results 
indicate that the static approximation implies a very large overestimate 
of the axial exchange current contribution. Nevertheless, as shown in 
Tables~\ref{pitab} and~\ref{tab3}, it serves to increase the calculated 
widths for pion emission. A complete calculation would also require 
consideration of the axial exchange currents associated with the 
short-range OGE or instanton induced interactions, which was not attempted 
in paper~{\bf II}.

\noindent
Both the scalar confining and OGE interactions contribute a two-quark 
operator to the axial charge amplitude in eq.~(\ref{piRIA}). These 
operators have been calculated in ref.~\cite{Kirch} and are of the form
\begin{equation}
A_{0a}^{\mathrm{Conf}} = \frac{g_A^q}{m_q^{\,2}}\:V_c(\vectr{$k$}_2)\:
\vectr{$\sigma$}_q\cdot\vectr{$P$}\:\tau_a,
\end{equation}
for the scalar confining interaction, and
\begin{equation}
A_{0a}^{\mathrm{Oge}} = \frac{g_A^q}{m_q\,M_{\bar Q}}\:V_g(\vectr{$k$}_2)\:
\left[\vectr{$\sigma$}_q\cdot\vectr{$P$}_{\bar Q} + 
\frac{1}{2}\vectr{$\sigma$}_q\times\vectr{$\sigma$}_{\bar 
Q}\cdot\vectr{$k$}_2\right]\:\tau_a,
\end{equation}
for the OGE interaction, which, when compared with the single quark 
axial charge
\begin{equation}
A_{0a} = - \frac{g_A^q}{m_q}\:\vectr{$\sigma$}_q\cdot\vectr{$P$}\:\tau_a
\end{equation}
reveals that the scalar confining interaction will tend to cancel out the 
axial charge component of the amplitude for pion emission. However, the 
calculations in paper~{\bf II} show that the OGE contribution is 
also large, although formally suppressed by a factor $m_q/M_{\bar Q}$. It 
has also been shown in paper~{\bf II} that a relativistic treatment of the 
axial exchange charge operators will weaken them significantly. Although 
qualitative, these conclusions are in line with the experimentally small 
width of the $L=1, J=1$ $D_1$ resonance.

\newpage

\section{The pionic widths of the $D$ mesons}

\noindent
The pionic decays of the $D^*$ mesons presented in Table~\ref{pitab} are 
intriguing since the emitted pions are extremely soft. Due to the very 
small phase space, the transition $D^{*0}\rightarrow D^\pm\pi^\mp$ is 
kinematically forbidden. The results appear to favor the value 
of $g_A^q = 0.87$, although that is accidental since only 
an upper limit on the width of the $D^{\pm *}$ was known~\cite{PDG} when 
paper~{\bf II} was published. For this value of $g_A^q$, the total widths 
of the $L=1$ $D$ mesons in Table~\ref{tab3} appear to be underpredicted, 
although $\pi\pi$ transitions may contribute significantly to these, as 
proposed in paper~{\bf III}. However, the empirical widths are still 
rather uncertain~\cite{NewPDG}, and have decreased over time.

\noindent
The results for the pionic widths of the excited $D$ mesons are 
rather similar to those of ref.~\cite{Goity}, especially for the 
transitions $D^*\rightarrow D\pi$, even though the Gross framework was 
employed in that paper. That calculation was restricted 
to the transitions allowed by the lowest order selection 
rules suggested by Heavy Quark Symmetry~(HQS)~\cite{HQS}, whereas the 
present work uses the $LS$-coupling scheme, which is more 
appropriate for equal mass quarkonia. The connection between the present 
calculation and the heavy quark limit remains as yet unexplored. 

\begin{table}[h!]
\centering{
\caption{The calculated and experimental~\cite{NewPDG,Dwidth} pionic 
widths in MeV for the $D^*$ mesons, corresponding to $g_A^q$ = 0.87.
The single quark approximation, with relativistic corrections
is denoted RIA, and the result obtained upon addition of the axial 
exchange current contribution is denoted RIA + EXCH. The net results are 
also shown for $g_A^q=1$.}
\vspace{.4cm}
\begin{tabular}{l|c|c|c|c|c}
\quad Transition & $\pi$ mom. & RIA & RIA + EXCH & $g_A^q=1$ &
Experiment \\ \hline\hline
 &&&& \\
$D^{*\pm} \rightarrow D^\pm \pi^0$ & 38.3 keV & 0.026 & {\bf 0.029} &
0.038 & $0.029\pm 0.008$ MeV \\
 &&&& \\
$D^{*\pm} \rightarrow D^0\pi^\pm$ & 39.6 keV & 0.056 & {\bf 0.064} &
0.084 & $0.065\pm 0.017$ MeV \\
 &&&& \\
$D^{*0}\rightarrow D^0\pi^0$ & 43.1 keV & 0.036 & {\bf 0.041} &
0.054 & $<1.3$ MeV \\
\end{tabular}
\label{pitab}}
\end{table}

\begin{table}[h!]
\centering{
\caption{Calculated and empirical pion decay widths of the $D_1$ and   
$D_2^*$ mesons driven by the axial current and charge
operators respectively, for $g_A^q = 0.87$. The empirical values are
total widths~\cite{NewPDG}, which should mainly be due to pionic 
transitions to the ground state. The numbers in parentheses are the widths 
obtained without the axial exchange current contribution. The calculated 
values are also shown for $g_A^q = 1$.}
\vspace{.4cm}
\begin{tabular}{c|c|c|c|c|c}
Transition & Current (RIA) & Charge & Total & $g_A^q=1$
& Experiment \\ \hline\hline
 &&&& \\
$D_1\rightarrow D^*\pi$ & 4.2 (3.4) & 6.1 & {\bf 10.3 MeV} & 13.6 & 
$18.9_{-3.5}^{+4.6}$ MeV \\ &&&&& \\
$D_2^*\rightarrow D\pi$ & 8.1 (6.7) & -- & {\bf 8.1 MeV} & 10.6
& -- \\ &&&& &\\
$D_2^*\rightarrow D^*\pi$ & 3.9 (3.1) & -- & {\bf 3.9 MeV} & 5.1
& -- \\ &&&& \\
$D_2^*\rightarrow D\pi +D^*\pi$ & 11.9 (9.9) & -- & {\bf 11.9 MeV} &
15.7 & 25$^{+8}_{-7}$ MeV \\
\end{tabular}
\label{tab3}}
\end{table}

\newpage

\section{$\pi^0$ and $\gamma$ transitions from the $D_s^*$ meson}

\noindent
Consideration of the coupling of the octet of light pseudoscalar mesons 
to the light ($u,d,s$) quarks yields, analogously to eq.~(\ref{lagr}) 
the couplings
\begin{equation}
{\cal L}_{qq\varphi}\:\:=\:\:i\frac{g_A^q}{2f_\varphi}\:\bar\psi_q\:
\gamma_5\gamma_\mu\partial_\mu\:\varphi_a\lambda_a\:\psi_q.
\label{lqq}
\end{equation}
For the pions and the $\eta$ meson, the empirical decay
constants are \mbox{$f_\pi$ = 93 MeV} and \mbox{$f_\eta$ = 112 MeV,}      
respectively, so at least for the decay constants $SU(3)$ flavor symmetry 
is broken only at the 10\% level. Combination of the chiral
coupling~(\ref{lqq}) with the representation
\begin{equation}
\varphi_a\lambda_a = \sqrt{2}
       \left[ \begin{array}{ccc}
        \frac{\pi^0}{\sqrt{2}} + \frac{\eta^0}{\sqrt{6}} & \pi^+ & K^+ \\
        \pi^- & -\frac{\pi^0}{\sqrt{2}} + \frac{\eta^0}{\sqrt{6}} & K^0 \\
        \bar K^- & \bar K^0 & -\sqrt{\frac{2}{3}}\eta^0
       \end{array} \right], \quad
\psi_q = \left( \begin{array}{c}
        u \\ d \\ s
       \end{array} \right)
\label{SU3matr}
\end{equation}
gives the following definitions for the quark-level pseudovector
coupling constants $f_{\varphi qq}$, 
\begin{equation}
f_{\pi qq} = \frac{m_\pi}{2f_\pi}g_A^q, \quad
f_{\eta qq} = \frac{m_\eta}{2\sqrt{3}f_\eta}g_A^q, \quad
f_{\eta ss} = -\frac{m_\eta}{\sqrt{3}f_\eta}g_A^q.   
\label{Treimanq}
\end{equation}
The above relations then suggest that the magnitude of the coupling of   
$\eta$ mesons to $u,d$ quarks should be one-half that of the $\eta$  
coupling to strange quarks, independently of the $\eta$ meson mass. In the 
static quark model the meson-quark coupling constants of 
eq.~(\ref{Treimanq}) are related to the meson-nucleon coupling constants 
as 
\begin{equation}
f_{\pi NN} = \frac{5}{3}\,f_{\pi qq}, \quad
f_{\eta NN} = f_{\eta qq}.
\label{qmod}
\end{equation}
Application of the relations~(\ref{Treimanq}) together with
eq.~(\ref{lqq}) yields a coupling of $\eta$ mesons to strange quarks in 
terms of $f_{\eta ss}$. The coupling of the $\pi^0$ meson to the strange
quark may thus be expressed in terms of the "effective" mixing angle
$\theta_m$, which corresponds to the sum of the $\pi^0\!-\!\eta$ and 
$\pi^0\!-\!\eta'$ contributions. The width for the process 
$D_s^*\rightarrow D_s\pi^0$ can then be 
conveniently obtained from the expression for $D^{\pm*}\rightarrow 
D^\pm\pi^0$ given in paper~{\bf II} by the replacement $g_A^q / 2f_\pi 
\rightarrow f_{\eta ss} \theta_m / m_\eta$, giving
\begin{equation}
\Gamma(D_s^*\rightarrow D_s\pi^0)={1\over 6\pi}
{M_{D_s}\over M_{D_s^*}}{f_{\eta ss}^2\over m_\eta^2}
\,\theta^2_m q_\pi^3\,|{\cal M}_\pi|^2,
\label{e6}
\end{equation}
if the $\pi^0$ emission takes place at the strange quark.
Here ${\cal M}_\pi$ is a radial matrix element for pion emission. On 
the other hand, the width for the radiative M1 transition $D_s^*\rightarrow 
D_s\gamma$ was obtained, in paper~{\bf V}, as
\begin{equation} 
\Gamma(D_s^*\rightarrow D_s\gamma) =
{16\over 3}{M_{D_s}\over M_{D_s^*}}\,\alpha
\,q_\gamma^3\, |{\cal M}_\gamma|^2,
\label{e8}
\end{equation}
where ${\cal M}_\gamma$ is a radial matrix element for M1 decay. By means 
of eqs.~(\ref{e6}) and~(\ref{e8}), the ratio of the $\pi^0$ and
$\gamma$ widths of the $D_s^*$ meson is then obtained as
\begin{equation}
\frac{\Gamma_\pi}{\Gamma_\gamma} = \frac{8}{9\pi}
\frac{f_{\eta ss}^2\theta^2_m}{m_\eta^2 \alpha}
\left(\frac{q_\pi}{q_\gamma}\right)^3
\left(\frac{|{\cal M}_\pi|}{|{\cal M}_\gamma|}\right)^2,
\label{e11}
\end{equation}
where the dimension of $|{\cal M}_\gamma|$ is [MeV]$^{-1}$. 
\pagebreak

\section{Estimation of $f_{\eta NN}$}

Through use of the empirical ratio of pion and photon
momenta~\cite{PDG} known to be approximately~139/48 and the $\eta$ meson
mass of 547 MeV one may solve for the coupling constant $f_{\eta ss}$ to
get
\begin{equation}
f_{\eta ss}^2 = {\theta^{-2}_m}\:\frac{\Gamma_\pi}{\Gamma_\gamma}
\:\left(\frac{|{\cal M}_\gamma|}{|{\cal M}_\pi|}\right)^2 \cdot
\,4.814\:\mathrm{fm^{-2}}.
\label{rat}
\end{equation}
As the ratio of the $\pi^0$ and $\gamma$ decay rates is experimentally
known, albeit with quite large errors, to be $0.062\pm 0.028$~\cite{PDG},
it is, given the rather well known value of $\theta_m$, possible to obtain
an estimate for the coupling constant $f_{\eta ss}$. However, if the charm 
quark also couples to $\pi^0$, which is suggested by the empirically 
detected transitions $\psi'\rightarrow J/\psi\,\eta$ and $\psi'\rightarrow 
J/\psi\,\pi^0$, then eq.~({\ref{rat}) should be modified. The appropriate 
modifications are given in paper~{\bf V}, where it was found that a 
significant coupling of the $\pi^0$ to the charm quark will increase the 
value of $f_{\eta NN}$. For that calculation, a matrix element
for $\pi^0$ emission by the charm quark is also required.

\noindent
The matrix elements required for eq.~(\ref{rat}) were in paper~{\bf V} 
found to be
\begin{center}
${\cal M}_\gamma = -1.22\cdot 10^{-2}\:\mathrm{fm}$
\end{center}
for the M1 transition $D_s^*\rightarrow D_s\,\gamma$, and
\begin{center}
${\cal M}_\pi^s = 0.794,\quad\quad {\cal M}_\pi^c = 0.949$,
\end{center}
for $\pi^0$ emission by the strange and charm quarks. Assuming that in the 
$\pi^0$ emission by the $D_s^*$, the pion couples mostly to the strange 
constituent quark, eq.~(\ref{rat}) may be used directly together with the 
above matrix elements for the $\pi^0$ and $\gamma$ transitions. Insertion
of those matrix elements for a value of the $\pi^0 - \eta$ mixing 
angle of $\theta_m \sim 0.012$ yields $|f_{\eta ss}| \sim 0.70$. If the 
uncertainties in the mixing angle and the empirical widths for the $\pi^0$ 
and $\gamma$ transitions are taken into account, then the best estimate of 
paper~{\bf V} for the coupling of the $\eta$ meson to strange constituent 
quarks is
\begin{center}
$f_{\eta ss}\:\:=\:\: -\,0.7\:^{+0.5}_{-0.3}$.
\end{center}
In the above result, the negative sign is suggested by the relations
in eq.~(\ref{Treimanq}). The static quark model then implies, through
eq.~(\ref{qmod}), that the magnitude of the corresponding pseudovector
$\eta$-nucleon coupling constant $f_{\eta NN}$ should be one half of this
value. Thus one obtains the following final result for the $\eta$-nucleon
coupling:
\begin{center}
\fbox{$ f_{\eta NN}\:\:=\:\: 0.35\:^{+0.15}_{-0.25}$}
\end{center}
This result should be compared with the value for $f_{\eta NN}$ or the
equivalent pseudoscalar coupling constant $g_{\eta NN} = (2 m_N/m_\eta)
f_{\eta NN}$, which has been determined by phenomenological model fits to
photoproduction of $\eta$ mesons on the nucleon~\cite{Tiator}. The latter 
value for $f_{\eta NN}$ is $\sim 0.64$. That value has also been found to 
be realistic in calculations of the cross section for $pp\rightarrow pp\eta$
near threshold~\cite{Pena}. Although the result obtained above for 
$f_{\eta NN}$ has quite large uncertainties which are mostly of empirical 
origin, it still appears to be significantly smaller. A larger value for 
$f_{\eta NN}$ could, of course, be obtained by decreasing the mixing angle 
$\theta_m$.  

\newpage

\chapter{Two-pion Transitions}

As the orbitally excited $L=1$ $D_1(2420)$ and $D_2^*(2460)$ charm meson 
states lie well above the threshold, not only for single pion but also for 
two pion decay, then it is likely that a significant fraction of their total 
widths are made up by $\pi\pi$ transitions to the ground state $D$ and $D^*$ 
mesons. It is thus particularly instructive to obtain theoretical
predictions and empirical information on the branching ratios
for the latter decay modes. At present, however, the total widths of 
the $D_1(2420)$ and $D_2^*(2460)$ states are known only within a very 
wide uncertainty range, and the remaining two members of the $L=1$ multiplet 
have not yet been discovered. Paper~{\bf III} reports a calculation of the 
$\pi\pi$ decay widths of the excited $L=1$ charm meson states, by 
extending a similar calculation of the widths of their single pion 
transitions in paper~{\bf II}. 

\noindent
Two pion emission from radially excited heavy quarkonium ($Q\bar Q$) states 
empirically constitutes a significant fraction of their total
decay widths~\cite{PDG}. Indeed, in the case of the $\psi'$ (or
$\psi(2S)$), the branching ratio is empirically as large as $\sim 50 \%$. 
As the charm quarks themselves do not couple to pions, the $\pi\pi$ 
coupling to heavy flavor mesons (or quarks) involves at least two gluons 
if not a glueball. A number of different theoretical approaches for the 
coupling of two-pions to heavy mesons have been proposed, from effective 
field theory descriptions~\cite{Savage} and directly QCD-motivated 
models~\cite{Shifman} to phenomenological models~\cite{Schwinger}. In 
ref.~\cite{Mannel}, a Lagrangian motivated by chiral perturbation theory 
has been fitted to experiment. In paper~{\bf IV}, the $\pi\pi$ transitions 
from excited $Q\bar Q$ states has been described by a derivative $Q\pi\pi$ 
coupling, mediated by a heavy scalar resonance.

\section{The width for a $\pi\pi$ transition}

The $\pi\pi$ width of an excited heavy flavor meson is of the form
\begin{equation}
\Gamma_{\pi\pi}\:=\:\: 
(2\pi)^4\int\frac{d^3k_a}{(2\pi)^3}\frac{d^3k_b}{(2\pi)^3}
\frac{d^3P_f}{(2\pi)^3} \frac{M_fM_i}{E_fE_i} \frac{|T_{fi}|^2}
{4\omega_a\omega_b}\:\delta^{(4)}(P_f + k_a + k_b - P_i),
\label{2piw}
\end{equation}
where $k_a$ and $k_b$ are the four-momenta of the emitted pions,
$P_i$ and $P_f$ are those of the initial and final state quarkonia, while 
$\omega_a$ and $\omega_b$ denote the energies of the
emitted pions, respectively. The factors $M/E$ are normalization factors
for the heavy meson states similar to those employed in ref.~\cite{Goity}.

\noindent
Evaluation of the above expression leads to the following form for the 
differential width of a $\pi\pi$ transition,
\begin{eqnarray}
\frac{d\Gamma_{\pi\pi}}{d\Omega_q} \:\:=\:\: \frac{1}{4}\frac{1}{(2\pi)^4} 
&& \hspace{-.6cm} \int_0^{q_f} dq\:q^2 \label{dec} \\
&& \hspace{-.6cm} \int_{-1}^1 dz\:\frac{Q_f^2(q,z)}{\omega_a(q,z)\left(Q_f + 
\frac{qz}{2}\right) + \omega_b(q,z) \left(Q_f - \frac{qz}{2}\right)}
\:\frac{M_f}{E_f(q)}\:|T_{fi}|^2, \nonumber 
\end{eqnarray}
where the variable $z$ is defined by $\vectr{$Q$}\cdot\vectr{$q$} = Qqz$, 
and the energies of the pions and the final state quarkonium are given by
\begin{eqnarray}
\omega_a &=& \sqrt{m_{\pi}^2 + Q_f^2 + \vectr{$q$}^2/4 - Q_fqz}, \\
\omega_b &=& \sqrt{m_{\pi}^2 + Q_f^2 + \vectr{$q$}^2/4 + Q_fqz}, \\
E_f &=& \sqrt{\vectr{$q$}^2 + M_f^2}. 
\end{eqnarray} 
In the above expressions, the relative momentum of the emitted pions has 
been fixed by the delta functions in eq.~(\ref{2piw}), and can be expressed 
as
\begin{equation}
Q_f^2 = \frac{(E_f - M_i)^4 - (4m_{\pi}^2 + \vectr{$q$}^2)(E_f - 
M_i)^2}{4(E_f - M_i)^2 - 4\vectr{$q$}^2z^2}.
\end{equation}
The integration limit $q_f$ corresponds to the maximal
momentum of any one of the final state particles, e.g. the final state 
quarkonium. Thus $q_f$ can be calculated as the q-value of a transition 
$A'\rightarrow AX$, where $X$ is a particle with mass $M_X = 2m_\pi$. 

\noindent
The above formalism is adapted for computation of a $\pi\pi$ width using a 
hadronic matrix element $T_{fi}$. However, since experiments 
generally measure the invariant mass $\sqrt{s_{\pi\pi}}$, then it is 
useful to define a dimensionless variable $x$,
\begin{equation}
x = \frac{m_{\pi\pi} - 2m_\pi}{\Delta M},
\label{xvar}
\end{equation}
where $m_{\pi\pi}$ denotes the invariant mass $\sqrt{s_{\pi\pi}}$ of the
two-pion system and $\Delta M = M_i - M_f - 2m_{\pi}$. The relation
between $q$ ($ = |\vectr{$q$}|$) and $m_{\pi\pi}$ may then be obtained 
as~\cite{PDG}
\begin{equation}
|\vectr{$q$}| = \frac{\left[M_i^2 - (M_f + m_{\pi\pi})^2\right]
\left[M_i^2 - (M_f - m_{\pi\pi})^2\right]}{4M_i^2}.
\label{qfunc}
\end{equation}
From this relation, the transformation Jacobian may be
obtained as
\begin{equation}
\frac{dq}{dx} = \frac{\Delta M}{|\vectr{$q$}|}
\left\{\left[\frac{M_i^2 - (M_f + m_{\pi\pi})^2}{4\,M_i^2}\right]
(M_f\!-\!m_{\pi\pi}) - \left[\frac{M_i^2 - (M_f - 
m_{\pi\pi})^2}{4\,M_i^2}\right](M_f\!+\!m_{\pi\pi})\right\},
\end{equation}
where $|\vectr{$q$}|$ is given in terms of $m_{\pi\pi}$ by 
eq.~(\ref{qfunc}). The width for a $\pi\pi$ transition may thus be 
calculated as 
\begin{equation}
\Gamma_{\pi\pi} = \int_0^{q_f} dq \frac{d\Gamma_{\pi\pi}}{dq} 
= - \int_0^1 dx
\frac{d\Gamma_{\pi\pi}}{dq} \frac{dq}{dx}.
\label{dwidth}
\end{equation}
In the above equation, the latter form turns out to be the most useful, 
since the experimental $\pi\pi$ spectra are presented either in terms of 
$\Gamma^{-1} d\Gamma/dx$ or $d\Gamma/dx$ versus $x$.
\pagebreak

\section{$\pi\pi$ transitions in heavy-light mesons}

The emission of two pions from the light constituent quarks is modeled in 
paper~{\bf III} in terms of the conventional chiral pion-quark pseudovector 
coupling model, augmented by a pointlike Weinberg-Tomozawa term. The only 
free parameter in this model is the axial coupling $g_A^q$ of the light 
constituent quarks, as the quark masses and other parameters of the 
Hamiltonian model were fitted to the $D$ meson spectrum. In addition to the 
single quark amplitude for two-pion emission, the exchange current 
contribution to the Weinberg-Tomozawa interaction was also considered, and 
found to interfere destructively with the single quark amplitude. 

\noindent
The pion-quark pseudovector coupling~(\ref{lagr}) gives rise to Born and 
crossed Born amplitudes of conventional form for the emission of two 
pions from an interacting constituent quark. The chiral model for the 
$\pi\pi$ emission amplitude is completed by the Weinberg-Tomozawa~(WT) 
interaction, which is described by the Lagrangian
\begin{equation}
{\cal L}_{\mathrm{WT}} =
-\frac{i}{4f_{\pi}^2}\:\bar\psi_q\,\gamma_{\mu}\,
\tau_a\:\pi_a\times\partial_{\mu}\pi_a\,\psi_q.
\label{Tomoz}
\end{equation}
The general isospin decomposition of the $\pi\pi$ emission
amplitude for constituent quarks is, in analogy with that for nucleons,   
\begin{equation}
T = \delta_{ab}T^+ + \frac{1}{2}[\tau_b,\tau_a] T^-,
\label{plmi}
\end{equation}  
while the general expression for the amplitudes $T^+$ and $T^-$ is
\begin{equation}
T^\pm = \bar u(\vectr{$p$}') \left(A^\pm - i\gamma_\mu Q_\mu B^\pm\right) 
u(\vectr{$p$}).
\label{ampl}
\end{equation}
In these expressions, $Q$ denotes the relative four-momentum $Q =
(k_b - k_a)/2$ of the emitted pions. In this notation the Born,
crossed Born and Weinberg-Tomozawa amplitudes are, respectively
\begin{eqnarray}
T_{\mathrm B} &=& \:\:\:i\left(\frac{g_A^q}{2f_{\pi}}\right)^2
\left[\gamma_\mu Q_\mu - 2im_q - 4m_q^2\frac{\gamma_\mu Q_\mu}{p_a^2 +
m_q^2}\right] \left(\delta_{ba} + 
\frac{1}{2}\left[\tau_b,\tau_a\right]\right),
\quad\quad \\
T_{\mathrm {CB}} &=& -i\left(\frac{g_A^q}{2f_{\pi}}\right)^2
\left[\gamma_\mu Q_\mu + 2im_q - 4m_q^2\frac{\gamma_\mu Q_\mu}{p_b^2 +
m_q^2}\right] \left(\delta_{ba} - 
\frac{1}{2}\left[\tau_b,\tau_a\right]\right), \\
T_{\mathrm{WT}} &=& - i\,\frac{\gamma_\mu Q_\mu}{4f_{\pi}^2}\:
[\tau_b,\tau_a].
\end{eqnarray}
Comparison of these amplitudes with eq.~(\ref{ampl}) yields the desired 
expressions for the sub-amplitudes $A^\pm$ and $B^\pm$, which are of the 
form
\begin{eqnarray}
A^+ &=& \left(\frac{g_A^q}{2f_{\pi}}\right)^2 4m_q, \\
A^- &=& 0, \\
B^+ &=& -\left(\frac{g_A^q}{2f_{\pi}}\right)^2 4m_q^2
\left[\frac{1}{s - m_q^2} - \frac{1}{u - m_q^2}\right], \\
B^- &=& -\left(\frac{g_A^q}{2f_{\pi}}\right)^2
\left(2 + 4m_q^2\left[\frac{1}{s - m_q^2} + \frac{1}{u - m_q^2}\right]
\right) + \frac{1}{2f_\pi^2}. \label{B-}
\end{eqnarray}
Here the identities $p_a^2 = -s$ and $p_b^2 = -u$, where
$s$ and $u$ are the invariant Mandelstam variables, have been used.
These results for   
the $A$ and $B$ amplitudes are formally equivalent to the
corresponding results for the two-pion emission
amplitude for nucleons~\cite{Hamilton}. Note that in
eq.~(\ref{B-}), the contribution from the Weinberg-Tomozawa interaction
tends to cancel the constant term in the $B^-$ amplitude that
arises from the Born terms. If the axial coupling constant $g_A^q$ is
taken to equal 1, then this cancellation is exact.

\noindent
For calculational purposes, eq.~(\ref{ampl}) was split in paper~{\bf III} 
into spin-independent and spin-dependent parts according to
\begin{equation}
T^{\pm} = \alpha^{\pm} + i\vectr{$\sigma$}_q\cdot\vectr{$\beta$}^{\pm}.
\label{ssum}
\end{equation}
The spin and isospin summed squared amplitude for a given $\pi\pi$ 
transition is then expressed in the form
\begin{equation}
|T|^2\:\:=\:\:|T|_{\alpha^+}^2 + |T|_{\alpha^-}^2 + |T|_{\beta^+}^2 + 
|T|_{\beta^-}^2,
\end{equation}
from which the $\pi\pi$ width is obtained by insertion into 
eq.~(\ref{dec}). The explicit expressions for the above squared 
amplitudes, given in paper~{\bf III}, are different for each transition 
because of the summation over spin and isospin. The numerical results for 
the above squared amplitudes are presented in Table~\ref{reltab}.

\noindent
It is instructive for the determination of the two-quark
contribution to the Weinberg-Tomozawa interaction to write the 
Lagrangian~(\ref{Tomoz}) in the form of a current-current coupling,
\begin{equation}
{\cal L}_\mathrm{WT} = -\frac{1}{4 f_\pi^2}\:V_{\mu a}\:
\pi_a\times\partial_\mu\pi_a,
\label{ex1}
\end{equation}
where $V_{\mu a} = i\bar\psi_q\,\gamma_{\mu}\,\psi_q\,\tau_a$ 
is the isovector current of the light constituent quark and $\pi_a
\times\partial_\mu\pi_a$ is the current of the $\pi\pi$ system. Given this 
expression, it becomes natural to describe the irreducible two-quark 
contribution to the $\pi\pi$ production operator by means of two-quark 
interaction current contributions to the isovector current $V_{\mu a}$.

\noindent
In the non-relativistic limit, the spatial term in the isovector 
current $V_{\mu a} = (\vectr{$V$}_a, iV_{0a})$ of the light constituent 
quark takes the form
\begin{equation}
\vectr{$V$}_a\:\:=\:\:\left[\frac{\vectr{$p$}'_q + \vectr{$p$}_q 
-i\vectr{$\sigma$}_q\times\vectr{$q$}}{2m_q}\right]\,\tau_a,
\label{ex2}   
\end{equation}
where $m_q$ is the light constituent quark mass and $\vectr{$p$}_q$ and
$\vectr{$p$}'_q$ the initial and final quark momenta, respectively. The 
expressions for the exchange current contributions for the scalar 
confining and OGE interactions have been calculated in 
ref.~\cite{Tsushima}, and may be expressed as
\begin{eqnarray}
\vectr{$V$}_a^{\mathrm{Conf}} &=& - 
\frac{V_c(\vectr{$k$}_2)}{m_q}\:\vectr{$V$}_a, 
\label{ex3} \\
\vectr{$V$}_a^{\mathrm{Oge}}  &=& - \frac{V_g(\vectr{$k$}_2)}{2 m_q^2}\left[
\frac{m_q}{M_Q}
\left(\vectr{$p$}'_Q + \vectr{$p$}_Q + i\vectr{$\sigma$}_Q\times\vectr{$ 
k$}_2\right) + i\vectr{$\sigma$}_q\times\vectr{$k$}_2\right]\:\tau_a.
\label{gluecur}
\end{eqnarray}
In paper~{\bf III}, the above operators contribute only to 
$|T|_{\alpha^-}^2$, and were found to reduce the widths for $\pi\pi$ 
transitions. However, it was also found that the nonrelativistic treatment 
of the two-quark operators is somewhat unrealistic because of the small 
mass of the light constituent quark. An approximate relativistic treatment 
akin to that employed for the single pion transitions in paper~{\bf II} 
was therefore used in paper~{\bf III}. The results obtained upon addition 
of the two-quark operators~(\ref{ex3}) and~(\ref{gluecur}) are shown in 
Table~\ref{tbtab}.

\pagebreak

\begin{table}[h!]
\centering 
\caption{Numerical results from paper~{\bf III} with single-quark amplitudes 
and relativistic matrix elements for the $\pi\pi$ widths of the spin 
triplet $D_2^*$, $D_1$, $D_0^*$ and the spin singlet $D_1^*$ mesons. The 
individual results from the spin independent ($\alpha^\pm$) and spin 
dependent amplitudes ($\beta^\pm$) are shown for $g_A^q$ = 1.}
\vspace{.3cm}
\small
\begin{tabular}{l|c|c|c|c|c|c}
\quad Transition & $|T|_{\alpha^+}^2$ & $|T|_{\alpha^-}^2$ & $|T|_{\beta^-}^2$
& $|T|_{\beta^+}^2$ & $g_A^q=1$ & $g_A^q=0.87$
\\ \hline\hline &&&&&&\\
$D_2^*\rightarrow D^*\pi\pi$ & 0.896 & 2.864 & $5.06\cdot 10^{-3}$ &  
$1.17\cdot 10^{-3}$ & {\bf 3.77 MeV} & 2.39 MeV \\
$D_2^*\rightarrow D \pi\pi$  & --    & --    & $6.45\cdot 10^{-2}$ &
$7.51\cdot 10^{-3}$ & {\bf 0.07 MeV} & 0.05 MeV \\
$D_1  \rightarrow D^*\pi\pi$ & 0.367 & 1.291 & $2.33\cdot 10^{-3}$ &
$7.85\cdot 10^{-4}$ & {\bf 1.66 MeV} & 1.05 MeV \\
$D_1^*\rightarrow D^*\pi\pi$ & --    & --    & $6.54\cdot 10^{-4}$ &
$3.30\cdot 10^{-4}$ & {\bf $\simeq 0$ MeV} & $\simeq 0$ MeV \\
$D_1^*\rightarrow D \pi\pi$  & 2.749 & 7.915 & --                  &
--                  & {\bf 10.7 MeV} & 6.80 MeV \\  
$D_0^*\rightarrow D^*\pi\pi$ & 0.020 & 0.098
& -- & --  & {\bf 0.12 MeV} & 0.07 MeV \\
$D_0^*\rightarrow D \pi\pi$ & --    & --    & $1.43\cdot 10^{-2}$ &   
$2.86\cdot 10^{-3}$ & {\bf 0.02 MeV} & 0.01 MeV 
\end{tabular} 
\label{reltab}
\end{table}

\begin{table}[h!]   
\centering 
\caption{Numerical results from paper~{\bf III} for the most important 
$\pi\pi$ transitions, upon addition of the two-quark contributions 
associated with the scalar confining and OGE interactions to the 
Weinberg-Tomozawa Lagrangian, for $g_A^q = 1$.}
\vspace{.3cm}
\small
\begin{tabular}{l|c|c|c||c|c}
& \multicolumn{3}{c||}{$|T|_{\alpha^-}^2$} & \multicolumn{2}{c}{Total} \\
\quad Transition & Rel & +Conf & +OGE & $g_A^q = 1$ & $g_A^q = 0.87$ \\
\hline\hline &&&&& \\
$D_2^*\rightarrow D^*\pi\pi$ & 2.864 & 2.377 & 2.144 & {\bf 3.05 MeV} & 1.82 MeV 
\\
$D_1\rightarrow D^*\pi\pi$   & 1.291 & 1.076 & 0.974 & {\bf 1.34 MeV} & 0.80 MeV 
\\
$D_1^*\rightarrow D\pi\pi$   & 7.915 & 6.535 & 5.872 & {\bf 8.62 MeV} & 5.17 MeV 
\\
\end{tabular}
\label{tbtab}
\end{table}

\noindent
The results for the $\pi\pi$ widths of the positive parity charm mesons with 
$L=1$ in Table~\ref{reltab} reveal a strong sensitivity to the value of 
$g_A^q$ as well as the large hyperfine splittings in the empirical $D$ meson 
spectrum. Consequently, some transitions are kinematically favored, while 
others, in particular $D_0^*\rightarrow D^*\pi\pi$, are strongly inhibited 
by the small phase space available. The $\pi\pi$ widths of the $L=1$ $D$
mesons are, therefore, also very sensitive to the spin-orbit structure of 
the $Q\bar q$ interaction, which was also found to be the case for the 
analogous single pion transitions in paper~{\bf II}. In the absence of 
empirical information and definite QCD lattice calculations~\cite{Dlattpap}, 
the energies of the as yet undiscovered $D_0^*$ and $D_1^*$ mesons were in 
paper~{\bf III} taken to equal those predicted by the calculation of 
ref.~\cite{Lahdeout}.

\noindent
The predicted total widths of the $L=1$ $D$ mesons may be obtained by adding 
the calculated $\pi\pi$ widths in Table~\ref{reltab} to those for the $\pi$ 
transitions obtained in paper~{\bf II}. With $g_A^q=1$ the total   
calculated $\pi\pi$ width in the single quark approximation of the 
$D_2^*(2460)$ is 3.8 MeV and that of the $D_1(2420)$ is 1.7 MeV. 
Such an addition brings the total calculated width of the $D_2^*(2460)$ to 
19.5 MeV, which is well within the uncertainty margin of 
the empirical value $25_{-7}^{+8}$ MeV~\cite{PDG}. In the case of the 
$D_1(2420)$ meson, the total calculated width for $\pi$ and $\pi\pi$ 
emission comes to 15.3 MeV, which is close to the empirical uncertainty 
margin of the total width $18.9_{-3.5}^{+4.6}$ MeV~\cite{PDG}. However, 
reduction of the value for $g_A^q$ to 0.87 brings the calculated widths 
for $\pi$ and $\pi\pi$ emission somewhat below the empirical values for 
the total widths. Likewise, employment of the two-quark contributions 
associated with the $Q\bar q$ interaction has the effect of reducing the 
calculated $\pi\pi$ widths, as shown in Table~\ref{tbtab}.

\pagebreak

\section{$\pi\pi$ transitions in heavy quarkonia}

Theoretical work on the $\pi\pi$ decays of excited heavy quarkonia has   
demonstrated that the empirical energy spectra of the emitted $\pi\pi$ 
system demands that the pions be derivatively coupled to the
heavy quarkonium states. This is consistent with the role of the pions as
Goldstone bosons of the spontaneously broken approximate chiral symmetry
of QCD. Most models~\cite{Savage,Shifman,Schwinger,Mannel} have dealt with 
the coupling of two-pions to the heavy meson as a whole rather than to its 
constituent quarks. The satisfactory description obtained suggests that the 
decay amplitude $T_{fi}$ at the quark level should be a smooth function of 
the $\pi\pi$ momentum \vectr{$q$}, which is dominated by single-quark 
mechanisms for $\pi\pi$ emission. However, the pion rescattering or pion 
exchange term that appears naturally as a consequence of the coupling of 
two-pions to constituent quarks has been found, in paper~{\bf IV}, to 
be dominant since it is not suppressed by the orthogonality of the 
quarkonium wave functions.

\noindent
It was shown in paper~{\bf IV} that an unrealistically large pion 
rescattering contribution may be avoided if the $Q\pi\pi$ vertex involves an 
intermediate, fairly light and broad $\sigma$ meson, in line with the 
phenomenological resonance model of ref.~\cite{Schwinger}. An intermediate 
$\sigma$ meson suppresses the contributions from pion rescattering 
mechanisms, while single quark amplitudes are but slightly affected. 
Together with a relativistic treatment of the single quark amplitude, this 
suppression of the pion rescattering contribution has been shown in 
paper~{\bf IV} to reproduce the expected smooth behavior of the transition 
amplitude.

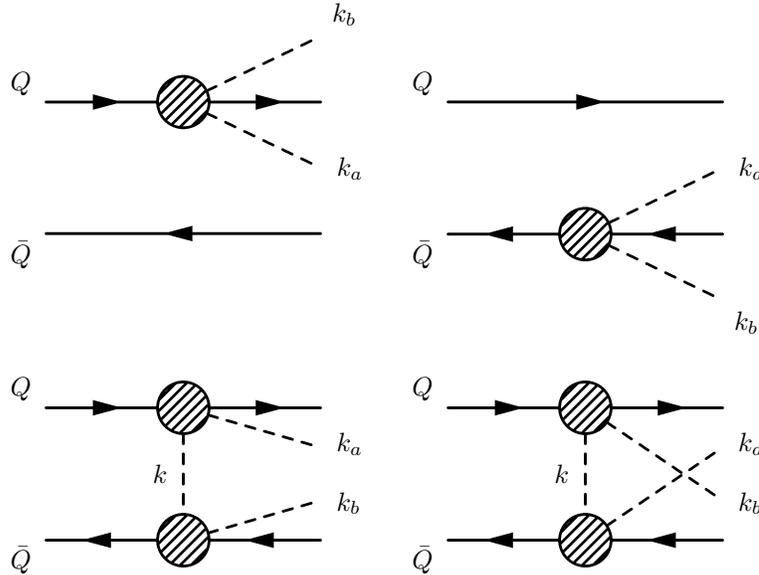
\begin{figure}[h!] 
\centering  
\begin{tabular}{c c}
\begin{fmffile}{2pisq1}
\begin{fmfgraph*}(130,100) \fmfpen{thin}
\fmfleft{i1,i2}
\fmfright{o1,o2}
\fmftop{o3,o4}
\fmf{fermion}{i2,v1,o2}
\fmf{fermion}{o1,i1}
\fmf{dashes}{v1,o3}
\fmf{dashes}{v1,o4}
\fmfblob{.15w}{v1}
\fmflabel{$Q$}{i2} \fmflabel{$\bar Q$}{i1}
\fmflabel{$k_b$}{o3} \fmflabel{$k_a$}{o4}
\fmfforce{(.1w,.75h)}{i2}
\fmfforce{(.1w,.25h)}{i1}
\fmfforce{(.9w,.75h)}{o2}
\fmfforce{(.9w,.25h)}{o1}  
\fmfforce{(.5w,.75h)}{v1}
\fmfforce{(.9w,h)}{o3}
\fmfforce{(.9w,.5h)}{o4}
\end{fmfgraph*}
\end{fmffile}
&
\begin{fmffile}{2pisq2}
\begin{fmfgraph*}(130,100) \fmfpen{thin}
\fmfleft{i1,i2} 
\fmfright{o1,o2}
\fmftop{o3,o4}
\fmf{fermion}{o1,v1,i1}
\fmf{fermion}{i2,o2}
\fmf{dashes}{v1,o3}
\fmf{dashes}{v1,o4}
\fmfblob{.15w}{v1}
\fmflabel{$Q$}{i2} \fmflabel{$\bar Q$}{i1}
\fmflabel{$k_b$}{o3} \fmflabel{$k_a$}{o4}
\fmfforce{(.1w,.75h)}{i2}
\fmfforce{(.1w,.25h)}{i1}
\fmfforce{(.9w,.75h)}{o2}
\fmfforce{(.9w,.25h)}{o1}
\fmfforce{(.5w,.25h)}{v1}
\fmfforce{(.9w,0)}{o3}   
\fmfforce{(.9w,.5h)}{o4} 
\end{fmfgraph*}
\end{fmffile}
\\ &\\
\begin{fmffile}{2piex1} 
\begin{fmfgraph*}(130,100) \fmfpen{thin}
\fmfleft{i1,i2}
\fmfright{o1,o2}
\fmftop{o3,o4}
\fmf{fermion}{i2,v1,o2}
\fmf{fermion}{o1,v2,i1}
\fmf{dashes}{v1,o3}
\fmf{dashes,label=$k$}{v1,v2}
\fmf{dashes}{v2,o4}
\fmfblob{.15w}{v1}  
\fmfblob{.15w}{v2} 
\fmflabel{$Q$}{i2} \fmflabel{$\bar Q$}{i1}
\fmflabel{$k_a$}{o3} \fmflabel{$k_b$}{o4}
\fmfforce{(.1w,.75h)}{i2}
\fmfforce{(.1w,.25h)}{i1}
\fmfforce{(.9w,.75h)}{o2}
\fmfforce{(.9w,.25h)}{o1}
\fmfforce{(.5w,.75h)}{v1}
\fmfforce{(.5w,.25h)}{v2}
\fmfforce{(.9w,.6h)}{o3} 
\fmfforce{(.9w,.4h)}{o4} 
\end{fmfgraph*}
\end{fmffile}  
&
\begin{fmffile}{2piex2}
\begin{fmfgraph*}(130,100) \fmfpen{thin}
\fmfleft{i1,i2}
\fmfright{o1,o2}
\fmftop{o3,o4}  
\fmf{fermion}{i2,v1,o2}
\fmf{fermion}{o1,v2,i1}
\fmf{dashes}{v1,o3}
\fmf{dashes,label=$k$}{v1,v2}
\fmf{dashes}{v2,o4}
\fmfblob{.15w}{v1} 
\fmfblob{.15w}{v2}  
\fmflabel{$Q$}{i2} \fmflabel{$\bar Q$}{i1}
\fmflabel{$k_b$}{o3} \fmflabel{$k_a$}{o4} 
\fmfforce{(.1w,.75h)}{i2}
\fmfforce{(.1w,.25h)}{i1}
\fmfforce{(.9w,.75h)}{o2}
\fmfforce{(.9w,.25h)}{o1}
\fmfforce{(.5w,.75h)}{v1}
\fmfforce{(.5w,.25h)}{v2}
\fmfforce{(.9w,.4h)}{o3} 
\fmfforce{(.9w,.6h)}{o4} 
\end{fmfgraph*}
\end{fmffile} \\
\end{tabular}  
\caption{Tree-level Feynman diagrams for the emission of two-pions by heavy
constituent quarks. The two upper diagrams correspond to the single-quark
amplitudes $T_Q$ and $T_{\bar Q}$ respectively, while the two lower
diagrams describe the pion rescattering amplitudes $T_{\mathrm{ex}}$ and 
$T_{\mathrm{exc}}$ which involve a pion exchange between the heavy 
constituent quarks.}
\label{Feynfig} 
\end{figure}
\pagebreak

\noindent
If the coupling of the pions to the constituent quark does not
involve derivatives of the pion field, then agreement with experiment is
excluded for the pion invariant mass distributions in the decays 
$\Upsilon'\rightarrow \Upsilon\,\pi\pi$ and $\psi'\rightarrow
J/\psi\,\pi\pi$~\cite{bbexp,Schwinger}. Derivative couplings for the pions
are also consistent with the role of pions as Goldstone bosons of the
spontaneously broken approximate chiral symmetry of QCD. The effctive 
$Q\pi\pi$ interaction Lagrangian is therefore expected to have the form
\begin{equation}
{\mathcal L}_{Q\pi\pi} = 4\pi\lambda\:\bar\psi_Q\:
\partial_\mu\pi_a\:\partial_\mu\pi_a\:\psi_Q,
\label{Q2pi}
\end{equation}
where $\lambda$ is a coupling constant of dimension~$[\mathrm{MeV}]^{-3}$
and $\psi_Q$,$\bar\psi_Q$ denote the heavy quark spinors. The total 
tree-level amplitude for $\pi\pi$ emission can then be expressed in 
terms of single quark and pion rescattering terms, as illustrated in 
Fig.~\ref{Feynfig}. The isospin dependence of the dipion-quark coupling 
then implies that
\begin{equation}
|T|^2_{\pi\pi} = 2|T|^2_{\pi^+\pi^-} + |T|^2_{\pi^0\pi^0}.
\end{equation}
As a consequence the width for emission of charged pions should be
twice that for emission of neutral pions, which is in fair agreement with 
what is found experimentally~\cite{PDG,bbexp}.

\noindent
The effect of the strong interaction in the $\pi\pi$ system may be 
approximately accounted for by inclusion of an intermediate scalar meson 
($\sigma$ or glueball) resonance in the vertex. This is brought about by 
modification of the coupling constant $\lambda$ with a relativistic scalar 
meson propagator of the Breit-Wigner type:
\begin{equation}
\lambda \rightarrow \lambda\left(\frac{M_\sigma^2 + \Gamma_\sigma^2/4}
{M_\sigma^2 + q^2 + \Gamma_\sigma^2/4}\right).
\label{sigmares}
\end{equation}
Here $M_\sigma$ denotes the pole position $m_\sigma-i\Gamma_\sigma/2$, and 
$q$ the four-momentum, of the effective scalar ($\sigma$) meson resonance. 
The $\sigma$ resonance appears by infinite
iteration of the four-pion vertex in the isospin 0 spin 0 channel.
Therefore, as pointed out in ref.~\cite{Schwinger}, it is natural to
describe the strongly interacting $\pi\pi$ state by a broad $\sigma$ pole
rather than by the driving term (4-pion vertex) alone. When the 
modification~(\ref{sigmares}) is taken into account, the
expression for the single-quark amplitude becomes
\begin{equation}
T_{\mathrm{1q}} =
-16\:\pi\lambda\left(\frac{M_\sigma^2 + \Gamma_\sigma^2/4}
{M_\sigma^2 + q^2 + \Gamma_\sigma^2/4}\right)\left[m_\pi^2 -
\frac{1}{2}\left((\omega_a + \omega_b)^2 - \vectr{$q$}^2\right)\right]
{\mathcal M}_{\mathrm{1q}}.
\label{1q2}
\end{equation}
The nonrelativistic approximation for the matrix element ${\mathcal 
M}_{\mathrm{1q}}$ is unreliable, as the radial $S$-wave quarkonium wave 
functions are orthogonal. A relativistic form for the matrix element 
${\mathcal M}_{\mathrm{1q}}$, where $P = |\vectr{$P$}|$, $q = 
|\vectr{$q$}|$ and $\vectr{$P$}\cdot\vectr{$q$} = Pqv$, may be obtained as
\begin{eqnarray}
{\mathcal M}_{\mathrm{1q}}^{\mathrm{rel}} &=& \frac{1}{\pi}
\int_0^\infty dr'\,r'\,u_f(r')\,\int_0^\infty dr\,r\,u_i(r)
\int_0^\infty dP\,P^2 \int_{-1}^1 dv\:\alpha(P,v,q) \nonumber \\
&& j_0\left(r'\sqrt{\vectr{$P$}^2 + \frac{\vectr{$q$}^2}{16} - 
\frac{Pqv}{2}}\:\right)\:
j_0\left(r\sqrt{\vectr{$P$}^2 + \frac{\vectr{$q$}^2}{16} + 
\frac{Pqv}{2}}\:\right),
\label{relmat}
\end{eqnarray} 
where $\alpha(P,v,q)$ is a factor which includes the quark spinors in the
coupling~(\ref{Q2pi}),
\begin{equation}
\alpha(P,v,q) = \sqrt{\frac{(E+M_Q)(E'+M_Q)}{4EE'}}
\left(1-\frac{\vectr{$P$}^2 - \vectr{$q$}^2/4}{(E'+M_Q)(E+M_Q)}\right).
\label{alpha}   
\end{equation}   
\pagebreak

\noindent
In principle the quark spinors in the coupling~(\ref{Q2pi}) also contain
a spin dependent part that is proportional to 
$\vectr{$q$}\times\vectr{$P$}$. For the present purposes that contribution 
turns out to be very tiny and may be safely neglected. It has been shown in 
paper~{\bf IV} that the relativistic modifications to the single-quark 
matrix element increase the magnitude of the single quark amplitudes 
for $\pi\pi$ decay. 

\begin{figure}[h!]
\parbox{0.48\textwidth}{ 
\caption{Modeling of the pion rescattering term in Fig.~\ref{Feynfig} in 
terms of intermediate $\sigma$ mesons. The four-momenta of the $\sigma$ 
mesons are defined as $k_1 = -k -k_a$ and $k_2 = k - k_b$. The crossed 
rescattering diagram in Fig.~\ref{Feynfig} can be obtained by 
interchanging $k_a$ and $k_b$. The physically reasonable approximations 
$\vectr{$k$}_1 \approx -\vectr{$k$}$, $\vectr{$k$}_2 \approx \vectr{$k$}$, 
$|k_1^0| \approx |k_2^0| \approx (\omega_a+\omega_b)/2$ and $k_0 \approx 
(\omega_b-\omega_a)/2$ were made in paper~{\bf IV}, to allow for a 
simpler treatment of the triple propagators in the pion rescattering 
amplitudes.}
\label{sigmafig}}
\parbox{0.48\textwidth}{\begin{center}
\begin{fmffile}{sigmaex}
\begin{fmfgraph*}(150,150) \fmfpen{thin}
\fmfcmd{%
 vardef port (expr t, p) =
  (direction t of p rotated 90)
   / abs (direction t of p)
 enddef;}
\fmfcmd{%
 vardef portpath (expr a, b, p) =
  save l; numeric l; l = length p;
  for t=0 step 0.1 until l+0.05:
   if t>0: .. fi point t of p
    shifted ((a+b*sind(180t/l))*port(t,p))
  endfor
  if cycle p: .. cycle fi
 enddef;}
\fmfcmd{%
 style_def brown_muck expr p =
  shadedraw(portpath(thick/2,2thick,p)
   ..reverse(portpath(-thick/2,-2thick,p))
   ..cycle)
 enddef;}
\fmfleft{i1,i2}
\fmfright{o1,o2}
\fmftop{o3,o4}   
\fmf{fermion}{i2,v1,o2}
\fmf{fermion}{o1,v2,i1} 
\fmf{dashes}{v3,o3}
\fmf{dashes,label=$\pi(k)$}{v3,v4}
\fmf{dashes}{v4,o4}
\fmfdot{v1,v2,v3,v4}
\fmf{brown_muck,lab.s=right,lab.d=4thick,lab=$\sigma(k_1)$}{v1,v3}
\fmf{brown_muck,lab.s=right,lab.d=4thick,lab=$\sigma(k_2)$}{v2,v4}
\fmflabel{$Q$}{i2} \fmflabel{$\bar Q$}{i1}
\fmflabel{$\pi(k_a)$}{o3} \fmflabel{$\pi(k_b)$}{o4}
\fmfforce{(.1w,.95h)}{i2}
\fmfforce{(.1w,.05h)}{i1}
\fmfforce{(.9w,.95h)}{o2}
\fmfforce{(.9w,.05h)}{o1}
\fmfforce{(.5w,.95h)}{v1}
\fmfforce{(.5w,.05h)}{v2}
\fmfforce{(.6w,.70h)}{v3}
\fmfforce{(.6w,.30h)}{v4}
\fmfforce{(.9w,.6h)}{o3}
\fmfforce{(.9w,.4h)}{o4}
\end{fmfgraph*}
\end{fmffile}\end{center}}
\end{figure}

\noindent
The pion rescattering amplitude that corresponds to Fig.~\ref{sigmafig} 
may be expressed as 
\begin{equation}
T_{\mathrm{ex}} = -64\:\pi^2\lambda^2\:(M_\sigma^2 + \Gamma_\sigma^2/4)^2
\frac{k_{a\mu}k_\mu\:k_{b\nu}k_\nu}{(k_1^2+M_\sigma^2 + \Gamma_\sigma^2/4)
(k^2+m_\pi^2)(k_2^2+M_\sigma^2 + \Gamma_\sigma^2/4)},
\label{3prop}
\end{equation}  
where the momenta are defined as in Fig.~\ref{sigmafig}. Upon simplification 
of the triple propagator according to the recipe of Fig.~\ref{sigmafig}, the 
crossed pion rescattering diagram gives an extra factor of 2, yielding the 
expression
\begin{eqnarray}
T_{\mathrm{2q}} &=& -128\:\pi^2\lambda^2\:
\left\{\frac{1}{3}\left(\frac{\vectr{$q$}^2}{4}-Q_f^2\right)
\left[{\mathcal M_{\mathrm e1}} - A^2({\mathcal M_{\mathrm e2}}-{\mathcal
M_{\mathrm e3}})\right] \right. \label{2q3} \\
&&\left.+\left(\frac{\vectr{$q$}^2z^2}{4} - \frac{2}{3}Q_f^2 - \frac{
\vectr{$q$}^2}{12}\right){\mathcal M_{\mathrm e4}}
+\frac{\omega_a\omega_b}{4}(\omega_a - \omega_b)^2
({\mathcal M_{\mathrm e2}}-{\mathcal M_{\mathrm e3}})\right\}, \nonumber
\end{eqnarray}
where the term which is proportional to the matrix element $\mathcal 
M_{\mathrm e4}$ represents an amplitude for $D$-wave $\pi\pi$ emission. The 
matrix elements in eq.~(\ref{2q3}) may, in the non-relativistic 
approximation, be expressed as
\begin{eqnarray}
{\mathcal M_{\mathrm e1}} \!\!\!&=&\!\!\!\! \int_0^\infty \!\!
dr\,u_f(r)u_i(r)\,j_0(Q_fr)\:
\frac{(M_\sigma^2 + \Gamma_\sigma^2/4)^2}{4\pi}
\left(\frac{e^{-Xr}}{2X}\right), \\
{\mathcal M_{\mathrm e2}} \!\!\!&=&\!\!\!\! \int_0^\infty \!\!
dr\,u_f(r)u_i(r)\,j_0(Q_fr)\:
\frac{(M_\sigma^2 + \Gamma_\sigma^2/4)^2}
{4\pi(X^2-A^2)^2}\:A\,Y_0(Ar), \label{M2}\\
{\mathcal M_{\mathrm e3}} \!\!\!&=&\!\!\!\! \int_0^\infty \!\!
dr\,u_f(r)u_i(r)\,j_0(Q_fr)\:
\frac{(M_\sigma^2 + \Gamma_\sigma^2/4)^2}
{4\pi(X^2-A^2)^2}\left[XY_0(Xr)\!+\! 
\frac{(X^2\!-\!A^2)}{2X}\,e^{-Xr}\right],\quad\quad \\
{\mathcal M_{\mathrm e4}} \!\!\!&=&\!\!\!\! \int_0^\infty \!\!
dr\,u_f(r)u_i(r)\,j_2(Q_fr)\:\frac{(M_\sigma^2 + \Gamma_\sigma^2/4)^2}
{4\pi(X^2-A^2)^2}\:F_2(r) \label{M4}.
\end{eqnarray}
In the above matrix elements, $X$ is defined as $X = \sqrt{M_\sigma^2 +
\Gamma_\sigma^2/4 - (\omega_a+\omega_b)^2/4}$, while $Y_0(r)$ denotes the
Yukawa function $e^{-r}/r$. Note that when the value of $k_0^2$ exceeds
$m_\pi^2$, the analytic continuation $A\rightarrow
-i\,\sqrt{k_0^2 - m_\pi^2}$~\cite{Chai} is employed for the matrix \mbox{
element (\ref{M2}).} Further, $u_f(r)$ and $u_i(r)$ denote the reduced
radial wave functions for the final and initial state heavy quarkonia,
respectively. The function $F_2(r)$, which is defined in paper~{\bf IV}, is 
closely related to, and in the limit $m_\sigma \rightarrow \infty$ 
actually reduces to, a Yukawa $Y_2$ function~\cite{Chai}. It turns out 
that the matrix 
element~(\ref{M4}) is numerically quite insignificant, because of the strong 
suppression caused by the $j_2$ function for small values of $Qr$. Also, the 
smallness of $k_0$ as compared with \vectr{$k$} precludes any terms 
proportional to $k_0$ or $k_0^2$ from playing any major role.

\noindent
In view of the many approximations involved in the above treatment of the 
pion rescattering terms, a check against an unapproximated calculation is 
desirable. This is possible since the triple propagator in eq.~(\ref{3prop}) 
may also be considered without any approximation in $k_1$ and $k_2$, at the 
price of numerically much more cumbersome expressions. By means of the 
Feynman parameterization
\begin{equation}
\frac{1}{ABC} = 2\int_0^1 dx\,x \int_0^1 dy
\frac{1}{[A(1-x)+Bxy+Cx(1-y)]^3},
\label{feynpar}
\end{equation}
the two-quark amplitudes of Fig.~\ref{sigmafig} may be cast in the form
\begin{eqnarray}
T_{\mathrm{2q}} &=& -(8\pi \lambda)^2
\left\{\frac{1}{3}\left(\frac{\vectr{$q$}^2}{4}-Q_f^2\right)
\left[\int_0^1 dx\,x\int_0^1 dy \left\{
{\mathcal M_{\mathrm I}} - A^2{\mathcal M_{\mathrm {II}}}\right\}\right]
\right. \nonumber \\
&& \quad\quad\quad\quad
+ \int_0^1 dx\,x\int_0^1 dy \left(-\frac{\vectr{$q$}^2}{4}(1-2x+xy) -
Q_f^2(1-xy) + qQ_fz(1-x)\right)
\nonumber \\
&& \hspace{3.1cm}{\mathcal M_{\mathrm {II}}}
\left(-\frac{\vectr{$q$}^2}{4}(1-2x+xy) + Q_f^2(1-xy) + qQ_fzx(1-y)\right)
\nonumber \\
&& \quad\quad\quad\:\:\:\:
\left. +\:\omega_a\omega_b\,k_0^2 \int_0^1 dx\,x\int_0^1 dy
\:{\mathcal M_{\mathrm {II}}}\right\} + T_{\mathrm{exc}},
\label{compl}
\end{eqnarray}
where the matrix elements are given by
\begin{eqnarray}
{\mathcal M_{\mathrm {I}}} &=& 2\int_0^\infty dr\,u_f(r)u_i(r)\,
\frac{(M_\sigma^2 + \Gamma_\sigma^2/4)^2}{8\pi A}\:
e^{-Ar} \nonumber \\
&& j_0\left(r\sqrt{\frac{\vectr{$q$}^2}{4}(1-2x+xy)^2
+Q_f^2x^2y^2 + qQ_fz(1-2x+xy)xy}\:\right), \\
{\mathcal M_{\mathrm {II}}} &=& 2\int_0^\infty dr\,u_f(r)u_i(r)\,
\frac{(M_\sigma^2 + \Gamma_\sigma^2/4)^2}{32\pi A^3}\:
e^{-Ar}(rA+1) \nonumber \\
&& j_0\left(r\sqrt{\frac{\vectr{$q$}^2}{4}(1-2x+xy)^2
+Q_f^2x^2y^2 + qQ_fz(1-2x+xy)xy}\:\right).
\end{eqnarray}
Here the term proportional to $k_0^2$ is again only of minor importance.
Note that in order to obtain the contribution $T_{\mathrm{exc}}$ to
eq.~(\ref{compl}), it is necessary to make the substitution $k_a  
\leftrightarrow k_b$, which implies $\omega_a \leftrightarrow \omega_b$
and $Q_f \rightarrow -Q_f$. In the above matrix elements, the quantity $A$
is now defined as $A = \sqrt{m_*^2 - K_0^2}$ and involves an effective mass
$m_*$ and an energy transfer variable $K_0$. These are 
abbreviations of the following expressions:
\pagebreak
\begin{eqnarray}
m_*^2 &=& \left(M_\sigma^2 + \frac{\Gamma_\sigma^2}{4}\right)(1-xy)
- m_\pi^{\,2}\:x\left(2(1-x)(1-y) -xy^2\right) \\ 
&&\hspace{3.4cm} +\,2\left(\frac{\vectr{$q$}^2}{4} - Q_f^2\right)
x(1-x)(1-y), \nonumber \\
K_0 &=& \omega_a(1-x) -\omega_b x(1-y) + k_0,
\end{eqnarray}
where $M_\sigma$ and $k_0$ are defined as $m_\sigma-i\Gamma_\sigma/2$ and 
$(\omega_b - \omega_a)/2$, respectively. Numerical comparison of the above 
formalism with the approximate model of paper~{\bf IV} indicates that 
eq.~(\ref{2q3}) is accurate to within $\sim 3\%$.

\noindent
All pion rescattering matrix elements have, in paper~{\bf IV}, been 
considered in the nonrelativistic limit, even though it was noted that 
that limit is not realistic in the case of the single quark amplitudes. 
This treatment is expected to be permissible in the case of two-quark amplitudes, 
since the Yukawa functions from the propagators of the exchanged 
pions and $\sigma$ mesons cancel the orthogonality of the radial $S$-wave 
quarkonium wave functions. Relativistic effects thus constitute only a 
correction to the pion rescattering matrix elements, which is expected to 
be rather small because of the large constituent masses of the charm and 
bottom quarks.

\section{The transitions $\Upsilon' \rightarrow \Upsilon\,\pi\pi$ and 
$\psi' \rightarrow J/\psi\,\pi\pi$}

If a $\sigma$ meson lighter than about 1 GeV is employed, together with a 
relativistic treatment of the single quark amplitudes, then the effects of 
pion rescattering  diagrams may be reduced, and they may become 
subdominant as 
compared to the single quark amplitudes, which allows for agreement with 
experiment. The results of paper~{\bf IV} indicate that a $\sigma$ mass of 
$\sim 500$ MeV gives a favorable description of the present experimental 
data on the $\pi\pi$ transitions from the $2S$ states of heavy quarkonia. 

\noindent
The calculated widths and $\pi\pi$ energy distributions were obtained by 
simultaneously optimizing the results for $\Upsilon' \rightarrow  
\Upsilon\,\pi^+\pi^-$ and $\psi' \rightarrow J/\psi\,\pi^+\pi^-$. The best 
results from paper~{\bf IV}, which yielded the $\sigma$ meson parameters 
$m_\sigma$ = 450 MeV and $\Gamma_\sigma$ = 550 MeV, are shown in 
Table~\ref{restab} and Fig.~\ref{expdat}. A coupling constant close 
to $\lambda = -0.02\:\mathrm{fm}^3$ was found to provide an optimal 
description of the 
$\pi\pi$ widths and energy spectra, where the sensitivity to the 
sign of $\lambda$ is due to the consideration of pion rescattering 
amplitudes. The heavy quark masses correspond to those of the $Q\bar Q$ 
Hamiltonian model in paper~{\bf IV}.

\begin{table}[h!]
\centering
\caption{Experimental data~\cite{PDG} and calculated widths for $\pi\pi$ 
transitions from the $\Upsilon'$ and $\psi'$ states, for $\lambda = 
-0.02\:\mathrm{fm}^3$, $m_\sigma$ = 450 MeV and $\Gamma_\sigma$ = 550 MeV, 
together with experimental widths, branching fractions, and 
maximal momenta for each transition.} 
\vspace{.3cm}
\small
\begin{tabular}{c||c|c|c|c|c}
Transition & $\Gamma_{\mathrm {tot}}$ & \% of $\Gamma_{\mathrm {tot}}$
& $\Gamma_{\mathrm {exp}}$ & $\Gamma_{\mathrm {calc}}$ & $q_{\mathrm 
{max}}$ \\ \hline\hline
& & & & & \\
$\Upsilon'\rightarrow \Upsilon\:\pi^+\pi^-$ & 44 $\pm$ 7 keV &
18.8 $\pm$ 0.6 \% & 8.3 $\pm$ 1.3 keV & {\bf 5.89 keV} & 475 MeV \\ 
$\Upsilon'\rightarrow \Upsilon\:\pi^0\pi^0$ &                &  
9.0 $\pm$ 0.8 \%  & 4.0 $\pm$ 0.8 keV & {\bf 3.07 keV} & 480 MeV \\
& & & & & \\
$\psi'\rightarrow J/\psi\:\pi^+\pi^-$ & 277 $\pm$ 31 keV     &
31.0 $\pm$ 2.8 \% & 86 $\pm$ 12 keV   & {\bf 53.5 keV} & 477 MeV \\
$\psi'\rightarrow J/\psi\:\pi^0\pi^0$ &                      &
18.2 $\pm$ 2.3 \% & 50 $\pm$ 10 keV   & {\bf 27.8 keV} & 481 MeV \\
\end{tabular}
\label{restab} 
\end{table}

\newpage

\begin{figure}[h!]
\centering
\epsfig{file = 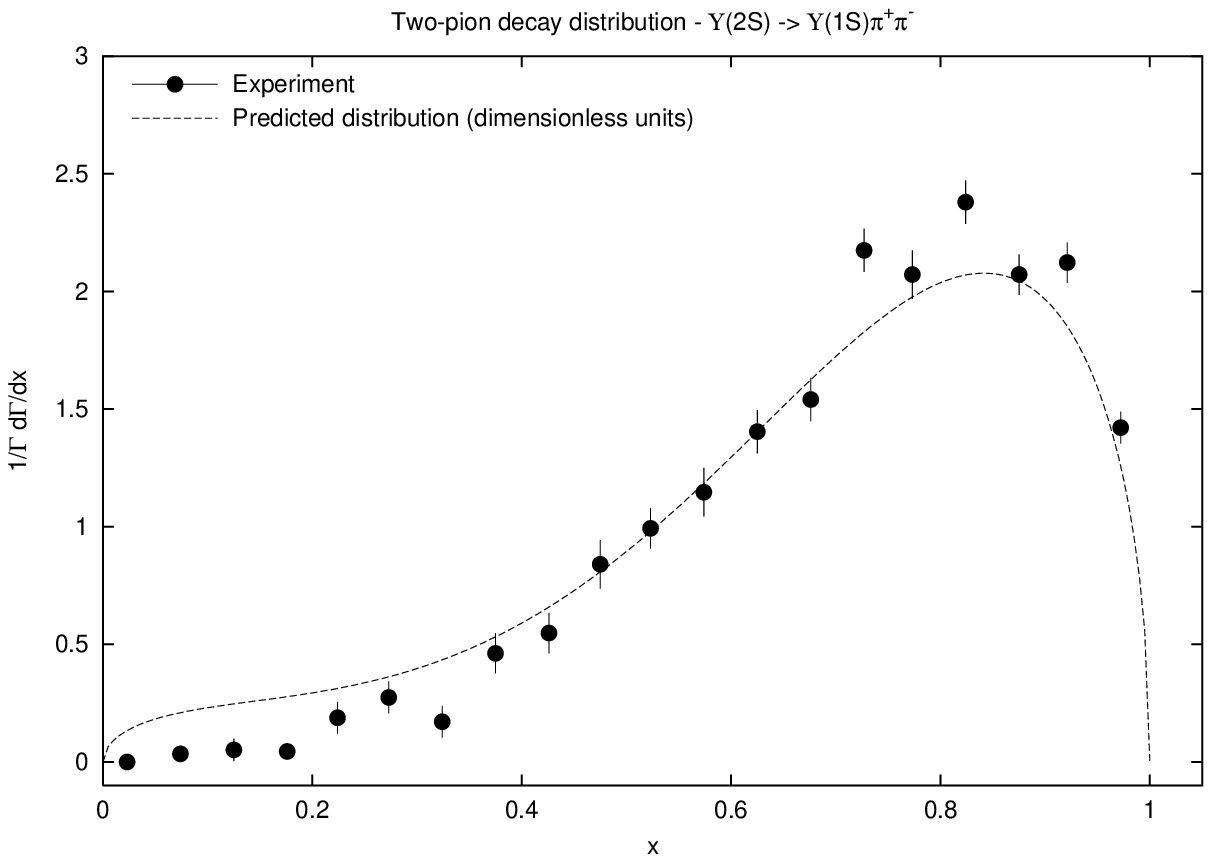, height = 7.5cm}
\epsfig{file = 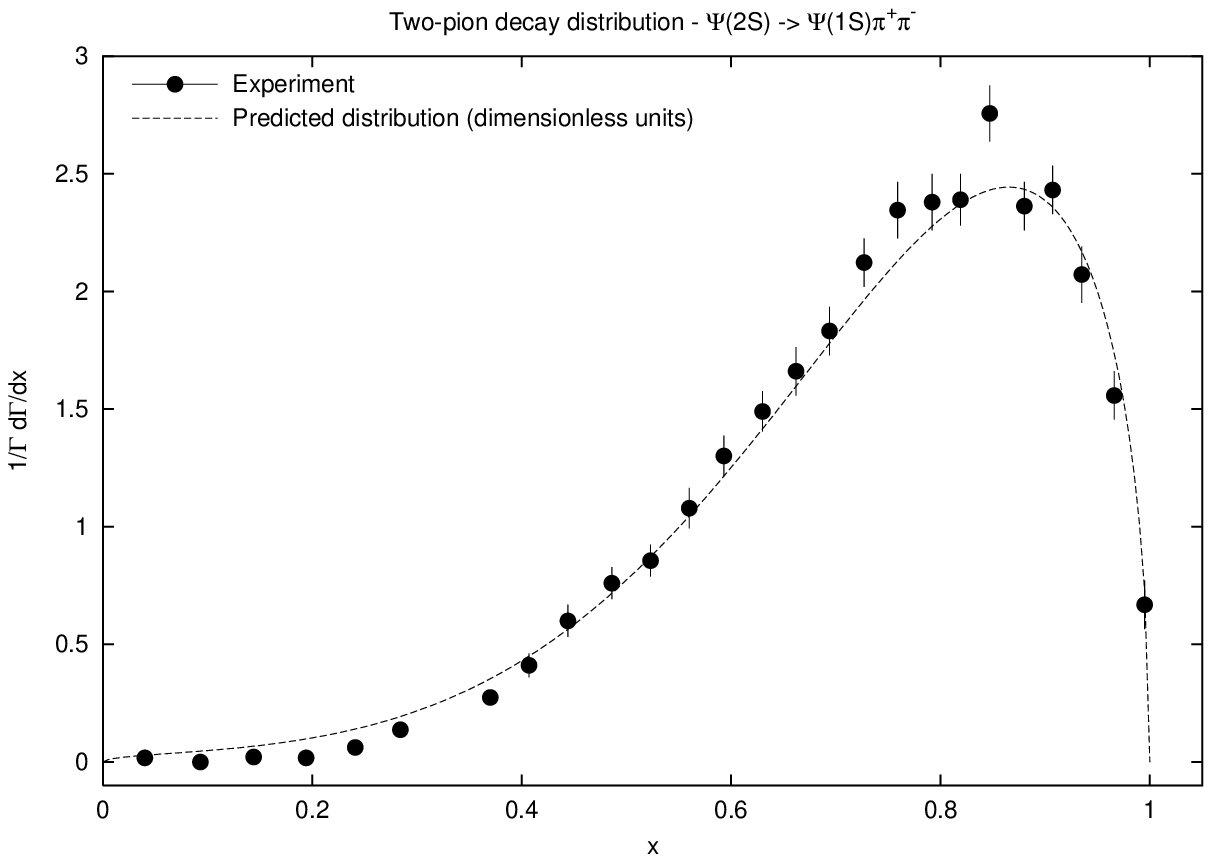, height = 7.5cm}
\caption{Comparison of calculated and experimental~\cite{bbexp} $\pi\pi$
energy \mbox{distributions} for
$\Upsilon' \rightarrow \Upsilon\pi^+\pi^-$ and $\psi' \rightarrow
J/\psi\,\pi^+\pi^-$, for $m_\sigma$ = 450 MeV, $\Gamma_\sigma$ = 550 MeV
and \mbox{$\lambda = -0.02\:\mathrm{fm}^3$.} The calculated width for
$\pi^+\pi^-$ decay is 5.89 keV for $b\bar b$ and 53.5 keV for $c\bar
c$. The scaled $\pi\pi$ invariant mass $x$ is defined in
eq.~(\ref{xvar}).}
\label{expdat}
\end{figure}

\noindent
The results presented in Fig.~\ref{expdat} indicate that the shapes of the 
experimental $\pi\pi$ spectra for $b\bar b$ and $c\bar c$ are slightly 
different. In particular, the peak at high $x$ appears somewhat 
lower for $b\bar b$, while the tail at low $x$ is more pronounced for $b\bar 
b$. It was therefore noted in ref.~\cite{bbexp} that the resonance model of 
ref.~\cite{Schwinger} cannot be simultaneously fitted to both the $b\bar b$ 
and $c\bar c$ data. This is because the shape of the $\pi\pi$ energy 
distribution is rather insensitive to the properties of the $\sigma$ meson 
when only amplitudes of the single quark type are considered. Thus widely 
different masses and widths of the $\sigma$ meson have to be employed 
to fit the empirical $\pi\pi$ energy spectra in the $c\bar c$ and $b\bar b$ 
systems.

\newpage

\noindent
However, if pion rescattering amplitudes are considered as well, then a 
$\sigma$ mass higher than about 450~MeV was shown in paper~{\bf IV} to lead 
to unrealistically large pion rescattering contributions, as the $\pi\pi$ 
energy spectrum then begins to develop a second peak at low $x$. As the 
pion rescattering effects are of a short-ranged character, then it 
turns out that they are significant only for the $\pi\pi$ spectrum in 
the bottomonium system. While the pion rescattering contribution is seen 
from Fig.~\ref{expdat} to account for the qualitative differences between 
the $\pi\pi$ spectra for $b\bar b$ and $c\bar c$, it nevertheless appears to 
be somewhat overpredicted. This problem can be traced to the nonrelativistic 
treatment of the pion rescattering contribution, and may be alleviated if 
relativistic matrix elements are employed, as discussed in paper~{\bf IV}.

\noindent
It may be seen from Table~\ref{restab} that the $\pi\pi$ widths of the 
$b\bar b$ system are somewhat underpredicted. This is because the presence 
of the pion rescattering terms precludes an independent fit of the width and 
the $\pi\pi$ spectrum. A larger value of $\lambda$, which would improve the 
$\pi\pi$ widths, would then worsen the description of the $\pi\pi$ spectrum 
presented in Fig.~\ref{expdat}. However, in the case of the transition 
$\psi'\rightarrow J/\psi\:\pi^+\pi^-$, it may actually be desirable to 
employ a $20-30\,\%$ larger value of $\lambda$, as was also suggested in 
ref.~\cite{Mannel}.

\begin{figure}[h!]
\centering
\epsfig{file = 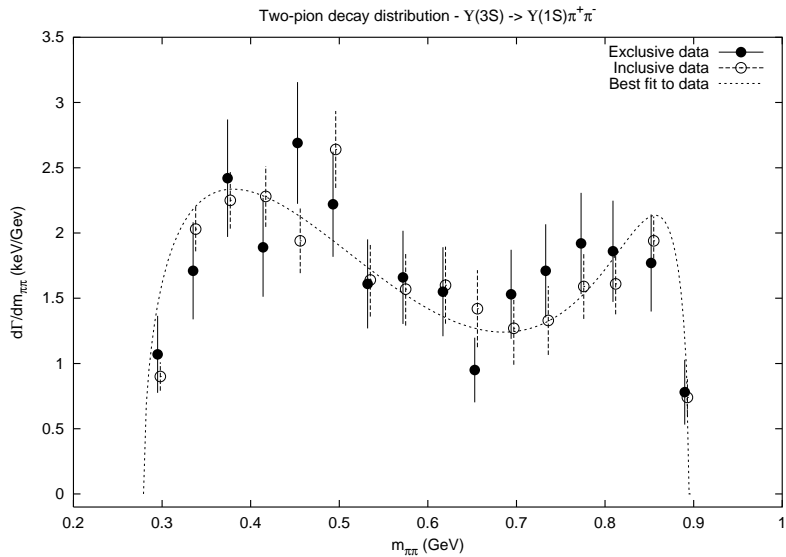, height = 7.5cm}
\caption{The empirical double-peaked $\Upsilon(3S)\rightarrow 
\Upsilon\,\pi\pi$ spectrum, fitted in paper~{\bf IV} by the parameters 
$m_\sigma$ = 1400~MeV, $\Gamma_\sigma$ = 100 MeV, $\lambda = 2.7\cdot      
10^{-3}$, yielding a width of $\Gamma_{\pi^+\pi^-}$ = 1.07 keV.}
\label{3Sfig}
\end{figure}

\noindent
An outstanding problem, lately verified by experimental 
reanalysis~\cite{Pedlar}, is the double-peaked structure of the empirical 
$\Upsilon(3S)\rightarrow \Upsilon\,\pi\pi$ spectrum, which cannot be 
explained by models dominated by single-quark amplitudes, such as the one 
employed in paper~{\bf IV}. However, it was also shown that single-quark 
+~pion rescattering models do in fact have sufficient freedom to accommodate 
a double-peaked $\pi\pi$ spectrum, as may be seen from Fig.~\ref{3Sfig}. As 
a much heavier scalar meson is employed, the contributions from the pion 
rescattering and single quark amplitudes are of equal magnitude. 
Incidentally, the best fit parameters fall within the range of the empirical 
scalar resonances $f_0$(1370) and $f_0$(1500)~\cite{PDG}, both of which 
possess a strong coupling to $\pi\pi$.

\newpage

\chapter{Conclusions}

The electromagnetic and pionic transitions in the heavy $Q\bar Q$ and 
heavy-light $Q\bar q$ mesons have been calculated within the 
framework of the covariant BSLT equation, with the assumption that the 
quark-antiquark interaction can be modeled in terms of a long-range 
confining interaction and a short-ranged OGE or instanton induced interaction. 
It has been demonstrated that a reasonable description of the empirical 
heavy meson spectra can be achieved within such an approach, which also 
yields reasonable values for the confining string tension, constituent quark 
masses and the parameters $\Lambda_{\mathrm{QCD}}$ and $m_g$ which control 
the strong coupling $\alpha_s$. However, the question of the effective 
Lorentz structure of the quark-antiquark interaction cannot be answered by 
the quarkonium spectra alone, since there are many models which produce 
equally satisfactory spectra using different assumptions for the Lorentz 
structure of the effective confining interaction. Numerical lattice QCD is 
unrevealing in this case, as the different components of the $Q\bar q$ 
interaction that can be measured on the lattice may also be well fitted by 
different assumptions for the effective confining interaction.

\noindent
In this situation, a study of electromagnetic and pionic transitions is 
instructive, as two-quark or negative energy contributions have been shown, 
within the framework of e.g. the Schr\"odinger and Gross equations, or the 
instantaneous approximation to the Bethe-Salpeter equation, to be 
significant for several of these transitions. As the 
computation of these effects requires an explicit assumption of the Lorentz 
structure of the confining interaction, then it is possible that the 
question may be settled in the future when adequate experimental data on 
electromagnetic and pionic transitions is available. For the time being, the 
M1 transitions in the charmonium system are the most instructive, as the 
two-quark effects for those transitions have been shown to be large. 
Furthermore, as the OGE interaction does not contribute any two-quark 
operator for M1 transitions in equal-mass quarkonia, then the transition 
$J/\psi\rightarrow\eta_c\gamma$ may provide direct insight into the Lorentz 
structure of the confining interaction. Such a calculation has been 
described in this thesis, where it is found that the two-quark operator 
associated with a scalar confining interaction can explain the observed 
width of about 1~keV.

\noindent
The theoretical importance of the M1 transition 
$J/\psi\rightarrow\eta_c\gamma$ suggests that a new and more accurate 
measurement of the width for that transition should be performed as soon as 
possible. The experimental situation is similar to that for the E1 
transitions $\chi_{cJ}\rightarrow J/\psi\,\gamma$, which were thought to be 
overpredicted by most model calculations for a long time, until the issue 
was resolved by the new data on these E1 transitions reported in the latest 
edition of the PDG~\cite{NewPDG}. Thus, in order to avoid prolonged 
speculation, a new and independent measurement of 
$J/\psi\rightarrow\eta_c\gamma$ is desirable. More empirical information is 
also needed on the total widths of the $b\bar b$ states and their 
E1 branching fractions, as the E1 widths are very sensitive to the $Q\bar Q$ 
wave functions, if not to the Lorentz structure of the quark-antiquark 
interaction. Progress has recently been made for the E1 transitions in the 
$b\bar b$ and $c\bar c$ systems as well, since models that employ 
nonperturbative hyperfine interactions have become available. It was found 
in the calculation reported in this thesis that many of the E1 transitions 
in the bottomonium system cannot be accurately modeled with spin-averaged 
wave functions.

\noindent
A realistic description of the heavy-light $Q\bar q$ mesons, most notably 
the charmed $D$ meson, presents much more serious theoretical challenges for 
a number of reasons, 
most notably the uncertain composition of the $Q\bar q$ interaction, the 
relativistic nature of the light constituent quark and the limited 
empirical knowledge of the excitation spectra. It has nevertheless been 
demonstrated in this thesis that the M1 transitions provide an instructive 
test for the Lorentz structure of the $Q\bar q$ interaction Hamiltonian. 
Promising results have been obtained for a Hamiltonian with scalar 
confining and vector OGE components, possibly augmented by an instanton 
induced interaction. The single pion transitions of the $D$ meson are 
likewise instructive in this 
sense since the two-quark contributions to the axial charge component of the 
transition amplitude are large. Consequently, transitions that involve 
$S$-wave pion emission are predicted to be suppressed, an effect which has 
also been observed within the framework of the Gross equation. 

\noindent
The recently measured total width of the $D^*$ vector meson has been 
shown to provide useful and constraining information on the pion-quark axial 
coupling constant $g_A^q$. Also, the flavor symmetry violating 
$D_s^*\rightarrow D_s\pi^0$ transition can not only provide constraining 
information on the magnitude of $\pi^0 - \eta$ mixing, but can also be used 
to estimate the magnitude of the $\eta$-quark, and in particular, the 
$\eta$-nucleon coupling $f_{\eta NN}$. In addition to single pion emission, 
$\pi\pi$ transitions have also been found to contribute significantly to the 
predicted total widths of the $L=1$ $D$ mesons, a situation which is similar 
to that in the well explored $K$ meson spectrum. These $\pi\pi$ transitions 
were found to be very sensitive, both to the value of $g_A^q$, as well as 
to the hyperfine splittings in the $D$ meson spectrum. In all likelihood, 
the $\pi\pi$ transitions will be found to contribute several MeV to the 
total widths for strong decay. The empirically strong $\pi\pi$ transitions 
in the charmonium and bottomonium systems have been investigated by means 
of a phenomenological model, where the $Q\pi\pi$ interaction is mediated 
by a broad scalar $\sigma$ meson. It has been shown that such a model can 
explain several features of the empirically observed $\pi\pi$ transitions 
in heavy quarkonia, although a completely satisfactory description was not 
achieved.

\noindent
An important conclusion reached in this thesis is that the most instructive 
test for a given $Q\bar Q$ or $Q\bar q$ interaction Hamiltonian is not the 
excitation spectrum but rather the radiative M1 transitions between the 
ground state vector and pseudoscalar mesons. Once the question of hyperfine 
splittings and total widths has been settled by experiment, then the pionic 
transitions from the $L=1$ $D$ mesons will provide a similar testing ground. 
So far, a pure scalar confining interaction has passed the above tests, 
although other conceivable forms have not been systematically ruled out.

\newpage

\chapter*{Svenskspr{\aa}kigt sammandrag}
\addcontentsline{toc}{chapter}{Svenskspr{\aa}kigt sammandrag}

Denna avhandling presenterar en utr\"akning av elektromagnetiska och 
starka \"overg{\aa}ngar i mesoner med en ($Q\bar q$) eller tv{\aa} ($Q\bar 
Q$) tunga kvarkar. Dessa mesoner har beskrivits med hj\"alp av den 
kovarianta Blankenbecler-Sugar (BSLT) ekvationen, under antagandet att 
v\"axelverkningen mellan kvarkarna kan beskrivas som summan av en effektiv
fj\"attrande v\"axelverkning med l{\aa}ng r\"ackvidd och en 
kort-r\"ackviddskomponent, vilken ger upphov till hyperfinstruktur i 
mesonernas excitationsspektrum. En dylik växelverkning ger en acceptabel 
beskrivning av det empiriska mass-spektret, ifall 
kort-r\"ackviddskomponenten beskrivs med hjälp av st\"orningsteoretiskt 
gluon-utbyte (OGE) eller en instanton-inducerad v\"axelverkning. 
Emellertid kan mass-spektret inte ensamt avg\"ora den relativistiska 
Lorentz-strukturen hos den fj\"attrande v\"axelverkningen, eftersom flera 
olika ansatser ger ekvivalenta beskrivningar av det empiriska spektret.

\noindent
I denna situation \"ar de elektromagnetiska och starka \"overg{\aa}ngarna 
av stor betydelse, eftersom dessa har visats vara k\"ansliga f\"or 
negativ-energi bidrag till \"overg{\aa}ngsamplituden, vilka i sin tur 
beror uttryckligen p{\aa}~v\"axelverkningens Lorentz-struktur. Dessa 
effekter har, till dags dato, p{\aa}visats inom ett antal teoretiska 
beskrivningar, d\"aribland Schr\"odinger- och Gross-ekvationerna. Eftersom 
utr\"akningen av dessa bidrag till s\"onderfallsvidderna kr\"aver en 
ansats f\"or kvark-antikvark v\"axelverkningens Lorentz-struktur, kan de 
starka och elektromagnetiska s\"onderfallen utg\"ora ett test f\"or olika 
modeller för den fj\"attrande v\"axelverkningen. F\"or \"ogonblicket \"ar 
den magnetiska (M1) dipol\"overg{\aa}ngen $J/\psi\rightarrow 
\eta_c\,\gamma$ av st\"orsta betydelse, eftersom den experimentellt 
upp\"atta vidden p{\aa}~$\sim 1$~keV beskrivs d{\aa}ligt av den 
icke-relativistiska kvarkmodellen, vilken leder till en tredubbel 
\"overuppskattning av detta resultat. Ett av nyckelresultaten i denna 
avhandling \"ar, att en skal\"art kopplad effektiv fj\"attrande 
v\"axelverkning kan f\"orklara den uppm\"atta vidden p{\aa}~$\sim 1$~keV. 
Emellertid b\"or en ny experimentell m\"atning utf\"oras innan en 
definitiv slutsats kan dras av det ovanst{\aa}ende resultatet.

\noindent
En annan slutsats, presenterad i denna avhandling, ang{\aa}r de 
elektriska (E1) dipols\"onderfallen i charmonium ($c\bar c$) och 
bottomonium ($b\bar b$). I detta fall visar det sig att Lorentz-struktren 
hos kvark-antikvark v\"axelverkningen i de flesta fall endast har en 
f\"orsvinnande liten effekt p{\aa}~\"overg{\aa}ngsamplituden. D\"aremot 
\"ar E1 \"overg{\aa}ngarna k\"ansliga f\"or sm{\aa} effekter i mesonernas 
v{\aa}gfunktioner. D\"armed kan en realistisk beskrivning av ett flertal 
s\"onderfall endast uppn{\aa}s, ifall kvark-antikvark v\"axelverkningens 
hyperfinstruktur behandlas fullst\"andigt, vilket \"ar m\"ojligt ifall 
mesonerna beskrivs med hj\"alp av BSLT-ekvationen.

\noindent
J\"amf\"ort med $Q\bar Q$ mesonerna st\"aller en realistisk beskrivning av 
$Q\bar q$ systemet mycket stora krav p{\aa}~de teoretiska modellerna, 
eftersom v\"axelverkningens form mellan tunga och l\"atta kvarker \"ar 
os\"aker, och den l\"atta kvarken i h\"og grad relativistisk. Dessutom 
kompliceras situationen ytterligare av den knapph\"andiga empiriska 
kunskapen om $Q\bar q$ mesonernas mass-spektrum. Inte desto mindre har det 
visats i denna avhandling, att M1 \"overg{\aa}ngarna i $D$ mesonerna 
utg\"or ett viktigt test f\"or $Q\bar q$ v\"axelverkningens 
Lorentz-struktur. Lovande resultat har erh{\aa}llits f\"or en skal\"ar 
fj\"attrande + OGE v\"axelverkning, m\"ojligen med tillsats av en 
instanton-inducerad komponent. De starka pion-s\"onderfallen inom $D$ 
mesonerna \"ar ocks{\aa}~av stort intresse eftersom negativ-energi 
bidragen till den axiala laddningsamplituden \"ar stora. En 
f\"oruts\"agelse presenterad i denna avhandling \"ar att 
pion-\"overg{\aa}ngar drivna av den axiala laddningsamplituden \"ar starkt 
f\"orhindrade, vilket nyligen ocks{\aa}~har observerats med hj\"alp av 
Gross-ekvationen.

\noindent
Den nyligen uppm\"atta totala vidden f\"or den exciterade $D^*$ mesonen 
\"ar av stor betydelse, eftersom den kan fixera v\"ardet p{\aa}~den axiala 
kopplingskonstanten $g_A^q$ f\"or l\"atta kvarkar. Dessutom kan det 
smak-symmetri brytande s\"onderfallet $D_s^*\rightarrow D_s\pi^0$ ge 
information om styrkan hos $\eta$ mesonens koppling till kvarkar och 
baryoner, f\"orutsatt av storleken hos $\eta - \pi^0$ blandningsvinkeln 
\"ar k\"and. Ut\"over s\"onderfall, i vilka endast en pion emitteras, kan 
\"aven tv{\aa}-pion ($\pi\pi$) s\"onderfall vara av betydelse i $D$ 
mesonerna. Denna \mbox{slutsats} \"overensst\"ammer med den experimentella 
situationen i de s\"ara $K$ mesonerna, d\"ar $\pi\pi$ s\"onderfallen \"ar 
v\"al uppm\"atta. I denna avhandling befanns $\pi\pi$ s\"onderfallen vara 
mycket k\"ansliga b{\aa}de f\"or v\"ardet p{\aa}~$g_A^q$ och det 
tillg\"angliga fasrummet. Det \"ar d\"arigenom sannolikt, att $\pi\pi$ 
s\"onderfallen utg\"or flera MeV av de totala vidderna f\"or starkt 
s\"onderfall i $D$ mesonerna. 

\noindent
De empiriskt betydande $\pi\pi$ \"overg{\aa}ngarna i charmonium ($c\bar 
c$) och bottomonium ($b\bar b$) har i denna avhandling unders\"okts med 
hj\"alp av en 
fenomenologisk modell, vilken beskriver kvark-pion v\"axelverkningen med 
hj\"alp av en skal\"ar $\sigma$ resonans. En dylik modell befanns ge en 
god beskrivning av flera egenskaper hos $\pi\pi$ \"overg{\aa}ngarna, 
\"aven om en fullst\"andigt tillfredst\"allande beskrivning inte 
uppn{\aa}ddes.

\noindent
Den viktigaste slutsatsen i denna avhandling \"ar att det mest 
betydelsefulla testet f\"or en given kvark-antikvark 
v\"axelverkningsmodell \"ar inte mass-spektret, utan snarare de 
elektromagnetiska M1 \"overg{\aa}ngarna d\"ar mesonernas 
grundtillst{\aa}nd byter spinn. I framtiden, n\"ar hyperfin-niv{\aa}erna 
och de totala vidderna hos de exciterade $D$ mesonerna \"ar experimentellt 
kartlagda, kommer dessa att utg\"ora ett ytterligare test f\"or de 
ovann\"amnda v\"axelverkningsmodellerna. Hittills har en skal\"ar 
fj\"attrande v\"axelverkning klarat dessa test, \"aven om andra t\"ankbara 
former inte har uteslutits systematiskt. \\ \\

\vspace{-.5cm}
\noindent
\hspace{7cm} Helsingfors, Oktober 2002 \\ 

\vspace{-.6cm}
\noindent
\hspace{7cm} {\it Timo L\"ahde}

\newpage

\chapter*{Suomenkielinen tiivistelm\"a}
\addcontentsline{toc}{chapter}{Suomenkielinen tiivistelm\"a}

T\"am\"a v\"ait\"oskirja k\"asittelee s\"ahk\"omagneettisia ja vahvoja 
siirtymi\"a mesoneissa, jotka koostuvat joko kahdesta raskaasta ($Q\bar 
Q$) tai yhdest\"a raskaasta ja yhdest\"a kevyest\"a ($Q\bar q$) kvarkista. 
N\"ait\"a mesoneja on kuvattu kovariantin Blankenbecler-Sugar (BSLT) 
yht\"al\"on ratkaisuina, olettaen ett\"a kvarkkien v\"alinen vuorovaikutus 
voidaan esitt\"a\"a efektiivisen, pitk\"an kantaman kvarkkeja kahlitsevan 
vuorovaikutuksen ja lyhyen kantaman ylihieno-vuorovaikutuksen summana. 
Mik\"ali lyhyen kantaman vuorovaikutus oletetaan joko \linebreak
h\"airi\"oteoreettiseksi gluonivaihdoksi taikka instantoni-indusoiduksi, 
on tuloksena laadultaan hyv\"aksytt\"av\"a malli mesonien kokeellisille 
viritysspektreille. Valitettavasti viritys\-spektrit eiv\"at yksin\"a\"an 
riit\"a antamaan ratkaisevaa tietoa kahlitsevan vuorovaikutuksen 
suhteellisuusteoreettisesta Lorentz-rakenteesta, sill\"a useampien eri 
oletusten on todettu johtavan samanlaatuisiin spektreihin.

\noindent
S\"ahk\"omagneettisten ja vahvojen vuorovaikutusten merkitys on t\"ass\"a 
tilanteessa eritt\"ain suuri, sill\"a on osoitettu n\"aiden riippuvan 
herk\"asti siirtym\"aamplitudin negatiivisen \linebreak energian 
komponenteista. 
N\"am\"a puolestaan riippuvat eksplisiittisesti kvarkkien v\"alisen 
vuorovaikutuksen Lorentz-rakenteesta. T\"ah\"an johtop\"a\"at\"okseen on 
p\"a\"adytty aikaisemmin m.m. Schr\"odingerin ja Grossin yht\"al\"oiden 
kautta. Koska siirtym\"aamplitudien laskeminen vaatii oletuksen kvarkkien 
v\"alisen vuorovaikutuksen Lorentz-rakenteesta, voivat n\"am\"a 
\linebreak siis toimia testin\"a, mill\"a voidaan vertailla eri mallien 
todenmukaisuutta my\"os silloin, kun viritysspektrien ennustukset ovat 
degeneroituneet. 
T\"all\"a hetkell\"a magneettisen (M1) dipolisiirtym\"an 
$J/\psi\rightarrow\eta_c\,\gamma$ merkitys on hyvin suuri, sill\"a sen 
kokeellinen viivaleveys $\sim 1$ keV ei ole sopusoinnussa 
ep\"arelativistisen ennustuksen kanssa, joka on kolme kertaa suurempi. 
Er\"as t\"am\"an teoksen t\"arkeimmist\"a tuloksista on, ett\"a 
skalaarinen kahlitseva vuorovaikutus tarjoaa mahdollisen selityksen 
yll\"amainitulle kokeelliselle viivaleveydelle. Lopullinen 
johtop\"a\"at\"os vaatii kuitenkin kokeellisen tuloksen varmistamisen 
uudella mittauksella.

\noindent
T\"ass\"a v\"ait\"oskirjassa on my\"os k\"asitelty raskaiden mesonien 
s\"ahk\"oisi\"a (E1) dipolisiirtymi\"a, jolloin on todettu ett\"a 
kvarkkien v\"alisen vuorovaikutuksen Lorentz-rakenteella on 
\linebreak useimmiten 
vain h\"avi\"av\"an pieni vaikutus teoreettisiin viivaleveyksiin. 
Toisaalta E1 siirtym\"at \mbox{ovat} herkki\"a pienille muutoksille mesonien 
aaltofunktioissa. N\"ain ollen onkin t\"ass\"a teoksessa osoitettu, ett\"a 
useita E1 siirtymi\"a voidaan kuvata onnistuneesti vain, mik\"ali 
\mbox{kvarkkien} v\"alinen ylihieno-vuorovaikutus otetaan huomioon 
k\"aytt\"am\"att\"a ensimm\"aisen kertaluvun h\"airi\"oteoriaa. 

\noindent
Mik\"ali mesoniin sis\"altyy kevyt kvarkki, on realistisen mallin 
rakentaminen heti huomattavasti vaikeampaa, koska raskaiden ja kevyiden 
kvarkkien v\"alisen vuorovaikutuksen muoto on eritt\"ain kyseenalainen. 
Lis\"aksi kevyt kvarkki on hyvin relativistinen, ja viritysspektrin 
kokeellinen tuntemus v\"ah\"ainen. T\"ast\"a huolimatta on t\"ass\"a 
teoksessa osoitettu, ett\"a $D$ mesonien M1 siirtym\"at tarjoavat 
mahdollisuuden tutkia $Q\bar q$-vuorovaikutuksen Lorentz-rakennetta, 
jolloin lupaavia tuloksia on saatu edell\"amainituilla 
vuorovaikutusmalleilla. $D$ mesonien vahvat pionisiirtym\"at ovat my\"os 
t\"ass\"a mieless\"a kiinnostavia, sill\"a pionin ja kevyen kvarkin 
v\"alisen vuorovaikutuksen aksiaalinen varauskomponentti riippuu 
vahvasti kvarkkien v\"alisen vuorovaikutuksen Lorentz-rakenteesta. 
T\"ass\"a v\"ait\"oskirjassa on havaittu, ett\"a aksiaalisesta 
varauskomponentista riippuvat siirtym\"at lienev\"at voimakkaasti 
estyneit\"a. Samankaltaiseen johtop\"a\"at\"okseen ollaan 
p\"a\"asty hiljattain my\"os Grossin yht\"al\"on kautta.

\noindent
Eksitoituneen $D^*$ mesonin viivaleveys, josta hiljattain saatiin 
ensimm\"ainen kokeellinen mittaus, on teoreettisesti hyvin t\"arke\"a 
suure, sill\"a sen avulla voidaan m\"a\"ar\"at\"a kevyiden kvarkkien 
aksiaalinen kytkinvakio $g_A^q$. Lis\"aksi voidaan makusymmetriaa rikkovan 
$D_s^*\rightarrow D_s\pi^0$ siirtym\"an avulla tutkia $\eta$ mesonin ja 
kvarkkien (tai baryonien) v\"alist\"a vuorovaikutusta. T\"am\"a 
edellytt\"a\"a, ett\"a neutraalin pionin ja $\eta$ mesonin v\"alinen 
sekoituskulma on tunnettu. Lis\"aksi on t\"ass\"a v\"ait\"oskirjassa 
tutkittu kahden pionin ($\pi\pi$) siirtymi\"a, jotka hyvin 
todenn\"ak\"oisesti ovat merkityksellisi\"a $D$ mesoneissa, mik\"a 
johtop\"a\"at\"os on yht\"apit\"av\"a oudosta $K$ mesonista saadun 
kokeellisen tiedon kanssa. T\"ass\"a teoksessa on osoitettu $D$ mesonien 
$\pi\pi$ siirtymien olevan herkki\"a $g_A^q$:n numeeriselle arvolle sek\"a 
k\"aytett\"aviss\"a olevalle faasiavaruudelle. T\"aten $\pi\pi$ 
siirtymien viivaleveydet ovat todenn\"ak\"oisesti muutaman MeV:n suuruisia.

\noindent
Charmonium ($c\bar c$) ja bottomonium ($b\bar b$) mesonien kokeellisesti 
merkitt\"avi\"a $\pi\pi$ siirtymi\"a on t\"ass\"a v\"ait\"oskirjassa 
tutkittu fenomenologisella mallilla, jossa raskaiden kvarkkien ja pionien 
v\"alist\"a vuorovaikutusta kuvataan skalaarin $\sigma$ resonanssin 
avulla. T\"all\"a mallilla saavutettiin tyydytt\"av\"a, joskaan ei 
t\"aydellinen, kuvaus raskaiden mesonien $\pi\pi$ siirtymist\"a.

\noindent
T\"am\"an v\"ait\"oskirjan t\"arkein johtop\"a\"at\"os on, ett\"a 
s\"ahk\"omagneettiset M1 siirtym\"at, joissa kvarkkien spin muuttuu, 
muodostavat viritysspektrist\"a riippumattoman testin kvarkkeja 
kahlitsevan vuorovaikutuksen Lorentz-rakenteelle. Tulevaisuudessa, 
ylihienorakenteen ja 
viivaleveyksien ollessa tunnettuja, tulevat $D$ mesonien M1 ja 
pionisiirtym\"at tarjoamaan vastaavan testin. T\"ah\"an menness\"a on 
skalaarinen yrite kvarkkeja kahlitsevalle vuorovaikutukselle osoittautunut 
menestyksekk\"a\"aksi, vaikka muita mahdollisia muotoja ei ole 
j\"arjestelm\"allisesti eliminoitu. \\ \\

\vspace{-.5cm}
\noindent
\hspace{7cm} Helsingiss\"a, Lokakuussa 2002 \\

\vspace{-.6cm}
\noindent
\hspace{7cm} {\it Timo L\"ahde}

\end{document}